\documentclass[aps,prd,amssymb,eqsecnum]{revtex4}

\usepackage{graphicx}% Include figure files
\usepackage{bm}% bold math
\usepackage{amsmath}

\begin{document}

\title{
Gravitational self-force on a particle in eccentric orbit \\
around a Schwarzschild black hole}

\author{Leor Barack and Norichika Sago}
\affiliation{
School of Mathematics, University of Southampton, Southampton,
SO17 1BJ, United Kingdom, \\
Yukawa Institute for Theoretical Physics, Kyoto University,
Kyoto 606-8502, Japan}

\date{March 16, 2010}% It is always \today, today,

\begin{abstract}
We present a numerical code for calculating the local
gravitational self-force acting on a pointlike particle
in a generic (bound) geodesic orbit around a Schwarzschild
black hole. The calculation is carried out in the Lorenz gauge:
For a given geodesic orbit, we decompose the Lorenz-gauge
metric perturbation equations (sourced by the delta-function
particle) into tensorial harmonics, and solve for each harmonic
using numerical evolution in the time domain (in 1+1 dimensions).
The physical self-force along the orbit is then obtained via
mode-sum regularization. The total self-force contains a
dissipative piece as well as a conservative piece, and we
describe a simple method for disentangling these two pieces
in a time-domain framework. The dissipative component is responsible
for the loss of orbital energy and angular momentum through
gravitational radiation; as a test of our code we demonstrate that 
the work done by the dissipative component of the computed force is 
precisely balanced by the asymptotic fluxes of energy and angular momentum,
which we extract independently from the wave-zone numerical
solutions. The conservative piece of the self-force does not
affect the time-averaged rate of energy and angular-momentum
loss, but it influences the evolution of the orbital phases;
this piece is calculated here for the first time in eccentric
strong-field orbits. As a first concrete application of our
code we recently reported the value of the shift in the
location and frequency of the innermost stable circular orbit
due to the conservative self-force
[Phys.\ Rev.\ Lett.\ {\bf 102}, 191101 (2009)].
Here we provide full details of this analysis, and discuss
future applications.
\end{abstract}

\maketitle

%%%%%%%%%%%%%%%%%%%%%%%%%%%%%%%%%%%%%%%%%%%%%%%%%%%%%%%%%%
\section{Introduction}
%%%%%%%%%%%%%%%%%%%%%%%%%%%%%%%%%%%%%%%%%%%%%%%%%%%%%%%%%%
The prospects for detecting gravitational waves from the
inspiral of compact objects into massive black holes have
motivated, over the past decade, research in effort to
understand the general-relativistic orbital evolution in
such systems. The underlying elementary theoretical problem
is that of a pointlike mass particle in a strong-field orbit
around a Kerr black hole of a much larger mass. The dynamics
of such systems can be described in a perturbative fashion in
terms of an effective gravitational self-force (SF)
\cite{Mino:1996nk,Quinn:1996am,Poisson:2003nc,Gralla:2008fg,Pound:2009sm};
knowledge of this force is a prerequisite for describing the
precise evolution of the orbit and the phasing of the emitted
gravitational waves. There is an active research program focused
on the development of computational methods and actual working
codes for the SF in Kerr spacetime \cite{Barack:2009ux}. This
research agenda is being pursued in incremental steps,
through exploration of a set of simplified model problems with
increasing complexity and physical relevance. Much of the
initial work has concentrated on a scalar-field toy model
\cite{Burko:2000xx,Barack:2000zq,Detweiler:2002gi,
DiazRivera:2004ik,Barack:2007jh,Haas:2007kz,Vega:2007mc},
but more recently workers have begun to tackle the gravitational case
\cite{Barack:2002ku,Barack:2007tm,Keidl:2006wk,Detweiler:2008ft,Berndtson:2009hp}.
The state of the art is represented by three independent calculations
of the gravitational SF for circular geodesic orbits in
Schwarzschild geometry \cite{Barack:2007tm,Detweiler:2008ft,Berndtson:2009hp}.
These calculations use different analytic and numerical methods
(and they even invoke different physical interpretations of the SF),
but they were shown to be fully consistent with each other
\cite{Sago:2008id,Berndtson:2009hp}. These calculations were also shown to be
consistent with results from post-Newtonian theory in the
weak-field limit \cite{Detweiler:2008ft,Blanchet:2009sd,Damour:2009sm}.

In the current work we extend the analysis of
Ref.\ \cite{Barack:2007tm} (hereafter ``Paper I'')
from the special class of circular geodesics to generic
(bound) geodesics of the Schwarzschild geometry.
This generalization is astrophysically relevant because real
inspirals often remain quite eccentric up until the eventual
plunge into the massive hole \cite{Barack:2003fp}.
At a more fundamental level, the generalization to eccentric
orbits is interesting because it allows us to start exploring
in earnest the conservative effects of the SF---for instance,
how it influences the orbital precession. Eccentric orbits
have already been considered in calculations of the scalar
\cite{Haas:2007kz} and electromagnetic (EM) \cite{Haas:Capra}
SFs by Haas.
While these calculations are of a less direct astrophysical
relevance, they offer an important test-ground for
computational techniques potentially applicable in the
gravitational problem too. Indeed, many elements of our
numerical method take their inspiration from Haas' work.

The numerical code we present here takes as input the two
orbital parameters of an eccentric Schwarzschild geodesic
(the semi-latus rectum and eccentricity, to be defined below),
and returns the value of the Lorenz-gauge gravitational SF
along this geodesic. The dissipative and conservative pieces
of the SF are returned separately. Here we do {\em not}
consider the evolution of the orbit under the effect of the
SF, but leave this important next step for future work.
We envisage using, to this end, a version of the ``osculating
geodesics'' method \cite{Pound:2007th}, which takes as input
the value of the SF along geodesics tangent to the actual
inspiral orbit. A systematic framework for analyzing the
long-term evolution of the inspiral orbits, using multiple-scale
perturbation methods, was recently developed by Hinderer
and Flanagan \cite{Hinderer:2008dm}
(cf.~Sec.~VII of Gralla and Wald \cite{Gralla:2008fg}).

Our strategy is similar to that of Paper I.
Its basic elements are (i) the Lorenz-gauge perturbation
formalism of Barack and Lousto \cite{Barack:2005nr},
(ii) a finite-difference algorithm for numerical
integration of the Lorenz-gauge perturbation equations
in the time domain, and (iii) mode-sum regularization
\cite{Barack:1999wf,Barack:2001bw,Barack:2001gx,Barack:2002bt}.
The perturbation formalism is based on a tensor-harmonic
decomposition of the perturbed Einstein equations in the
Lorenz gauge. The equations are augmented with
``gauge damping'' terms designed to suppress gauge violations
\cite{Barack:2005nr}, and are written as a set of 10 hyperbolic
equations (for certain linear combinations of metric components)
which do not couple at their principal parts. These equations
are sourced by the (tensor-harmonic modes of the) particle's
energy-momentum, modeled with a delta-function distribution
along the specified eccentric geodesic. The equations are
solved numerically mode by mode in the time domain using
characteristic coordinates on a uniform 1+1-dimensions mesh.
The non-radiative monopole and dipole modes cannot be evolved
stably in this manner; instead, we solve for these two modes
separately in the frequency domain, using the
recently introduced ``extended homogeneous solutions''
technique \cite{Barack:2008ms} to cure the irregularity
of the Fourier sum near the particle. The code records
the value of the perturbation modes and their derivatives
along the orbit (each mode has a $C^0$ behavior at the
particle and hence a well-defined value there, as well
as a well-defined ``one-sided'' derivatives). These values
are then fed into the ``mode-sum formula''
\cite{Barack:2001gx}, which returns the physical SF through
mode-by-mode regularization.

One of the primary advantages of the time-domain approach
is that eccentric orbits---even ones with large
eccentricity---are essentially ``as easy'' to deal with as
circular orbits, with computational cost being only a weak
function of the eccentricity \cite{Barton:2008eb}.
Also, a time-domain code for circular orbits can be upgraded
with relative ease to accommodate eccentric orbits
(such a generalization is radically less straightforward in the
frequency domain). Still, there are several important technical
issues which arise in the time-domain upgrade from circular
to eccentric orbits, and need to be addressed. We list some
of these issues below.

\begin{itemize}
\item Most obvious, the computational burden increases
significantly because the parameter space for geodesics
turns from 1D (circular) to 2D (eccentric).
Moreover, for each given geodesic parameters the SF becomes
a function along the orbit (it has a constant value along
a circular geodesic), and one is required to obtain this
function over an entire radial period. The latter becomes
a technical hurdle in situations where the radial period
is very large---e.g., close to the last stable orbit,
or for orbits with very large radii.

\item In paper I we were able to improve the convergence
rate of our finite-difference algorithm using a Richardson-type
extrapolation to the limit of a vanishing numerical
grid-cell size. That was possible because in the circular-orbit
case the numerical mesh could be easily arranged such that the
local discretization error varied smoothly along the orbit.
This cannot be achieved in any simple way when the orbit
is eccentric, and as a result one cannot
implement a similar Richardson extrapolation. The practical
upshot is that one is forced to implement a higher-order
finite-difference scheme: a 2nd-order-convergent
algorithm (as in Paper I) proves insufficient in practice.
For this work we developed an algorithm with a 4th-order
global convergence. The algorithm takes a rather complicated
form near the particle's trajectory, where the field
(the Lorenz-gauge metric perturbation) has discontinuous
derivatives. To somewhat lessen this complexity
(and reduce the number of grid points needed as input for
the finite-difference formula) the algorithm makes use of
suitable junction conditions across the orbit.
The eventual numerical scheme is considerably more
sophisticated---and involved---compared to that of Paper I.

\item
In the mode-sum scheme one first calculates the contribution
to the ``full'' (pre-regularization) force from each
tensorial-harmonic mode of the perturbation, and then
decomposes this into {\it spherical} harmonics. The necessary
input data for the mode-sum formula are the individual
spherical-harmonic contributions. This procedure involves
the implementation of a tensor--scalar coupling formula,
whose details depend on the orbit in question. The coupling
formula simplifies considerably in the circular-orbit case;
it reverts to its full complicated form [Eq.\ (\ref{eq:Flfull})
with Appendix \ref{app:Ffull}] when eccentric orbits are considered.

\item
The computation of the monopole and dipole contributions
to the SF (which we perform in the frequency domain, as
mentioned above) becomes much more involved in the
eccentric-orbit case. First, the spectrum of the orbital
motion now includes all harmonics of the radial frequency,
and one has to calculate and add up sufficiently many of
these harmonics. A second, more technically challenging
complication arises from the fact that the perturbation
becomes a non-smooth function of time across the orbit
(at a given radius), which disrupts the high-frequency
convergence of the Fourier sum at the particle
(a behavior reminiscent of the Gibbs phenomenon).
A general method for circumventing this problem in
frequency-domain calculations was devised recently in
Ref.\ \cite{Barack:2008ms}, and we implement it here
for the first time.

\item
In exploring the physical consequences of the SF it is
useful to split the SF into its dissipative and conservative
pieces, and discuss their corresponding effects in separate.
This splitting is straightforward in the circular-orbit case:
The conservative piece is precisely the (Schwarzschild)
$r$ component of the SF, while the (Schwarzschild)
$t,\varphi$ components exactly account for the entire
dissipative effect. This is no longer true for eccentric
orbits, where each of the Schwarzschild components mixes
up both dissipative and conservative pieces, and it is not
immediately obvious how to extract these pieces individually.
Here we suggest and implement a simple new method for
constructing the dissipative and conservative pieces out of
the computed Schwarzschild components of the SF (without
resorting to a calculation of the advanced perturbation).
The method takes advantage of the general symmetries of
Schwarzschild geodesics.

\end{itemize}

With the computational framework in place, we can start to
explore the physical effects of the gravitational SF.
In this article we concentrate on two such effects.
First, we calculate the loss of orbital energy and angular
momentum, over one radial period, due to the dissipative
piece of the SF. We extract these quantities directly from
the computed SF along the geodesic orbit (for a sample of
orbital parameters). These ``lost'' energy and angular
momentum must be balanced by the total amount of energy
and angular momentum in the gravitational waves radiated
to spatial infinity and into the black hole over a radial
period. We derive formulas for extracting these quantities
from the far-zone and near-horizon numerical Lorenz-gauge
solutions, and demonstrate numerically that they agree well
with the values computed from the local SF. Our values for
the energy and angular momentum losses also agree with those
previously obtained by others using other methods.

The second effect we consider is conservative, and cannot
be inferred indirectly from the asymptotic gravitational
waves: It is the conservative shift in the location and
frequency of the Innermost Stable Circular Orbit (ISCO).
The analysis of the ISCO shift requires knowledge of the
SF along slightly eccentric geodesics near the last stable
orbit, and our code provides the necessary SF data for the
first time. We reported the results in a recent Letter
\cite{Barack:2009ey}, and here we describe our analysis
in full detail. The quantitative determination of the ISCO
shift is important in that it provides a strong-field
benchmark for calibration of approximate
(e.g., post-Newtonian) descriptions of binary inspirals.
Our result for the ISCO frequency shift has already been
incorporated by Lousto {\it et al.}~in their ``empirical''
fitting formula for predicting the remnant mass and spin
parameters in binary mergers \cite{Lousto:2009mf,Lousto:2009ka};
and by Damour \cite{Damour:2009sm} for breaking the degeneracy
between certain unknown parameters of the Effective One Body
(EOB) formalism.

Perhaps of a more direct relevance to the problem of the
phase evolution in binaries with extreme mass-ratio is
the effect of the SF on the periapsis precession of the
eccentric orbit---also a conservative effect. SF corrections
to the precession rate have been analyzed for weak-field
orbits and within the toy model of the EM
SF \cite{Pound:2005fs,Pound:2007ti}, but never before for
the gravitational problem in strong field. Our code generates
the SF data necessary to tackle this problem for the first
time. We leave the detailed analysis of SF precession
effects to a forthcoming paper.

The paper is organized as follows. In Sec.\ II we review
the relevant theoretical background: bound geodesics in
Schwarzschild geometry, the Lorenz-gauge metric perturbation
formulation, and the construction of the SF via the mode-sum
formula. Section III describes our numerical method in detail,
and in Sec.\ IV we present numerical results for a few sample
eccentric orbits, including a ``zoom--whirl'' orbit.
We explain how the dissipative and conservative pieces of the
computed SF can be extracted from the numerical data, and
present these two pieces separately in a few sample cases.
We also analyze the dissipative effect of the SF and
demonstrate the consistency between the dissipated energy
and angular momentum inferred from the local SF, and that
extracted from the asymptotic gravitational waves.
Section V covers the ISCO-shift analysis, and in Sec.\ VI we
summarize and discuss future applications of our code.

Throughout this work we use standard geometrized units
(with $c=G=1$), metric signature ${-}{+}{+}{+}$, and
(unless indicated otherwise)
Schwarzschild coordinates $x^\mu = (t,r,\theta,\varphi)$.

%%%%%%%%%%%%%%%%%%%%%%%%%%%%%%%%%%%%%%%%%%%%%%%%%%%%%%%%%%
\section{Theoretical background}
%%%%%%%%%%%%%%%%%%%%%%%%%%%%%%%%%%%%%%%%%%%%%%%%%%%%%%%%%%

%%%%%%%%%%%%%%%%%%%%%%%%%%%%%%%%%%%%%%%%%%%%%%%%%%%%%%%%%%
\subsection{Eccentric geodesics in Schwarzschild geometry}\label{subsec:orbit}
%%%%%%%%%%%%%%%%%%%%%%%%%%%%%%%%%%%%%%%%%%%%%%%%%%%%%%%%%%
In this work we consider a pointlike particle with mass $\mu$ in a bound
orbit around a Schwarzschild black hole
%whose metric is given by
%\begin{equation}
%ds^2 = g_{\mu\nu}dx^\mu dx^\nu =
%- f(r)dt^2 + f^{-1}(r)dr^2
%+ r^2(d\theta^2 + \sin^2\theta d\varphi^2),
%\end{equation}
%where $f(r)=1-2M/r$ and
of mass $M\gg\mu$. %is the mass of the black hole.
In the limit $\mu\rightarrow 0$ the trajectory of the particle
is a timelike geodesic of the background Schwarzschild spacetime.
We parameterize this geodesic by proper time $\tau$, in the form
$x_{\rm p}^\mu(\tau)=(t_{\rm p}(\tau),r_{\rm p}(\tau),\theta_{\rm p}(\tau),
\varphi_{\rm p}(\tau))$, with corresponding four-velocity
$u^\mu=dx_{\rm p}^\mu/d\tau$.
Without loss of generality we take $\theta_{\rm p}(\tau)=\pi/2$.
The geodesic equations of the particle are given in integrated form as
\begin{equation}
\frac{dt_{\rm p}}{d\tau} = \frac{{\cal E}}{f(r_{\rm p})}, \quad\quad
\frac{d\varphi_{\rm p}}{d\tau} = \frac{{\cal L}}{r_{\rm p}^2}, \quad
\label{eq:geodesic}
\end{equation}
\begin{equation}
\left(\frac{dr_{\rm p}}{d\tau}\right)^2
= {\cal E}^2 - R(r_{\rm p},{\cal L}^2), \quad\quad
R(r,{\cal L}^2) \equiv
f(r)\left(1+\frac{{\cal L}^2}{r^2}\right),
\label{eq:geodesic2}
\end{equation}
where $f\equiv 1-2M/r$, and ${\cal E}\equiv -u_t$ and ${\cal L}\equiv u_{\varphi}$
are the integrals of motion corresponding to the particle's specific energy and
angular momentum.

\begin{figure}[htb]
\includegraphics[width=7cm]{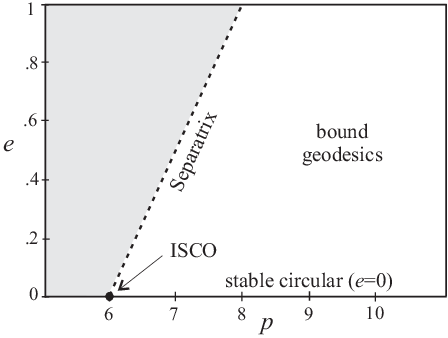}
\caption{Parameter space for bound geodesics in Schwarzschild spacetime.
The (dimensionless) ``semi-latus rectum'' $p$ and ``eccentricity'' $e$ are
defined in Eq.\ (\ref{eq:def-pe}). Bound geodesics have $e\geq 0$ and $p>6+2e$.
Points along the separatrix $p=6+2e$ represent marginally unstable orbits.
Stable circular orbits lie along the axis $e=0$ for $p\geq 6$. The point
$(p,e)=(6,0)$ is the ISCO.
}
\label{fig:pe}
\end{figure}

When ${\cal L}^2>12M^2$ the effective potential of the
radial motion, $R(r,{\cal L}^2)$, has a maximum and a minimum
and hence eccentric (bound) orbits exist.
These orbits can be parametrized by the two values of $r_{\rm p}$ at the
turning points, $r_{\rm min}$ and $r_{\rm max}$ (``periastron''
and ``apastron'', respectively). We may alternatively parameterize
the orbits by the (dimensionless) semi-latus rectum $p$ and eccentricity
$e$, defined through
\begin{equation}
p \equiv
\frac{2r_{{\rm min}}r_{{\rm max}}}
{M(r_{{\rm min}}+r_{{\rm max}})}, \quad
e \equiv
\frac{r_{{\rm max}}-r_{{\rm min}}}
{r_{{\rm max}}+r_{{\rm min}}}.
\label{eq:def-pe}
\end{equation}
From the two conditions $R(r_{\rm min})=R(r_{\rm max})={\cal E}^2$,
one readily obtains
\begin{equation}
{\cal E}^2 =
\frac{(p-2-2e)(p-2+2e)}{p(p-3-e^2)}, \quad
{\cal L}^2 =
\frac{p^2M^2}{p-3-e^2}.
\label{eq:EL_geodesic}
\end{equation}
Bound geodesics have $0\leq e<1$ and $p>6+2e$ \cite{Cutler:1994pb}. Points along the {\it separatrix} $p=6+2e$ (where ${\cal E}^2$ equals the maximum of the effective potential) represent marginally unstable orbits. Stable circular orbits are those with $e=0$ and $p\geq 6$, for which ${\cal E}^2$ equals the minimum of the effective potential. The point $(p,e)=(6,0)$ in the $e$--$p$ plane, where the separatrix intersects the $e=0$ axis,
is referred to as the {\em innermost stable circular orbit}
(ISCO)---see Fig.\ \ref{fig:pe}.

Following Ref.~\cite{Cutler:1994pb}, we introduce the monotonically increasing
``radial phase'' parameter $\chi$, defined so that the radial motion obeys
\begin{equation}
r_{\rm p}(\chi) = \frac{pM}{1+e\cos\chi}.
\label{eq:r-motion}
\end{equation}
Note $\chi=2\pi n$ ($n$ integer) correspond to periastron passages.
In terms of $\chi$, the $t$ and $\varphi$ components of the geodesic equations
(\ref{eq:geodesic}) are re-expressed as
\begin{eqnarray}
\frac{dt_{\rm p}}{d\chi} &=&
\frac{Mp^2}{(p-2-2e\cos\chi)(1+e\cos\chi)^2}
\sqrt{\frac{(p-2-2e)(p-2+2e)}{p-6-2e\cos\chi}},
\label{eq:dt_dchi} \\
\frac{d\varphi_{\rm p}}{d\chi} &=&
\sqrt{\frac{p}{p-6-2e\cos\chi}},
\label{eq:dphi_dchi}
\end{eqnarray}
and the radial velocity reads
\begin{equation}\label{ur}
u^r =
{\cal E} e \sin\chi
\sqrt{\frac{p-6-2e\cos\chi}{(p-2-2e)(p-2+2e)}}.
\end{equation}
The period of the radial motion can be derived by integrating
Eq.~(\ref{eq:dt_dchi}) with respect to $\chi$:
\begin{equation}
T_r \equiv \int_0^{2\pi} \frac{dt_{\rm p}}{d\chi}\,d\chi.
\end{equation}

With the initial conditions $t_{\rm p}=\varphi_{\rm p}=0$ at $\chi=0$, the particle's geodesic
trajectory is fully specified by Eqs.~(\ref{eq:r-motion}), (\ref{eq:dt_dchi})
and (\ref{eq:dphi_dchi}). The functions $t_{\rm p}(\chi)$ and $\varphi_{\rm p}(\chi)$ cannot
be written explicitly in analytic form, but it is easy to obtain them
numerically, for given $p$ and $e$, at any desired accuracy.

%%%%%%%%%%%%%%%%%%%%%%%%%%%%%%%%%%%%%%%%%%%%%%%%%%%%%%%%%%
\subsection{Gravitational self-force via mode-sum regularization}
%%%%%%%%%%%%%%%%%%%%%%%%%%%%%%%%%%%%%%%%%%%%%%%%%%%%%%%%%%
When $\mu$ is finite (yet still much smaller than $M$), the particle
experiences a gravitational SF, $F^{\alpha}[\sim O(\mu^2)]$, and the
equation of motion is formally given by
%~~~~~~~~~~~~~~~~~~~~~~~~~~~~~~~~~~~~~~~~~~~~~~~~~~~~~~~~~~~~~~~~
\begin{equation} \label{eq:EOM}
\mu \frac{D^2\tilde{x}_{\rm p}^{\alpha}}{D\tilde\tau^2}=
\mu \frac{D\tilde{u}^{\alpha}}{D\tilde\tau}=
F^{\alpha}(\tilde\tau).
\end{equation}
%~~~~~~~~~~~~~~~~~~~~~~~~~~~~~~~~~~~~~~~~~~~~~~~~~~~~~~~~~~~~~~~~
Here we use $\tilde{x}_{\rm p}^{\alpha}(\tilde\tau)$ to denote the (non-geodesic) trajectory
under the effect of the SF, with $\tilde\tau$ representing proper time
along this trajectory and $\tilde{u}_{\rm p}^{\alpha}\equiv d\tilde{x}_{\rm p}^{\alpha}/
d\tilde\tau$. The covariant derivatives $D/D\tilde\tau$ are taken with
respect to the background geometry. From symmetry we have $F^{\theta}=0$.
Furthermore, assuming the normalization $\tilde{u}_{\alpha}\tilde{u}^{\alpha}
=-1$, we have the orthogonality condition $\tilde{u}_{\alpha}F^{\alpha}=0$,
which interrelates the remaining 3 components of the SF. All in all, then,
there are two non-trivial independent components of the SF to be determined.

In this work we do not consider the evolution of the orbit
under the effect of the SF, i.e., we do not seek to obtain
consistent solutions of Eq.\ (\ref{eq:EOM}). Rather, we are
interested in calculating the value of the SF $F^{\alpha}(\tau)$
along a fixed, {\em geodesic} orbit $x_{\rm p}^{\alpha}(\tau)$,
with given values of $p,e$. We envisage that the SF information $F^{\alpha}(\tau;p,e)$
(calculated over the space of $p,e$) could be used, in a second step, to calculate
the orbital evolution in situations where at any given time the orbit deviates
only very slightly from a geodesic of the background, and the evolution takes
place over a timescale much longer than the radial period (``adiabatic approximation'').
Here, however, we concentrate on the first step, leaving the investigation of the
orbital evolution for future work.

The gravitational SF acting on the particle at any a point along
the geodesic $x_{\rm p}^{\alpha}(\tau;p,e)$ is calculated using the
mode-sum formula \cite{Barack:1999wf,Barack:2001bw,Barack:2001gx}
\begin{equation}
F^{\alpha} =
\sum_{l=0}^{\infty}\left[F_{{\rm full}\pm}^{\alpha l}
-A^{\alpha}_{\pm}L-B^{\alpha}\right]\equiv
\sum_{l=0}^{\infty} F_{\rm reg}^{\alpha l}.
\label{eq:ModeSum}
\end{equation}
Here $L\equiv l+1/2$, and $F_{\rm full}^{\alpha l}$ are the
multipole modes of the ``full'' force field constructed from
the (physical, retarded) Lorenz-gauge metric perturbation as
prescribed in Sec.~\ref{sec:const_Ffull} below.
The subscript $\pm$ refers to the two possible values of
$F_{\rm full}^{\alpha l}$ at $x_{\rm p}^\alpha$, resulting from
taking one-sided radial derivatives of the metric perturbation
from either $r\to r_{\rm p}^+$ or $r\to r_{\rm p}^-$.
$A^{\alpha}_{\pm}$ and $B^{\alpha}$ are the
``regularization parameters'', given by \cite{Barack:2001gx,Barack:2002bt}
%$A_{\pm}^\theta=B^\theta=0$,
\begin{equation}\label{eq:A}
A_{\pm}^t =
\mp\frac{\mu^2 u^r}{r_{\rm p}^2 f_{\rm p} U},
\quad
A_{\pm}^r =
\mp\frac{\mu^2 {\cal E}}{r_{\rm p}^2 U},
\quad
A_{\pm}^\varphi = 0,
\end{equation}
\begin{eqnarray}\label{eq:B}
B^t &=&
\frac{\mu^2 {\cal E} u^r}{\pi r_{\rm p}^2 f_{\rm p} U^{3/2}}
\left[
 - \hat{K}(w) + 2 ( 1 - U ) \hat{E}(w)
\right], \nonumber \\
B^r &=&
-\frac{\mu^2}{\pi r_{\rm p}^2 U^{3/2}}
\left[
  \left( {\cal E}^2 + f_{\rm p} U \right) \hat{K}(w)
 -\left[ 2 {\cal E}^2 ( 1 - U )
        -f_{\rm p} U ( 1 - 2U )
  \right] \hat{E}(w)
\right], \nonumber \\
B^\varphi &=&
\frac{\mu^2 u^r}{\pi {\cal L} r_{\rm p}^2 \sqrt{U}}
\left[
 \hat{K}(w)
 - \left(1+2\frac{{\cal L}^2}{r_{\rm p}^2}\right)\hat{E}(w)
\right],
\end{eqnarray}
where $\hat K(w)\equiv \int_{0}^{\pi/2}(1-w\sin^2x)^{-1/2}dx$
and $\hat E(w)\equiv \int_{0}^{\pi/2}(1-w\sin^2x)^{1/2}dx$
are complete elliptic integrals of the first and second kind,
respectively, $f_{\rm p}\equiv f(r_{\rm p})=1-2M/r_{\rm p}$, and
\begin{equation}
w=\frac{{\cal L}^2}{r_{\rm p}^2+{\cal L}^2}, \quad\quad
U=1+\frac{{\cal L}^2}{r_{\rm p}^2}.
\end{equation}

It is important to remember that the gravitational SF [as also the trajectory
$x_{\rm p}(\tau)$ itself] is a gauge-dependent entity \cite{Barack:2001ph}.
The mode-sum formula (\ref{eq:ModeSum}) is formulated in the Lorenz-gauge,
and requires as input the modes $F_{\rm full}^{\alpha l}$ derived from
the Lorenz-gauge metric perturbation. In our approach this perturbation
is obtained by tackling the linearized Lorenz-gauge Einstein equations
directly, in the time domain. We proceed by reviewing the relevant
Lorenz-gauge perturbation formalism.

%%%%%%%%%%%%%%%%%%%%%%%%%%%%%%%%%%%%%%%%%%%%%%%%%%%%%%%%%%
\subsection{Metric perturbation in Lorenz gauge}
%%%%%%%%%%%%%%%%%%%%%%%%%%%%%%%%%%%%%%%%%%%%%%%%%%%%%%%%%%

Ref.\ \cite{Barack:2005nr} presented a formulation of the
Lorenz-gauge perturbation equations in Schwarzschild spacetime,
amenable to numerical treatment in the time domain.
In Paper I we applied this formulation (with some minor
modifications) in our study of circular orbits. Here we shall use
the same Lorenz-gauge formulation to obtain our metric perturbation,
and we describe it here as applied to generic eccentric orbits.

Let $g_{\mu\nu}$ be the metric of the background Schwarzschild
geometry, and $h_{\mu\nu}$ be the physical (retarded) metric
perturbation due to the particle moving on the geodesic
$x_{\rm p}^{\alpha}(\tau;p,e)$. We assume $h_{\mu\nu}$ is given
in the Lorenz gauge, i.e., it satisfies
\begin{equation}
\bar{h}_{\mu\nu}{}^{;\nu} \equiv Z_{\mu} = 0,
\label{eq:Lgauge-condition}
\end{equation}
where $\bar{h}_{\mu\nu}=h_{\mu\nu}-(1/2)g_{\mu\nu}h$ is
the trace-reversed perturbation. The corresponding Einstein equations,
linearized in $h_{\mu\nu}$ over $g_{\mu\nu}$, take the compact form
\begin{equation}
\bar{h}_{\mu\nu}{}^{;\alpha}{}_{;\alpha}
+ 2R^\alpha{}_\mu{}^\beta{}_\nu \bar{h}_{\alpha\beta}
= -16\pi T_{\mu\nu},
\label{eq:Lgauge-MPeqs}
\end{equation}
where a semicolon denotes covariant differentiation with respect to
$g_{\mu\nu}$, and $h\equiv g^{\mu\nu}h_{\mu\nu}$ is the trace of $h_{\mu\nu}$.
$T_{\mu\nu}$ on the right-hand side is the energy-momentum tensor
associated with the particle, given by the distribution
\begin{equation}
T_{\mu\nu} =
\mu \int_{-\infty}^{\infty}
\frac{u_{\mu}u_{\nu}\delta^{(4)}(x^\alpha-x_{\rm p}^\alpha)}
{\sqrt{-g}}\,d\tau,
\end{equation}
where $g$ is the determinant of $g_{\mu\nu}$.

It is well known that the hyperbolic set (\ref{eq:Lgauge-MPeqs}) admits
a well-posed initial-value formulation, and that the gauge conditions
(\ref{eq:Lgauge-condition}) are satisfied automatically if only they are
satisfied on the initial (Cauchy or characteristic) surface. However, in
a time-domain numerical implementation of Eqs.\ (\ref{eq:Lgauge-MPeqs})
it is usually impossible to satisfy the gauge conditions (\ref{eq:Lgauge-condition})
{\em precisely} on the initial surface, and, even if one succeeded to do so,
finite differencing errors would usually lead to an uncontrollable violation
of the gauge conditions (unless one somehow actively imposes the gauge
conditions during the evolution). This problem, and its resolution, were
discussed in Ref.\ \cite{Barack:2005nr}, and we follow here the same method.
To the original field equations (\ref{eq:Lgauge-MPeqs}) we add
``divergence dissipation'' terms, in the form
\begin{equation}
\bar{h}_{\mu\nu}{}^{;\alpha}{}_{;\alpha}
+ 2R^\alpha{}_\mu{}^\beta{}_\mu \bar{h}_{\alpha\beta}
+ f'(t_\mu \tilde{Z}_\nu
     + t_\nu \tilde{Z}_\mu)
= -16\pi T_{\mu\nu},
\label{eq:Lgauge-MPeqs2}
\end{equation}
where $t_\mu=(1,f^{-1},0,0)$ and
$\tilde{Z}_\mu=(f Z_r,Z_r,Z_\theta,Z_\varphi)$.
While the inclusion of these extra terms does not (in principle) affect the
solutions of the equations, it guarantees that violations of the gauge
condition are efficiently damped during the numerical time evolution.
[This can be shown by considering the divergence of Eq.\ (\ref{eq:Lgauge-MPeqs2}),
noticing that this yields a hyperbolic equation for $Z_{\mu}$, with a manifest
dissipation term \cite{Barack:2005nr}.]

Owing to the spherical symmetry of the background geometry, the field equations
(\ref{eq:Lgauge-MPeqs2}) are separable into tensorial spherical-harmonics using
the substitution
%~~~~~~~~~~~~~~~~~~~~~~~~~~~~~~~~~~~~~~~~~~~~~~~~~~~~~~~~~~~~~~~~~~~~~~~
\begin{equation} \label{eq:decomp}
\bar h_{\mu\nu}=
\frac{\mu}{r}
\sum_{l=0}^{\infty}\sum_{m=-l}^l\sum_{i=1}^{10}
a_l^{(i)} \bar h^{(i)lm}(r,t)
Y^{(i)lm}_{\mu\nu}(\theta,\varphi;r)
\end{equation}
%~~~~~~~~~~~~~~~~~~~~~~~~~~~~~~~~~~~~~~~~~~~~~~~~~~~~~~~~~~~~~~~~~~~~~~~
(and similarly for the source $T_{\mu\nu}$). The tensorial-harmonic basis
$Y^{(i)lm}_{\mu\nu}$ and normalization factors $a_l^{(i)}$
($i=1,\ldots,10$) are the ones defined in Ref.\
\cite{Barack:2005nr}, except that here (as also in Paper I) we take
$Y^{(3)lm}_{\mu\nu}\to f(r)\times Y^{(3)lm}_{\mu\nu}$
(this modification is needed to achieve $\bar h^{(3)lm}\to$const as $r\to 2M$, in line
with the behavior of all other functions $\bar h^{(i)lm}$). In Appendix
\ref{app:reconst} we give explicit formulae for reconstruction of the various
Schwarzschild components $\bar h_{\mu\nu}$ out of the 10 scalar-like
functions $\bar h^{(i)lm}(r,t)$.

The above substitution reduces the field equations (\ref{eq:Lgauge-MPeqs2})
to the coupled set of 2-dimensional hyperbolic equations
%~~~~~~~~~~~~~~~~~~~~~~~~~~~~~~~~~~~~~~~~~~~~~~~~~~~~~~~~~~~~~~~~~~~~~~~
\begin{equation} \label{eq:field-eqs}
\square \bar h^{(i)lm}+
{\cal M}^{(i)l}_{\;(j)}\bar h^{(j)lm}
= S^{(i)lm}\delta(r-r_{\rm p}) \quad (i=1,\ldots,10).
\end{equation}
%~~~~~~~~~~~~~~~~~~~~~~~~~~~~~~~~~~~~~~~~~~~~~~~~~~~~~~~~~~~~~~~~~~~~~~~
Here a box represents the 2-dimensional scalar-field wave operator
\begin{equation} \label{eq:box}
\square = \partial_{uv}+ V(r), \quad\quad
V(r) = \frac{f}{4r^2}\left[l(l+1) + \frac{2M}{r}\right],
\end{equation}
where $v$ and $u$ are the standard Eddington--Finkelstein null coordinates, defined through $v=t+r_*$ and $u=t-r_*$, with $r_*=r+2M\ln[r/(2M)-1]$.
The terms ${\cal M}^{(i)l}_{\;(j)}\bar h^{(j)lm}$ (summation over $j$ implied)
involve first derivatives of the $\bar h^{(j)lm}$'s at most---hence the principal
part of the set (\ref{eq:field-eqs}) is entirely contained in the term
$\square \bar h^{(i)lm}$.  $S^{(i)lm}$ are the source terms for the point
particle, constructed from the tensor-harmonic coefficients of $T_{\mu\nu}$.
In Appendix \ref{app:field-eqs} we give explicit expressions for both
${\cal M}^{(i)l}_{\;(j)}\bar h^{(j)lm}$ and $S^{(i)lm}$.
The time-radial functions $\bar h^{(i)lm}$ also satisfy four elliptic
equations, which arise from the Lorenz gauge conditions (\ref{eq:Lgauge-condition}).
These relations, too, are given in Appendix \ref{app:field-eqs}.

%%%%%%%%%%%%%%%%%%%%%%%%%%%%%%%%%%%%%%%%%%%%%%%%%
\subsection{Construction of the full force mode}
\label{sec:const_Ffull}
%%%%%%%%%%%%%%%%%%%%%%%%%%%%%%%%%%%%%%%%%%%%%%%%%

Given the Lorenz-gauge metric perturbation $\bar h_{\alpha\beta}$,
the full force modes $F_{{\rm full}\pm}^{\alpha l}$ appearing in the
mode-sum formula (\ref{eq:ModeSum}) are formally constructed as we now prescribe.

First, following \cite{Barack:2001bw}, we define the ``full-force field''
as a tensor field at arbitrary spacetime points $x$, for a given (fixed)
worldline point $x_{\rm p}$ (where the SF is to be calculated):
\begin{equation}\label{eq:Ffull}
F_{\rm full}^{\alpha}(x;x_{\rm p}) =
\mu k^{\alpha\beta\gamma\delta}(x;x_{\rm p})
\bar{h}_{\beta\gamma;\delta}.
\end{equation}
Here the trace-reversed metric perturbation, $\bar h_{\alpha\beta}$,
is evaluated at $x$, and
\begin{equation}\label{eq:k-operator}
k^{\alpha\beta\gamma\delta}(x;x_{\rm p})=
 g^{\alpha\delta}u^{\beta}u^{\gamma}/2
 -g^{\alpha\beta}u^{\gamma}u^{\delta}
 -u^{\alpha}u^{\beta}u^{\gamma}u^{\delta}/2
 +u^{\alpha}g^{\beta\gamma}u^{\delta}/4
 +g^{\alpha\delta}g^{\beta\gamma}/4,
\end{equation}
where $g^{\alpha\delta}$ is the background metric at $x$,
and $u^{\alpha}$ are the values of the contravariant Schwarzschild
components of the four-velocity at $x_{\rm p}$ (treated as fixed coefficients).
In principle, one can choose to extend the quantity $k^{\alpha\beta\gamma\delta}$
off the worldline in any one of many different ways (a few natural choices
are discussed in Ref.\ \cite{Barack:2001gx}). The specific choice made here
is advantageous in that it guarantees a finite mode-coupling in Eq.\
(\ref{eq:Flfull}) below. Our choice of extension does not correspond to any
of the choices made in \cite{Barack:2001gx}, but it can be shown (using the
methods of \cite{Barack:2001gx}) that the regularization parameters associated
with our extension are the same as those of the ``fixed contravariant components''
extension defined in \cite{Barack:2001gx}---these are the parameters whose
values we state above in Eqs.\ (\ref{eq:A}) and (\ref{eq:B}).

In the next step we expand $\bar h_{\alpha\beta}$ in tensor harmonics as in Eq.\
(\ref{eq:decomp}) and substitute in Eq.~(\ref{eq:Ffull}).
Taking the limits $r\to r_{\rm p}$ and $t\to t_{\rm p}$
(but maintaining the $\theta,\varphi$ dependence),
the full force takes the form
\begin{eqnarray} \label{eq:Ffull2}
\left[F_{\rm full}^{\alpha}(\theta,\varphi;r_{\rm p},t_{\rm p})\right]_{\pm}
&=&
\frac{\mu^2}{r_{\rm p}^2}\sum_{l=0}^{\infty}\sum_{m=-l}^{l}
\left\{
f_{0\pm}^{\alpha lm} Y^{lm}
+f_{1\pm}^{\alpha lm} \sin^2\theta\, Y^{lm}
+f_{2\pm}^{\alpha lm} \cos\theta\sin\theta\, Y^{lm}_{,\theta}\right.
\nonumber \\ &&
+\left.f_{3\pm}^{\alpha lm} \sin^2\theta\, Y^{lm}_{,\theta\theta}
+f_{4\pm}^{\alpha}
 (\cos\theta Y^{lm}-\sin\theta Y^{lm}_{,\theta})\right.
\nonumber \\ &&
+\left.f_{5\pm}^{\alpha lm} \sin\theta\, Y^{lm}_{,\theta}
+f_{6\pm}^{\alpha lm} \sin^3\theta\, Y^{lm}_{,\theta}
+f_{7\pm}^{\alpha lm} \cos\theta\sin^2\theta\, Y^{lm}_{,\theta\theta}
\right\},
\end{eqnarray}
where $Y^{lm}(\theta,\varphi)$ are the spherical harmonics,
and the coefficients $f_{n\pm}^{\alpha lm}$ are constructed
from the perturbation fields $\bar h^{(i)lm}$ and their first
$r$ and $t$ derivatives, all evaluated at $x_{\rm p}$.
The labels $+/-$ correspond to taking one-sided derivatives
from $r\to r_{\rm p}^+$ or $r_{\rm p}^-$, respectively.
The explicit expressions for the $f_{n\pm}^{\alpha lm}$'s
are shown in Appendix~\ref{app:Ffull}.

Since the mode-sum formula (\ref{eq:ModeSum}) requires as input the
{\em spherical} harmonic modes of the full force, we must now re-expand
Eq.~(\ref{eq:Ffull2}) in terms of spherical harmonics. With the help of
the identities given in Appendix~\ref{app:iden}, we obtain
%~~~~~~~~~~~~~~~~~~~~~~~~~~~~~~~~~~~~~~~~~~~~~~~~~~~~~~~~~~~~~~~~~~~~~~~
\begin{equation}\label{eq:Flfull}
F_{{\rm full}\pm}^{\alpha l} =
\frac{\mu^2}{r_{\rm p}^2}\sum_{m=-l}^{l} Y^{lm}(\theta_{\rm p},\varphi_{\rm p})
\times\left\{
{\cal F}^{\alpha l-3,m}_{(-3)}
+{\cal F}^{\alpha l-2,m}_{(-2)}
+{\cal F}^{\alpha l-1,m}_{(-1)}
+{\cal F}^{\alpha lm}_{(0)}
+{\cal F}^{\alpha l+1,m}_{(+1)}
+{\cal F}^{\alpha l+2,m}_{(+2)}
+{\cal F}^{\alpha l+3,m}_{(+3)}
\right\}_{\pm},
\end{equation}
%~~~~~~~~~~~~~~~~~~~~~~~~~~~~~~~~~~~~~~~~~~~~~~~~~~~~~~~~~~~~~~~~~~~~~~~
where each of the functions ${\cal F}^{\alpha lm}_{(i)}$ is a certain linear
combination of the $f_{n\pm}^{\alpha lm}$'s (with the same $l,m$)---the explicit
relations are given in Appendix~\ref{app:Ffull}. Hence, in general, a given
full-force mode $F_{{\rm full}\pm}^{\alpha l}$ carries contributions from
tensorial-harmonic functions $\bar{h}_{l'm}^{(i)}$ with $l-3\le l' \le l+3$.
This coupling arises, of course, from the decomposition of the tensor-harmonic
contributions into spherical harmonics.

\subsection{Conservative and dissipative pieces of the SF}
\label{subsec:consdiss}

In analyzing the physical consequences of the SF---as we start to do in
Sec.\ \ref{Sec:ISCO}---it is useful to consider separately its conservative and
dissipative pieces. We therefore now define these two pieces, and obtain
a separate mode-sum formula for each. To this end, we first introduce the
notation $F^{\alpha}\equiv F^{\alpha}_{\rm ret}$, reminding us that the
SF $F^{\alpha}$ is derived from the physical, {\em retarded} metric
perturbation $\bar h_{\alpha\beta}\equiv \bar h_{\alpha\beta}^{\rm ret}$.
We then define the ``advanced'' SF through
\begin{equation}
F^{\alpha}_{\rm adv} =
\sum_{l=0}^{\infty}\left[F_{{\rm full}\pm}^{\alpha l}(\bar h_{\alpha\beta}^{\rm adv})
-A^{\alpha}_{\pm}L-B^{\alpha}\right],
\label{eq:ModeSumAdv}
\end{equation}
where the modes $F_{{\rm full}\pm}$ are constructed precisely as prescribed in
the previous subsection, only this time using the multipole modes of the
{\em advanced} metric perturbation, $\bar h_{\alpha\beta}^{\rm adv}$.
The regularization parameters $A^{\alpha}$ and $B^{\alpha}$ are the same as those
given above for the retarded SF. Since $\bar h_{\alpha\beta}^{\rm ret}$
and $\bar h_{\alpha\beta}^{\rm adv}$ have the same local singular behavior
near the particle \cite{Detweiler:2002mi,Mino:2003yg}, the sum in Eq.\ (\ref{eq:ModeSumAdv}) is guaranteed to
converge.

Following Hinderer and Flanagan \cite{Hinderer:2008dm},
we define the SF's conservative and dissipative components as the parts of the SF which are (correspondingly) 
symmetric and anti-symmetric under ret$\leftrightarrow$adv:
\begin{equation} \label{eq:split}
F^{\alpha}(\equiv F^{\alpha}_{\rm ret})=F_{\rm cons}^\alpha +F_{\rm diss}^\alpha,
\end{equation}
where
\begin{equation} \label{eq:def-con-dis}
F_{\rm cons}^\alpha \equiv
\frac{1}{2}\left(F_{\rm ret}^\alpha + F_{\rm adv}^\alpha\right),
\quad\quad
F_{\rm diss}^\alpha \equiv
\frac{1}{2}\left(F_{\rm ret}^\alpha - F_{\rm adv}^\alpha\right).
\end{equation}
Substituting from Eqs.\ (\ref{eq:ModeSum}) (with $F^{\alpha}\to F^{\alpha}_{\rm ret}$)
and (\ref{eq:ModeSumAdv}), we obtain the mode-sum formulas
\begin{equation} \label{eq:ModeSum-con}
F_{\rm cons}^{\alpha} =
\sum_{l=0}^{\infty}\left[
F_{\rm full(cons)\pm}^{\alpha l}
-A^{\alpha}_{\pm}L-B^{\alpha}\right]
\equiv \sum_{l=0}^{\infty} F_{\rm reg(cons)}^{\alpha l},
\end{equation}
\begin{equation} \label{eq:ModeSum-dis}
F_{\rm diss}^{\alpha} =
\sum_{l=0}^{\infty}F_{\rm full(diss)\pm}^{\alpha l}
\equiv \sum_{l=0}^{\infty} F_{\rm reg(diss)}^{\alpha l},
\end{equation}
where
\begin{equation} \label{eq:Flsplit}
F_{\rm full(cons)\pm}^{\alpha l} \equiv
\frac{1}{2}\left[F_{{\rm full}\pm}^{\alpha l}(\bar h_{\alpha\beta}^{\rm ret})
+F_{{\rm full}\pm}^{\alpha l}(\bar h_{\alpha\beta}^{\rm adv})\right],
\quad\quad
F_{\rm full(diss)\pm}^{\alpha l} \equiv
\frac{1}{2}\left[F_{{\rm full}\pm}^{\alpha l}(\bar h_{\alpha\beta}^{\rm ret})
-F_{{\rm full}\pm}^{\alpha l}(\bar h_{\alpha\beta}^{\rm adv})\right].
\end{equation}
Notice that the dissipative piece of the SF requires no regularization
within the mode-sum scheme.

In Eqs.\ (\ref{eq:ModeSum-con}) and (\ref{eq:ModeSum-dis}) the splitting of
the SF into its conservative and dissipative pieces is performed mode-by-mode.
This is useful in practice, because the $l$-modes of the two pieces exhibit
a rather different large-$l$ behaviour: While $F_{\rm full(cons)\pm}^{\alpha l}$
normally admit an asymptotic power series in $1/l$ [starting at $O(l)$],
the modes of $F_{\rm full(diss)\pm}^{\alpha l}$ die off at large $l$
faster than any power of $l$. We will come back to this issue in
Sec.~\ref{sec:implement-ModeSum}.

The extraction of the conservative and dissipative pieces using Eqs.\
(\ref{eq:ModeSum-con}) and (\ref{eq:ModeSum-dis}) entails a calculation of both
retarded and advanced metric perturbations. This would normally double the
computation time, as it requires one to solve the perturbation equations
twice, changing the boundary conditions in order to obtain
$\bar h_{\alpha\beta}^{\rm adv}$. Fortunately, in the case of a Schwarzschild
background we can avoid this extra computational burden using a simple trick.
For a given eccentric geodesic, we think of the SF as a function of $\tau$
along the orbit. Without loss of generality we take $\tau=0$ to correspond
to a certain periapsis passage [i.e., $r_{\rm p}(\tau=0)=r_{\rm min}$]. Then we
have the following symmetry relation, immediately following from Eq.~(2.80) of
\cite{Hinderer:2008dm}:
\begin{equation} \label{eq:symmetry}
F_{\rm adv}^\alpha(\tau) = \epsilon_{(\alpha)} F_{\rm ret}^\alpha(-\tau)
\end{equation}
(no summation over $\alpha$),
where $\epsilon_{(\alpha)}=(-1,1,1,-1)$ in Schwarzschild coordinates.
This relation can be used to re-express Eqs.\ (\ref{eq:def-con-dis})
in terms of the retarded SF alone, in the form
\begin{equation} \label{eq:trick}
F_{\rm cons}^\alpha(\tau) =
\frac{1}{2}\left[F_{\rm ret}^\alpha(\tau)
     +\epsilon_{(\alpha)} F_{\rm ret}^\alpha(-\tau)\right],
\quad\quad
F_{\rm diss}^\alpha(\tau) =
\frac{1}{2}\left[F_{\rm ret}^\alpha(\tau)
     -\epsilon_{(\alpha)} F_{\rm ret}^\alpha(-\tau)\right].
\end{equation}
Similar relations are applicable to the $l$ modes
$F_{\rm full(cons)\pm}^{\alpha l}$ and $F_{\rm full(cons)\pm}^{\alpha l}$
as well. These relations allow us to extract the conservative and
dissipative pieces of the SF in practice without resorting to
a calculation of the advanced perturbation: All that is required is
knowledge of the physical (retarded) SF along the orbit.\footnote{
The applicability of Eq.\ (\ref{eq:trick}) relies on the validity
of the symmetry relation (\ref{eq:symmetry}); the latter holds whenever
the orbit is periodic with one intrinsic frequency. This is the case
for all bound geodesics in Schwarzschild, and also for equatorial (and
possibly eccentric) or circular (and possibly inclined) orbits in Kerr.
However, the symmetry relation and Eq.\ (\ref{eq:trick}) do not apply
in the Kerr case when the orbit is both eccentric and inclined.}

%%%%%%%%%%%%%%%%%%%%%%%%%%%%%%%%%%%%%%%%%%%%%%%%%%%%%%%%%%
\section{Numerical method}
%%%%%%%%%%%%%%%%%%%%%%%%%%%%%%%%%%%%%%%%%%%%%%%%%%%%%%%%%%
In this section we describe the numerical method used to solve
the field equations (\ref{eq:field-eqs}) and to construct the
local SF along the eccentric orbit. Our numerical scheme is basically similar to that of Paper I---we still use
finite differencing on a characteristic grid in 1+1
dimensions---but we have modified our code in several important
aspects. Most importantly, we abandon the use of a Richardson
extrapolation over the grid size: this technique relies crucially on
the uniformity of the local discretization error along the orbit, which
can no longer be guaranteed in any simple way when dealing with eccentric
orbits. To accelerate the numerical convergence we have instead
upgraded our finite-difference scheme from second-order to fourth-order.
This introduces a significant amount of additional complexity,
especially in the treatment of grid cells traversed by the particle.
Our method inherits from Lousto~\cite{Lousto:2005ip} (fourth-order scheme
for 1+1D evolution with a particle source) and Haas~\cite{Haas:2007kz}
(implementation for a scalar field), but we deviate from their methods
in several aspects.

\subsection{Numerical domain and initial data} \label{sec:domain}

\begin{figure}[htb]
\includegraphics[width=6cm]{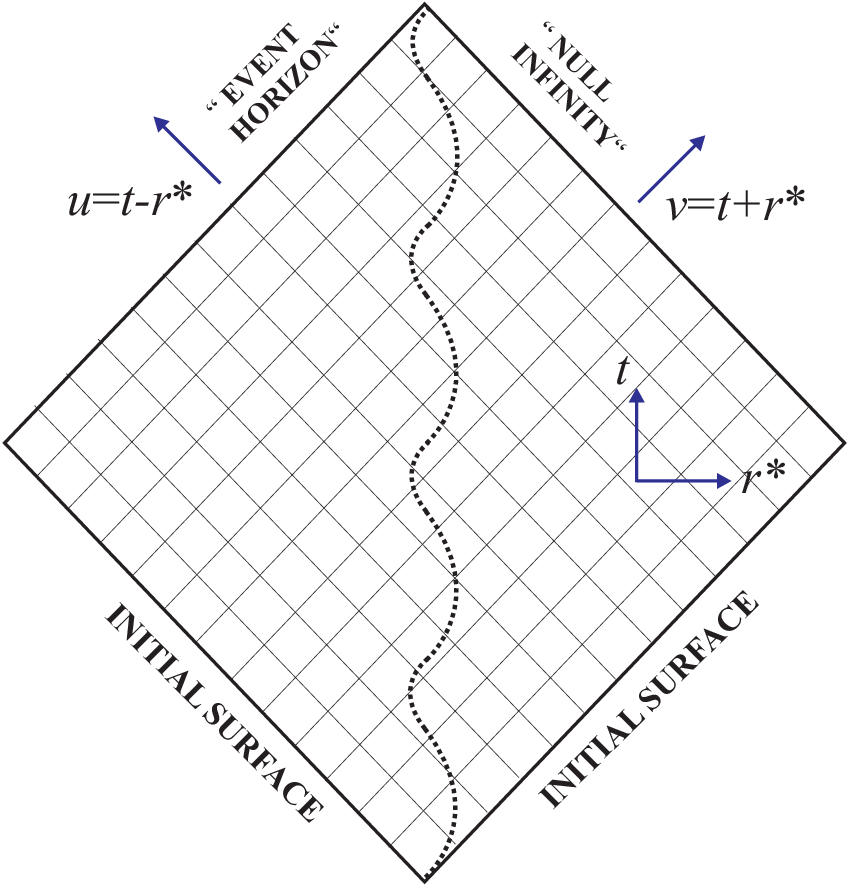}
\caption{Numerical domain: a staggered 1+1-dimensional mesh in null coordinates
$v,u$. $r_*$ is the standard Schwarzschild `tortoise' radial coordinate. The
dotted line represents the trajectory of a typical eccentric orbit. In actual
implementation the mesh is, of course, much finer than it is depicted here.}
\label{fig:domain}
\end{figure}

Our integration domain is discreterized using a two-dimensional uniform
mesh based on the double-null coordinates $v$ and $u$,
as depicted in Fig.~\ref{fig:domain}. The numerical evolution starts with
characteristic initial data
\begin{equation} \label{eq:initial-values}
\bar{h}^{(i)}(u=u_0,v)=\bar{h}^{(i)}(u,v=v_0)=0\quad \text{for all $i$},
\end{equation}
where the vertex $(v_0,u_0)$ corresponds to the particle's location at
$t=0$ [so $v_0=-u_0=r_{*p}(t=0)$ where $r_{*p}=r_*(r_{\rm p})$].
The early stage of the evolution will be dominated by spurious radiation
resulting from the imperfection of the initial data. However, as demonstrated in
\cite{Barack:2005nr} (also in Paper I), these spurious waves damp down rapidly,
and the error related to this behavior becomes negligible at late time.
Our numerical algorithm monitors the residual error from spurious initial
waves by comparing the SF values recorded at regular intervals along
the orbit. We then make sure to evolve long enough for this
error to drop below a set threshold. For a fractional error threshold of
$10^{-4}$ in the final SF we find that the error from the spurious radiation can be
safely ignored after $\sim 2$--$3\times T_r$ of evolution (depending primarily
on the value of $p$; larger $p$ requires a longer evolution).
%~~~~~~~~~~~~~~~~~~~~~~~~~~~~~~~~~~~~~~~~~~~~~~~~~~~~~~~~~~~~~~~~~~~~~~~~~~
%~~~~~~~~~~~~~~~~~~~~~~~~~~~~~~~~~~~~~~~~~~~~~~~~~~~~~~~~~~~~~~~~~~~~~~~~~~

Note that, in our setup, the numerical domain has no causal boundaries.
Therefore, no boundary conditions need be imposed.

\subsection{Finite-difference scheme} \label{subsec:FDS}
\begin{figure}[htb]
\includegraphics[width=7cm]{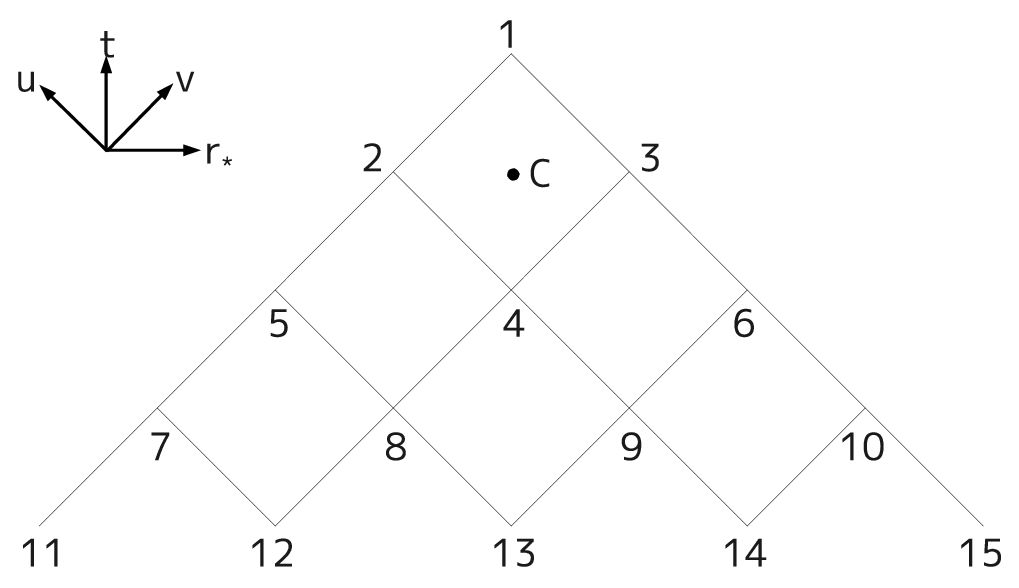}
\caption{Grid points involved in constructing our finite-difference scheme.
The point C is the center of the cell over which we integrate the field
equations, as described in the text.
%Our 2-D grid is based on characteristic
%(Eddington--Finkelstein) coordinates $v$ and
%$u$. These are related to the Schwarzschild coordinates through
%$v=t+r_*$ and $u=t-r_*$, where $r_*=r+2M\ln[r/(2M)-1]$.
The dimensions of each grid cell are $\Delta v\times\Delta u=h\times h$.}
\label{fig:grid}
\end{figure}

To derive our
finite-difference equations, let us focus on a grid cell of dimensions
$\Delta v\times\Delta u=h\times h$---say, the one in Fig.~\ref{fig:grid}
with center C and vertices 1, 2, 3 and 4. We shall assume that the numerical
values of $h^{(i)}$ at points 2--15 are already known, and we wish to derive
the value of $h^{(i)}$ at point 1. To this end we consider the
integral of the field equations (\ref{eq:field-eqs}) over the cell with
center C. The $uv$-derivative term on the left-hand side is integrated in
{\em exact} form to give
%~~~~~~~~~~~~~~~~~~~~~~~~~~~~~~~~~~~~~~~~~~~~~~~~~~~~~~~~~~~~~~~~~~~~~~~~~~
\begin{equation}
\int_{\rm cell} \bar h^{(i)}_{,uv}\, dudv
=
\bar{h}_1^{(i)} - \bar{h}_2^{(i)}
- \bar{h}_3^{(i)} + \bar{h}_4^{(i)},
\label{eq:FDM-uv}
\end{equation}
%~~~~~~~~~~~~~~~~~~~~~~~~~~~~~~~~~~~~~~~~~~~~~~~~~~~~~~~~~~~~~~~~~~~~~~~~~~
where $\bar{h}_n^{(i)}$ denotes the value of $\bar{h}^{(i)}$ at the grid
point labeled `$n$' in Fig.~\ref{fig:grid}.
The integral of the source term on the right-hand side of the field equations
is expressed as
%~~~~~~~~~~~~~~~~~~~~~~~~~~~~~~~~~~~~~~~~~~~~~~~~~~~~~~~~~~~~~~~~~~~~~~~~~~
\begin{eqnarray}
{\cal S}^{(i)} \equiv
\int_{\rm cell} S_{lm}^{(i)}\delta(r-r_{\rm p})\, dudv
&=&
\left\{\begin{array}{ll}
2\int_{t_i}^{t_f}dt f_{\rm p}^{-1}S_{lm}^{(i)}(x_{\rm p}(t)),
   &  \text{(orbit crosses cell)}, \\
0, & \text{(orbit outside cell)}, \\
\end{array} \right.
\label{eq:FDM-source}
\end{eqnarray}
%~~~~~~~~~~~~~~~~~~~~~~~~~~~~~~~~~~~~~~~~~~~~~~~~~~~~~~~~~~~~~~~~~~~~~~~~~~
where $t_i$ and $t_f$ are the values of $t$ at which, correspondingly, the
particle enters and leaves the cell in question. Since the integrand  on the
right-hand side
depends only on the known trajectory of the particle (obtained in advance
by solving the geodesic equation numerically), the integral can be
evaluated in exact form. Our code implements a 5-point closed Newton-Cotes
formula (``Boole's rule'') to evaluate this integral at each grid cell
crossed by the particle.
%~~~~~~~~~~~~~~~~~~~~~~~~~~~~~~~~~~~~~~~~~~~~~~~~~~~~~~~~~~~~~~~~~~~~~~~~~~
%~~~~~~~~~~~~~~~~~~~~~~~~~~~~~~~~~~~~~~~~~~~~~~~~~~~~~~~~~~~~~~~~~~~~~~~~~~

The rest of the terms appearing in the field equations (\ref{eq:field-eqs})
[cf.\ Eqs.\ (\ref{eq:M-term1})--(\ref{eq:M-term10})]
can each be expressed schematically as either $H \equiv \tilde f(r)\bar{h}^{(j)lm}$,
$H_{,v}$ or $H_{,r_*}$, where $\tilde f(r)$ is some known function of $r$,
$j=1,\ldots,10$, and the indices $(j)lm$ are suppressed for brevity.
To complete the formulation of the finite-difference scheme we need to obtain
finite-difference expressions for the integrals
%~~~~~~~~~~~~~~~~~~~~~~~~~~~~~~~~~~~~~~~~~~~~~~~~~~~~~~~~~~~~~~~~~~~~~~~~~~
\begin{equation} \label{eq:Is}
I\equiv \int_{\rm cell} H\, dudv, \quad\quad
I_v\equiv \int_{\rm cell} H_{,v}\, dudv, \quad \text{and}\quad
I_{r_*}\equiv \int_{\rm cell} H_{,r_*}\, dudv.
\end{equation}
%~~~~~~~~~~~~~~~~~~~~~~~~~~~~~~~~~~~~~~~~~~~~~~~~~~~~~~~~~~~~~~~~~~~~~~~~~~
We are aiming here to achieve a quartic [$O(h^4)$] global numerical error
in the fields  $\bar h^{(i)lm}$. Hence, the local finite-difference error for
each of the integrals $I$, $I_v$ and $I_{r_*}$ should not exceed $O(h^6)$.
To formulate the necessary finite-difference relations we will need to
consider separately the following 3 cases (referring again to the
grid cell with center C shown in Fig.~\ref{fig:grid}):
(1) The orbit does not cross the triangular region shown in the figure (``vacuum cell'');
(2) The orbit crosses either the segment 2--11 or the segment 3--15 (``near-orbit cell'');
(3) The orbit crosses either the segment 1--2 or the segment 1--3 (``orbit cell'').

\subsubsection{Vacuum cell}

Consider the formal two-variable Taylor expansion of a typical term $H$
about the center of the cell in question---point C in Fig.~\ref{fig:grid},
with coordinates $v=v_c$ and $u=u_c$. We have
%~~~~~~~~~~~~~~~~~~~~~~~~~~~~~~~~~~~~~~~~~~~~~~~~~~~~~~~~~~~~~~~~~~~~~~~~~~
\begin{equation} \label{eq:F-Taylor}
H(u,v) =
\sum_{a+b=0}^{N}
\frac{c_{ab}}{a!b!} (u-u_c)^a (v-v_c)^b + O(h^{N+1}),
\end{equation}
%~~~~~~~~~~~~~~~~~~~~~~~~~~~~~~~~~~~~~~~~~~~~~~~~~~~~~~~~~~~~~~~~~~~~~~~~~~
where $a$, $b$ and $N$ are non-negative integers (the latter to be specified
below), and $c_{ab}$ are constant coefficients.
Since the desired error in $I$ (the integral of $H$ over the 2D cell) is $O(h^6)$,
we are allowed an error of $O(h^4)$ in $H$ and hence take $N=3$ in
Eq.\ (\ref{eq:F-Taylor}). This leaves us with 10 expansion coefficients $c_{ab}$,
which we can solve for, using Eq.\ (\ref{eq:F-Taylor}), in terms of the values of
$H$ at the 10 points $n=1$--$10$ indicated in Fig.~\ref{fig:grid}.
Substituting the values of these coefficients back in Eq.\ (\ref{eq:F-Taylor})
and integrating over the grid cell, we obtain
%~~~~~~~~~~~~~~~~~~~~~~~~~~~~~~~~~~~~~~~~~~~~~~~~~~~~~~~~~~~~~~~~~~~~~~~~~~
\begin{equation} \label{eq:FDS-MtermI}
I =
\frac{h^2}{24} \left[
2H_1 + 10(H_2+H_3+H_4) -4(H_5+H_6)+(H_7-H_8-H_9+H_{10})
\right] + O(h^6)\quad [\text{vacuum}],
\end{equation}
%~~~~~~~~~~~~~~~~~~~~~~~~~~~~~~~~~~~~~~~~~~~~~~~~~~~~~~~~~~~~~~~~~~~~~~~~~~
where $H_n$ denotes the value of $H$ at grid point $n$.
To evaluate $I_v$ and $I_{r_*}$ at local error $O(h^6)$, we need instead
truncate the Taylor series (\ref{eq:F-Taylor}) at $N=4$, now leaving us with $15$
coefficients $c_{ab}$. These are determined by solving Eq.\ (\ref{eq:F-Taylor})
given the values $H_n$ at the 15 points $n=1$--$15$. Taking $\partial_v$
and $\partial_{r_*}$ in Eq.\ (\ref{eq:F-Taylor}) and integrating over the
grid cell gives
%~~~~~~~~~~~~~~~~~~~~~~~~~~~~~~~~~~~~~~~~~~~~~~~~~~~~~~~~~~~~~~~~~~~~~~~~~~
\begin{eqnarray} \label{eq:FDS-MtermIv}
I_{v}&=&
\frac{h}{24} \left[
9(H_1-H_2) + 19(H_3-H_4)
- 5(H_6-H_9) + (H_{10}-H_{14})
\right] + O(h^6)\quad [\text{vacuum}], \\
I_{r_*}&=& \label{eq:FDS-MtermIx}
\frac{h}{24} \left[
28(H_3-H_2) - 5(H_6-H_5-H_9+H_8) + (H_{10}-H_7-H_{14}+H_{12})
\right] + O(h^6)\quad [\text{vacuum}],
\end{eqnarray}
%~~~~~~~~~~~~~~~~~~~~~~~~~~~~~~~~~~~~~~~~~~~~~~~~~~~~~~~~~~~~~~~~~~~~~~~~~~
which are the desired finite-difference expressions for $I_{v}$ and $I_{r_*}$.

\subsubsection{Near-orbit cell}
%~~~~~~~~~~~~~~~~~~~~~~~~~~~~~~~~~~~~~~~~~~~~~~~~~~~~~~~~~~~~~~~~~~~~~~~~~~
\begin{figure}[htb]
\includegraphics[width=7cm]{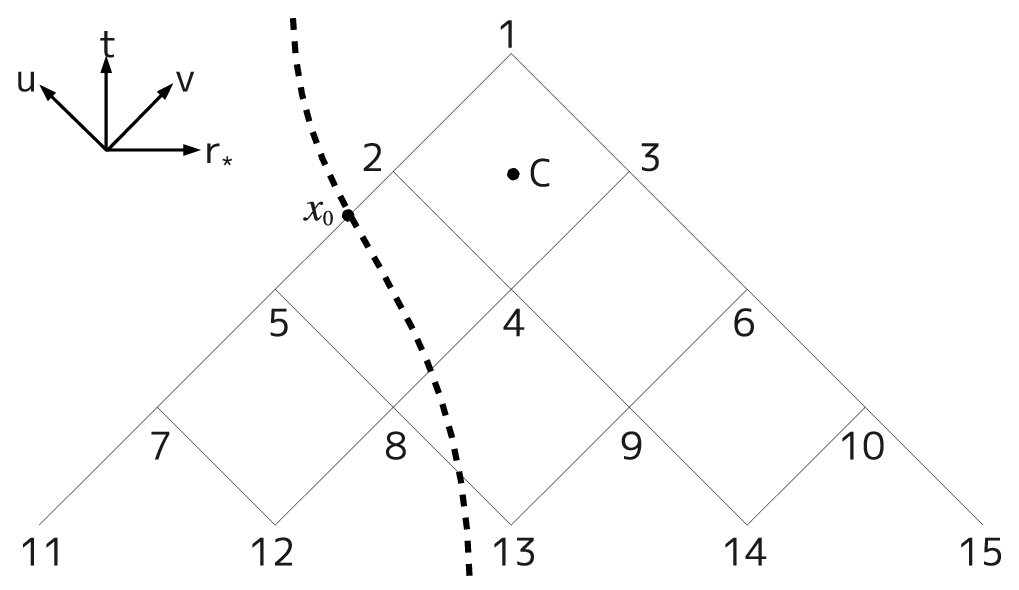}
\caption{Same as in Fig.~\ref{fig:grid}, but now point C is located near
the particle's worldline, represented by the dashed line. The finite-difference
scheme for this case is described in the text under ``near-orbit cell''.}
\label{fig:neighbor-cell}
\end{figure}
%~~~~~~~~~~~~~~~~~~~~~~~~~~~~~~~~~~~~~~~~~~~~~~~~~~~~~~~~~~~~~~~~~~~~~~~~~~
The above derivation assumes that the fields $\bar h^{(i)lm}$ (and hence $H$)
are sufficiently smooth across the region shown in Fig.~\ref{fig:grid}.
This is no longer the case if the worldline of the particle traverses
this region, since $H$ is generally non-differentiable across the worldline,
and the above Taylor-series-based method would fail. Let us first consider
the simpler case, where the particle crosses either the segment 2--11 or
the segment 3--15 (i.e., it does not traverse the integration cell itself),
as demonstrated in Fig.~\ref{fig:neighbor-cell}. In this case the function
$H(u,v)$ can still be expressed as a formal Taylor series,
%~~~~~~~~~~~~~~~~~~~~~~~~~~~~~~~~~~~~~~~~~~~~~~~~~~~~~~~~~~~~~~~~~~~~~~~~~~
\begin{equation} \label{eq:F-Taylor2}
H(u,v) =
\sum_{a+b=0}^{N}
\frac{c_{ab}^{\pm}}{a!b!} (u-u_c)^a (v-v_c)^b
+ O(h^{(N+1)}),
\end{equation}
%~~~~~~~~~~~~~~~~~~~~~~~~~~~~~~~~~~~~~~~~~~~~~~~~~~~~~~~~~~~~~~~~~~~~~~~~~~
but we now have two different sets of expansion coefficients, $c_{ab}^+$ and
$c_{ab}^-$, depending on whether $r(u,v)>r_{\rm p}(t)$ or $r(u,v)<r_{\rm p}(t)$, respectively.
As in the vacuum cell case, the value of $H$ at the grid points 1--15
provides 15 independent equations for the unknown coefficients $c_{ab}^\pm$,
which, however, are now 30 in number. The necessary additional 15 relations
between the various $c_{ab}^\pm$'s are obtained by utilizing
explicit junction conditions for $H$ and its derivatives across the
particle's orbit, as we now describe.

In Appendix \ref{app:jump-condition} we derive explicit jump conditions
for the perturbation $\bar h^{(i)lm}$ and its derivatives at a generic point
$x_0$ along the (known) geodesic worldline. Specifically (and using the
notation of Appendix \ref{app:jump-condition}), we calculate
the two jumps $[\bar h^{(i)lm}_{,u}]_0$ and $[\bar h^{(i)lm}_{,v}]_0$,
as well as the three jumps $[\bar h^{(i)lm}_{,uu}]_0$,
$[\bar h^{(i)lm}_{,uv}]_0$ and $[\bar h^{(i)lm}_{,vv}]_0$,
the 4 jumps in the various third derivatives, and the 5 jumps in
the various fourth derivatives. Together with the continuity condition
$[\bar h^{(i)lm}]_0=0$, we hence obtain 15 jump conditions in total.
Now referring back to our near-orbit cell scenario and to
Fig.~\ref{fig:neighbor-cell}, we take $x_0$ to be the intersection of the
worldline with the past light cone of point 1, as demonstrated in the
figure. The 15 jump conditions for $\bar h^{(i)lm}$ and its derivatives at
$x_0$ readily translate into 15 jump conditions for $H$ and its derivatives
at that point. Imposing these conditions in Eq.\ (\ref{eq:F-Taylor2})
yields the required additional 15 independent linear equations
for the coefficients $c_{ab}^\pm$. Our algorithm solves the total of 30
equations for $c_{ab}^\pm$ numerically, given the numerical values of
$H_n$ at points 1--15.

Once the coefficients $c_{ab}^\pm$ have been calculated, Eq.\ (\ref{eq:F-Taylor2})
can be integrated over the cell of centre C, giving
%~~~~~~~~~~~~~~~~~~~~~~~~~~~~~~~~~~~~~~~~~~~~~~~~~~~~~~~~~~~~~~~~~~~~~~~~~~
\begin{eqnarray}
I &=&
h^2\left[
c_{00}^{\pm}
+ \frac{h^2}{24}
  \left( c_{20}^{\pm} + c_{02}^{\pm} \right)
\right] + O(h^6)\quad [\text{near-orbit}], \\
I_v &=&
h^2\left[
c_{01}^{\pm}
+ \frac{h^2}{24}
  \left( c_{21}^{\pm} + c_{03}^{\pm} \right)
\right] + O(h^6)\quad [\text{near-orbit}], \\
I_{r_*} &=&
h^2\left[
c_{01}^{\pm} - c_{10}^{\pm}
+ \frac{h^2}{24}
  \left( c_{21}^{\pm} - c_{30}^{\pm}
       + c_{03}^{\pm} - c_{12}^{\pm} \right)
\right] + O(h^6)\quad [\text{near-orbit}],
\end{eqnarray}
%~~~~~~~~~~~~~~~~~~~~~~~~~~~~~~~~~~~~~~~~~~~~~~~~~~~~~~~~~~~~~~~~~~~~~~~~~~
where the values $c_{ab}^{+}$ apply if C is located at $r>r_{\rm p}(t)$
and the values $c_{ab}^{-}$ apply if C lies in the region $r<r_{\rm p}(t)$.

\subsubsection{Orbit cell}

\begin{figure}[htb]
\includegraphics[width=10cm]{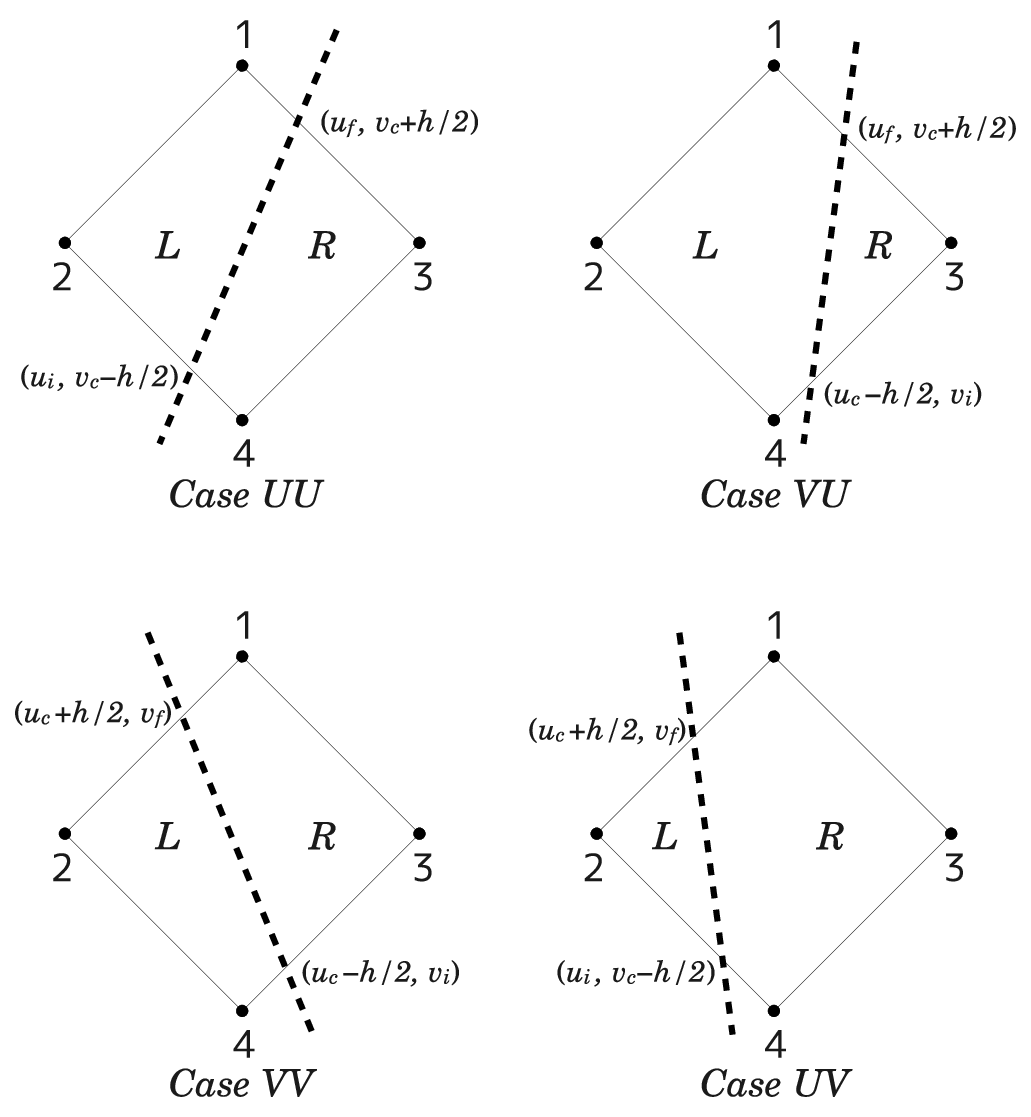}
\caption{Illustration of the four cases considered in formulating
the finite-difference equation for grid cells traversed by the particle's
worldline (here represented by dashed lines). The center of the cell
is at $(u,v)=(u_c,v_c)$, and in each case we indicate the $u,v$ coordinates
of the two points where the particle enters/leaves the cell. }
\label{fig:sourced-cell}
\end{figure}

The procedure described above for calculating the Taylor coefficients
$c_{ab}^{\pm}$ is applicable even when the particle's worldline crosses
the considered grid cell.
However, the integration of $H$ and its derivatives over the grid cell
then becomes slightly more involved, because the cell is
divided by the trajectory into two parts, in each of which the
coefficients take different values.
Following Lousto \cite{Lousto:2005ip}, we consider separately the
four cases illustrated in Fig.~\ref{fig:sourced-cell}.
%~~~~~~~~~~~~~~~~~~~~~~~~~~~~~~~~~~~~~~~~~~~~~~~~~~~~~~~~~~~~~~~~~~~~~~~~~~
%~~~~~~~~~~~~~~~~~~~~~~~~~~~~~~~~~~~~~~~~~~~~~~~~~~~~~~~~~~~~~~~~~~~~~~~~~~

{\it ``Case UU''} (top left in Fig.~\ref{fig:sourced-cell}):
the particle enters the cell crossing the $v$=const segment 2--4,
and leaves it crossing the $v$=const segment 1--3.
The worldline splits the cell into two bits, ``left'' and ``right'',
respectively labeled $L$ and $R$ in the figure. Denoting the corresponding
contributions to $I$ by $I^L$ and $I^R$, we have $I=I^L+I^R$, where,
using the expansion (\ref{eq:F-Taylor2}),
\begin{eqnarray}
I^L &=&
\int_{-h/2}^{h/2} d\bar{v}
\int_{\bar{u}_{\rm p}(\bar{v})}^{h/2} d\bar{u}\, H
=
\sum_{a+b=0}^3 \frac{c_{ab}^-}{(a+1)!b!}
\int_{-h/2}^{h/2} d\bar{v}\,
\bar{v}^b \left[
\left(h/2\right)^{a+1}
- \left(\bar{u}_{\rm p}(\bar{v})\right)^{a+1}
\right] + O(h^6), \\
I^R &=&
\int_{-h/2}^{h/2} d\bar{v}
\int_{-h/2}^{\bar{u}_{\rm p}(\bar{v})} d\bar{u}\, H
=
\sum_{a+b=0}^3 \frac{c_{ab}^+}{(a+1)!b!}
\int_{-h/2}^{h/2} d\bar{v}\,
\bar{v}^b \left[
\left(\bar{u}_{\rm p}(\bar{v})\right)^{a+1}
- \left(-h/2\right)^{a+1}
\right] + O(h^6),
\end{eqnarray}
Here $\bar{u}\equiv u-u_c$ and $\bar{v}\equiv v-v_c$, and
$\bar{u}_{\rm p}(\bar{v})$ represents the value of $\bar{u}$ on the trajectory,
viewed as a function of $\bar{v}$. Similar expressions are easily obtained
for $I_v$ and $I_{r_*}$.

{\it ``Case VU''} (top right in Fig.~\ref{fig:sourced-cell}):
the particle enters the cell crossing the $u$=const segment 3--4,
and leaves it crossing the $v$=const segment 1--3. We denote the
entry $v$-value by $v_i$ and the exit $u$-value by $u_f$, and further
denote $\bar v_i=v_i-v_c$ and $\bar u_f=u_f-u_c$. We obtain, in this
case,
\begin{eqnarray}
I^L &=&
\int_{-h/2}^{h/2} d\bar{v}
\int_{\bar{u}_f}^{h/2} d\bar{u} \, H
+ \int_{-h/2}^{\bar{u}_f} d\bar{u}
  \int_{-h/2}^{\bar{v}_{\rm p}(\bar{u})} d\bar{v} \, H
\nonumber \\ &=&
\sum_{a+b=0}^3 \frac{c_{ab}^-}{(a+1)!(b+1)!}
\Bigg\{
\left(h/2\right)^{b+1}
\left[ \left(h/2\right)^{a+1}
     - \left(\bar{u}_f \right)^{a+1} \right]
\left[ 1+(-1)^b \right]
\nonumber \\ && \hspace*{3cm}
+(a+1)\int_{-h/2}^{\bar{u}_f} d\bar{u}\,
\bar{u}^a \left[
\left(\bar{v}_{\rm p}(\bar{u})\right)^{b+1}
- \left(-h/2\right)^{b+1}
\right] \Bigg\} + O(h^6), \\
I^R &=&
\int_{-h/2}^{\bar{u}_f} d\bar{u}
\int_{\bar{v}_{\rm p}(\bar{u})}^{h/2} d\bar{v} \, H
=
\sum_{a+b=0}^3 \frac{c_{ab}^+}{a!(b+1)!}
\int_{-h/2}^{\bar{u}_f} d\bar{u}\,
\bar{u}^a \left[
\left(h/2\right)^{b+1}
-\left(\bar{v}_{\rm p}(\bar{u})\right)^{b+1}
\right] + O(h^6),
\end{eqnarray}
where $\bar{v}_{\rm p}(\bar{u})$ is the value of $\bar{v}$ on the
trajectory, expressed as a function of $\bar{u}$.
Once again, similar expressions can be obtained for $I_v$ and
$I_{r_*}$.

{\it ``Case VV''} (bottom left in Fig.~\ref{fig:sourced-cell}):
the particle enters the cell crossing the $u$=const segment 3--4,
and leaves it crossing the $u$=const segment 1--2.
In this case we obtain
\begin{eqnarray}
I^L &=&
\int_{-h/2}^{h/2} d\bar{u}
\int_{-h/2}^{\bar{v}_{\rm p}(\bar{u})} d\bar{v}\, H
=
\sum_{a+b=0}^3 \frac{c_{ab}^-}{a!(b+1)!}
\int_{-h/2}^{h/2} d\bar{u}\,
\bar{u}^a \left[
\left(\bar{v}_{\rm p}(\bar{u})\right)^{b+1}
- \left(-h/2\right)^{b+1}
\right], \\
I^R &=&
\int_{-h/2}^{h/2} d\bar{u}\,
\int_{\bar{v}_{\rm p}(\bar{u})}^{h/2} d\bar{v} \, H
=
\sum_{a+b=0}^3 \frac{c_{ab}^+}{a!(b+1)!}
\int_{-h/2}^{h/2} d\bar{u}\,
\bar{u}^a \left[
\left(h/2\right)^{b+1}
- \left(\bar{v}_{\rm p}(\bar{u})\right)^{b+1}
\right],
\end{eqnarray}
and similar expressions for $I_v$ and $I_{r_*}$.

{\it ``Case UV''} (bottom right in Fig.~\ref{fig:sourced-cell}):
the particle enters the cell crossing the $v$=const segment 2--4,
and leaves it crossing the $u$=const segment 1--2.
We denote the entry $u$-value by $u_i$ and the exit $v$-value by $v_f$,
with $\bar u_i=u_i-u_c$ and $\bar v_f=v_f-v_c$.
In this final case we have
\begin{eqnarray}
I^L &=&
\int_{-h/2}^{\bar{v}_f} d\bar{v}
\int_{\bar{u}_{\rm p}(\bar{v})}^{h/2} d\bar{u} \, H
=
\sum_{a+b=0}^3 \frac{c_{ab}^-}{(a+1)!b!}
\int_{-h/2}^{\bar{v}_f} d\bar{v}\,
\bar{v}^b \left[
\left(h/2\right)^{a+1}
-\left(\bar{u}_{\rm p}(\bar{v})\right)^{a+1}
\right], \\
I^R &=&
\int_{\bar{v}_f}^{h/2} d\bar{v}
\int_{-h/2}^{h/2} d\bar{u} \, H
+ \int_{-h/2}^{\bar{v}_f} d\bar{v}
  \int_{-h/2}^{\bar{u}_{\rm p}(\bar{v})} d\bar{u} \, H
\nonumber \\ &=&
\sum_{a+b=0}^3 \frac{c_{ab}^+}{(a+1)!(b+1)!}
\Bigg\{
\left(h/2\right)^{a+1}
\left[ 1+(-1)^a \right]
\left[ \left(h/2\right)^{b+1}
     - \left(\bar{v}_f \right)^{b+1} \right]
\nonumber \\ && \hspace*{3cm}
+(b+1)\int_{-h/2}^{\bar{v}_f} d\bar{v}\,
\bar{v}^b \left[
\left(\bar{u}_{\rm p}(\bar{v})\right)^{a+1}
- \left(-h/2\right)^{a+1}
\right] \Bigg\},
\end{eqnarray}
with similar expressions for $I_v$ and $I_{r_*}$.

\subsubsection{Predictor-corrector method}

In summary, recalling Eq.\ (\ref{eq:FDM-uv}) and with reference
to Fig.\ \ref{fig:grid}, our basic finite-difference formula takes
the form
\begin{equation} \label{fds}
\bar{h}_1^{(i)} =
\bar{h}_2^{(i)} + \bar{h}_3^{(i)} - \bar{h}_4^{(i)}
- \int_{\rm cell} dudv
  \left[ V(r)\bar{h}^{(i)}
         + {\cal M}^{(i)}_{(j)} \bar{h}^{(j)} \right]
+ {\cal S}^{(i)},
\end{equation}
where ${\cal S}^{(i)}$ is given in Eq.\ (\ref{eq:FDM-source}) and the
integral (over the cell with center C in the figure) is evaluated with
local error $O(h^6)$ as discussed above. A complication arises, since our
finite-difference expressions for the integral in Eq.\ (\ref{fds})
involve the value of the perturbation at point 1 itself
[as in, e.g., Eq.\ (\ref{eq:FDS-MtermI})], which is the very
unknown value we wish to compute.

We overcome this difficulty using a type of predictor-corrector
algorithm, whereby we first approximate the value of the field
at the point in question (our point 1) using extrapolation, and
then apply our finite-difference formula (\ref{fds}) iteratively,
until the required accuracy is achieved. Specifically, we use the
values of the perturbation $H$ at the four points 3, 6, 10, and 15
(see Fig.\ \ref{fig:grid}) to extrapolate the value $H_1$ with
an error of $O(h^4)$.
%[One cannot hope to achieve a better extrapolation
%accuracy using more points, because the {\em accumulated} evolution
%error in each of the four points used for the extrapolation is already
%$O(h^4)$---see below.]
If the particle's worldline happens to cross
the null segment 1--15 we instead use the points 2, 5, 7, and 11
for this extrapolation. We then use the value of $H_1$ thus obtained as input
in Eq.\ (\ref{fds}). Since the extrapolated terms enter Eq.\ (\ref{fds})
multiplied by at least one power of $h$ [see, e.g., Eqs.\
(\ref{eq:FDS-MtermI}) and (\ref{eq:FDS-MtermIv})], the resulting value
of $H_1$ would have a local error of $O(h^5)$. We then apply Eq.\ (\ref{fds})
once more, with the new value of $H_1$ as input. The output value
of $H_1$ following this second iteration should now have a local error
of $O(h^6)$ as desired.

The above finite-difference scheme is designed to yield a local error
of $O(h^6)$ in $\bar{h}_1^{(i)}$ at each grid point. Since the
overall number of grid points contributing to the accumulated error
scales as $\sim h^{-2}$, we expect our scheme to show a
quartic [i.e., $O(h^4)$] numerical convergence.

\subsection{Monopole and dipole modes}

In the monopole case ($l=0$) the system of 10 field equations
(\ref{eq:field-eqs}) reduces to 4 equations only; it reduces
to 6 equations for each of the two dipole modes $(l,m)=(1,\pm 1)$
and to a single equation for the dipole mode $(l,m)=(1,0)$.
One may attempt to apply the above numerical evolution scheme
to these non-radiative modes as well. However, numerical
experimentation suggests to us that the monopole and dipole cannot be
evolved stably using this scheme. A naive application of the
evolution scheme yields exponentially growing solutions, and,
since our scheme gives us no handle on the boundary conditions,
the occurrence of these unphysical solutions is difficult to control.

Instead, we deal with the two modes $l=0,1$ using the standard
frequency-domain method, just as in Paper I. The physical Lorenz-gauge
monolpole and dipole are constructed from a basis of homogeneous
frequency-mode solutions of the underlying ordinary differential
equations. These homogeneous solutions are obtained numerically.
However, unlike in the circular-orbit case dealt with in Paper I,
here we face the ``Gibbs phenomenon'', since for an eccentric orbit
the perturbation is a non-differentiable function of coordinate time
$t$ at the particle's location. We have discussed this problem
in depth in Ref.\ \cite{Barack:2008ms}, and proposed an elegant
solution, whereby the correct physical perturbation is constructed as
a sum of ``fake'' frequency-mode solutions whose Fourier sum
converges exponentially even at the particle's position.
Here we apply this method in order to obtain the physical
Lorenz-gauge monopole and dipole modes. A full description of
this construction will be given in a forthcoming paper
\cite{BGS:lowL}.

\subsection{Implementation of the mode-sum scheme}
\label{sec:implement-ModeSum}

Once we obtain the numerical values of the functions $\bar{h}^{(i)}$
and their derivatives along the orbit (over a complete radial period),
we can construct the full-force modes $F^{\alpha l}_{{\rm full}\pm}$
at any point along the orbit through the procedure described in
Sec.~\ref{sec:const_Ffull}. The mode-sum formula (\ref{eq:ModeSum})
then gives the physical SF at that point. The application of the
mode-sum formula involves summation over contributions from an
infinite number of modes, from $l=0$ to $l=\infty$. In reality,
of course, we are only able to compute a small number of low multipole
modes---not least because the numerical calculation
becomes increasingly more demanding with larger $l$.
Since the mode-sum scheme converges rather slowly (as $\sim 1/l$),
a calculation of the SF with even a modest accuracy requires that
we take a proper account of the contribution from the truncated tail
of the mode sum.

Let us denote by $\bar l$ the highest spherical-harmonic $l$-mode
calculated numerically (note this would entail calculating all
{\it tensor-harmonic} modes from $l=0$ to $l=\bar l+3$).
Recalling the notation of Eq.\ (\ref{eq:ModeSum}), we express the SF as
\begin{equation}
F^\alpha =\sum_{l=0}^{\bar l} F_{\rm reg}^{\alpha l}+
\sum_{l=\bar l+1}^{\infty} F_{\rm reg}^{\alpha l}\equiv
F^\alpha_{l\le \bar l} + F^\alpha_{l>\bar l},
\end{equation}
where $F^\alpha_{l\le \bar l}$ is the part calculated numerically,
and $F^\alpha_{l>\bar l}$ is the truncated tail we need to estimate.
Here it is beneficial to consider separately the conservative and
dissipative pieces of the SF. We remind that these can be constructed
individually using the mode-sum formulas (\ref{eq:ModeSum-con}) and
(\ref{eq:ModeSum-dis}) via the procedure described
in Sec.\ \ref{subsec:consdiss}.

Consider the conservative piece first. The regularized force modes in this
case admit the large-$l$ expansion
\begin{equation}\label{largel}
F_{\rm reg(cons)}^{\alpha l} = D_{-2}^{\alpha}L^{-2}
+D_{-4}^{\alpha}L^{-4}+\ldots,
%\sum_{n=1}^N \frac{D^\alpha_{2n}}{L^{2n}},
\end{equation}
where $D^\alpha_{-2n}$ are $l$-independent coefficients. An approximation for these coefficients
can be obtained by fitting a large-$l$ subset of numerical data to Eq.\ (\ref{largel}).
In practice we take $\bar l=15$ and find the two coefficients $D_{-2}^{\alpha}$
and $D_{-4}^{\alpha}$ using the numerically-derived modes $10 \le l \le 15$.
The large-$l$ tail piece of $F_{\rm reg(cons)}^{\alpha l}$ is then approximated
as
\begin{equation}
F^\alpha_{{\rm cons}, l>\bar l}\equiv
\sum_{l=\bar l+1}^{\infty} F_{\rm reg(cons)}^{\alpha l}
\cong D_{-2}^\alpha\Gamma_1(\bar l+3/2)
+D_{-4}^\alpha\Gamma_3(\bar l+3/2)/3!,
\end{equation}
where $\Gamma_n(x)$ is the polygamma function of order $n$,
defined in terms of the derivatives of the standard gamma function
as $\Gamma_n(x) = d^{n+1}[\log\Gamma(x)]/dx^{n+1}$.
Since the leading term omitted in Eq.\ (\ref{largel}) is of $O(L^{-6})$,
we expect the absolute error in our estimation of
$F^\alpha_{{\rm cons}, l>\bar l}$ to be of $O(\bar l^{-5})$,
or $O(\bar l^{-4})$ fractionally. With $\bar l=15$ this amounts
to a $\sim 10^{-5}$ fractional error, which we can afford to tolerate
in this work. 

Now consider the dissipative piece. We have that the magnitude of
$F_{\rm reg(diss)}^{\alpha l}$ falls off faster than any power of $1/l$
at large $l$. For the range of orbital parameters explored in this work we
find that the numerical value of $F_{\rm reg(diss)}^{\alpha l}$ drops
below the round-off error at $l\sim 7$--$12$, and so the large-$l$ tail,
estimated as $F^\alpha_{{\rm diss}, l>\bar l}\sim F_{\rm reg(diss)}^{\alpha\bar l}$,
can be safely neglected taking $\bar l=12$. In practice, to avoid adding up spurious
round-off contributions, we truncate the mode-sum series at
$\bar l=\min\{\hat l,15\}$, where $\hat l$ is the first value of $l$
above $l=7$ for which we find $\left|F_{\rm reg(diss)}^{\alpha \hat{l}}\right|>
\left|F_{\rm reg(diss)}^{\alpha, \hat{l}-1}\right|$.

Our procedure for estimating the numerical error is as follows.
First, we estimate the discretization error in each
of the computed $l$-mode contributions (as a time series along the orbit) by
repeating our numerical evolution at a coarser resolution and using the
difference between the high-and low-resolution data sets as a crude error estimator.
[For example, to obtain the data in
Tables~\ref{table:F_p7e02}--\ref{table:F_p15e03} below
we applied a cell size of
$\Delta u\times \Delta v=(0.02M)^2$ for our highest resolution runs,
then $\Delta u\times \Delta v=(0.04M)^2$ for error estimation.] The total
error in the numerically-computed part of the SF is then taken
(conservatively) as the sum of the errors (in absolute value) from
the various modes. In the case of the conservative piece we add to
this the estimated standard fitting error for the large-$l$ tail.
For the fitting itself we use the high-resolution $l$-mode data points,
weighted by their estimated discretization errors. This procedure yields
a conservative estimate for the numerical error in each of
$F^\alpha_{\rm cons}(\chi)$ and $F^\alpha_{\rm diss}(\chi)$.
It is this error that we quote in the next section when presenting
our results.

\subsection{Code validation and performance}

Using a few test runs with a range of sample parameters, we tested our
code (i) for 4th-order numerical convergence; (ii) against the SF
results of Paper I in the circular-orbit case ($e=0$); (iii) by extracting
the flux of energy and angular momentum carried by the gravitational waves
and comparing with results in the literature (see Sec.\ \ref{subsec:fluxes});
and (iv) by verifying that the dissipative component of the computed SF
precisely balances the above flux (Sec.\ \ref{subsec:fluxes}).
Our code seems to perform well, at a standard fractional accuracy of
$< 10^{-4}$ in the final SF, across the parameter range
$0\leq e\lesssim 0.5$ and $p\lesssim 20M$.
For larger eccentricities and/or larger values of $p$ the long evolution
time required begins to play a prohibitive role. Our code is still fully
functional at (say) $(p,e)=(50,0.9)$, although in this case it becomes computationally impractical to achieve our standard $10^{-4}$ accuracy running on a standard (single) desktop computer.

In developing and testing our code we used a desktop workstation
with a 2.5GHz 64-bit processor and 8Gb of RAM.
A typical computation of the SF over a complete radial period, with given parameters in the above ``workable'' range (and at the above accuracy standard), demands $4$--$8$ days of CPU time on this machine.

%The computation time depends almost on time-domain evolution for $l>=2$.
%If we take the grid as $10000 \times 10000$, it takes about 6 hours to
%calculate all (tensor harmonic) modes in $2<=l<=18$.
%For example, in the case with $(p,e)=(7,0.4)$ shown in
%Table~\ref{table:F_p7e04}, $40000 \times 40000$ grids are required for
%the calculation of the SF with the resolution of $h=0.02$. Therefore,
%it takes about 4 days to finish the computation.

%%%%%%%%%%%%%%%%%%%%%%%%%%%%%%%%%%%%%%%%%%%%%%%%%%%%%%%%%%
\section{Sample results}\label{Sec:results}
%%%%%%%%%%%%%%%%%%%%%%%%%%%%%%%%%%%%%%%%%%%%%%%%%%%%%%%%%%
%\subsection{Sample results}
%In this subsection, we present some sample results of the
%self-force by using the scheme explained in the previous section.

%As mentioned in Sec.~\ref{sec:domain}, some unphysical oscillation
%appear at the early stage of the evolution due to the incompleteness
%of the initial data. To get rid of the spurious waves, we put a
%pre-evolution stage lasting $T_{\rm pre}$, before the main part of
%the calculation. For the calculation of the self-force (except for
%the flux calclation) we take $T_{\rm pre}=350M$.

\begin{figure}[htb]
\includegraphics[width=8cm]{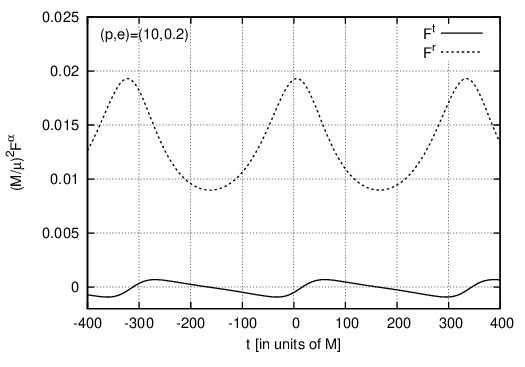}
\includegraphics[width=8cm]{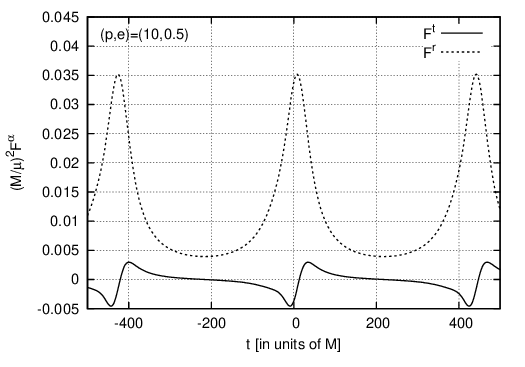} \\
\includegraphics[width=8cm]{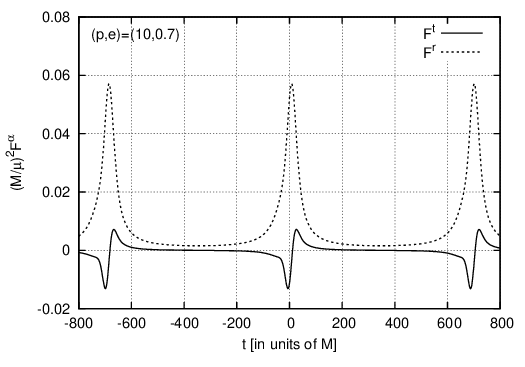}
\caption{The gravitational SF for a sample of eccentric orbits.
We plot the temporal and radial components of the SF as
functions of $t$ for the sample parameters $(p,e)=(10,0.2)$,
$(10,0.5)$ and  $(10,0.7)$. The solid and dashed lines show $F^t$ and
$F^r$, respectively. The $\varphi$ component of the SF can be trivially
obtained using the orthogonality condition $F^{\alpha}u_{\alpha}=0$.
In all graphs $t=0$ corresponds to a periapsis passage. The radial periods
for the $e=0.2,0.5,0.7$ plots are, respectively, $T_r\simeq 328M,434M,693M$.
We have cut out from these plots the early part of the numerical solution,
where non-physical initial spurious waves dominate; the plots display only
the later, stationary part of the solutions, after the spurious waves
have dissipated away. Note the small retardation manifest in the amplitude
of the total SF (with reference to the orbital phase). A similar retardation
is observed in the scalar and EM cases \cite{Haas:2007kz,Haas:Capra}.}
\label{fig:SF-plot}
\end{figure}

Figure \ref{fig:SF-plot} displays SF results for the sample parameters
$(p,e)=(10,0.2)$, $(10,0.5)$ and $(10,0.7)$. We plot the temporal and
radial components of the (total) SF along the geodesic orbit as functions
of coordinate time $t$. The third non-trivial component, $F^{\varphi}$,
can be obtained using the orthogonality condition $F^{\alpha}u_{\alpha}=0$.
In these plots $t=0$ corresponds to a periapsis passage (where $r=r_{\rm min}$),
so that, in accordance with the discussion at the end of Sec.\ \ref{subsec:consdiss},
the dissipative/conservative pieces of $F^t$ are described by the
symmetric/anti-symmetric parts of the $F^t$-graph, and conversely for $F^r$.
%In practice, we calculated the force
%for only one radial orbital period (plus the pre-evolution time).
%We obtained these plots by pasting the one-period data repeatedly.

In Tables~\ref{table:F_p7e02}--\ref{table:F_p15e03} we present numerical SF
results for the sets of orbital parameters $(p,e)=(7,0.2)$, $(7,0.4)$, $(10,0.3)$
and $(15,0.3)$, for a few sample values of the radial phase $\chi$. These numerical
values can be used as a reference for testing future calculations of the SF.
In the tables we display the conservative and dissipative pieces of the SF
separately. We only display values for $F^t$ and $F^r$---once again, the
azimuthal component can be obtained trivially using $F^{\alpha}u_{\alpha}=0$.
Also, we only give values for the ``outbound'' half of the orbital period
($0\leq\chi\leq\pi$, where $\dot r_{\rm p}\geq 0$); the values for the ``inbound''
half ($\pi\leq\chi\leq 2\pi$) can be obtained immediately based on the symmetry
relations (\ref{eq:trick}).
\begin{table}[htb]
\begin{tabular}{c||c|c||c|c}
\hline\hline
$\chi$ & $\eta^{-2}\,F^t_{\rm cons}$ & $\eta^{-2}\,F^t_{\rm diss}$
       & $\eta^{-2}\,F^r_{\rm cons}$ & $\eta^{-2}\,F^r_{\rm diss}$ \\
\hline
 $0$ 
 & $0$
 & $-4.06330(3) \times 10^{-3}$
 & $3.35760(5) \times 10^{-2}$
 & $0$ \\
 $\pi/8$ 
 & $5.3846(3) \times 10^{-4}$
 & $-3.47962(2) \times 10^{-3}$
 & $3.23228(4) \times 10^{-2}$
 & $3.148027(5) \times 10^{-3}$ \\
 $\pi/4$ 
 & $8.6422(6) \times 10^{-4}$
 & $-2.15692(2) \times 10^{-3}$
 & $2.90989(5) \times 10^{-2}$
 & $4.73496(1) \times 10^{-3}$ \\
 $3\pi/8$ 
 & $9.2840(3) \times 10^{-4}$
 & $-9.2831(1) \times 10^{-4}$
 & $2.50709(3) \times 10^{-2}$
 & $4.47010(1) \times 10^{-3}$ \\
 $\pi/2$ 
 & $8.2846(4) \times 10^{-4}$
 & $-2.5168(3) \times 10^{-4}$
 & $2.12504(2) \times 10^{-2}$
 & $3.204188(5) \times 10^{-3}$ \\
 $5\pi/8$ 
 & $6.61185(8) \times 10^{-4}$
 & $-3.385(1) \times 10^{-5}$
 & $1.81454(2) \times 10^{-2}$
 & $1.893665(3) \times 10^{-3}$ \\
 $3\pi/4$ 
 & $4.60708(2) \times 10^{-4}$
 & $-1.1241(3) \times 10^{-5}$
 & $1.590157(5) \times 10^{-2}$
 & $9.63374(2) \times 10^{-4}$ \\
 $7\pi/8$ 
 & $2.36409(2) \times 10^{-4}$
 & $-2.7138(1) \times 10^{-5}$
 & $1.454424(7) \times 10^{-2}$
 & $3.90516(1) \times 10^{-4}$ \\
 $\pi$ 
 & $0$
 & $-3.4614(2) \times 10^{-5}$
 & $1.40888(2) \times 10^{-2}$
 & $0$ \\
\hline\hline
\end{tabular}
\caption{The gravitational SF at selected points along an eccentric geodesic
with $(p,e)=(7,0.2)$. $\eta\equiv \mu/M$.
In the first column $\chi$ is the radial phase along
the orbit [cf.\ Eq.\ (\ref{eq:r-motion})], with $\chi=0,\pi$ corresponding
to periapsis and apoaspsis, respectively. Subsequent columns display, in
order, the conservative and dissipative pieces of $F^t$ and the conservative
and dissipative pieces of $F^r$. Values in brackets are estimates of
the uncertainty (due to numerical error) in the last displayed decimal place.
For example, $5.3844(3)\times 10^{-4}$ stands for $(5.3844\pm 0.0003)\times 10^{-4}$.
The values for $F^{\varphi}$ can be obtained from $F^{\alpha}u_{\alpha}=0$.
Values for the SF along the ``inbound'' half of the radial period
($\pi\leq\chi\leq 2\pi$) can be deduced based on the symmetry
relations (\ref{eq:trick}).
}
\label{table:F_p7e02}
\end{table}

\begin{table}[htb]
\begin{tabular}{c||c|c|c|c}
\hline\hline
$\chi$ & $\eta^{-2}\,F^t_{\rm cons}$ & $\eta^{-2}\,F^t_{\rm diss}$
       & $\eta^{-2}\,F^r_{\rm cons}$ & $\eta^{-2}\,F^r_{\rm diss}$ \\
\hline
 $0$ 
 & $0$
 & $-1.509113(8) \times 10^{-2}$
 & $5.56081(6) \times 10^{-2}$
 & $0$ \\
 $\pi/8$ 
 & $1.8006(1) \times 10^{-3}$
 & $-1.142319(3) \times 10^{-2}$
 & $5.05974(8) \times 10^{-2}$
 & $1.828802(3) \times 10^{-2}$ \\
 $\pi/4$ 
 & $2.5896(2) \times 10^{-3}$
 & $-3.67901(9) \times 10^{-3}$
 & $3.90832(9) \times 10^{-2}$
 & $2.28485(2) \times 10^{-2}$ \\
 $3\pi/8$ 
 & $1.9915(2) \times 10^{-3}$
 & $1.26976(5) \times 10^{-3}$
 & $2.77839(3) \times 10^{-2}$
 & $1.605261(1) \times 10^{-2}$ \\
 $\pi/2$ 
 & $1.30621(8) \times 10^{-3}$
 & $1.74587(2) \times 10^{-3}$
 & $1.98195(4) \times 10^{-2}$
 & $7.59798(5) \times 10^{-3}$ \\
 $5\pi/8$ 
 & $9.8662(4) \times 10^{-4}$
 & $8.03401(6) \times 10^{-4}$
 & $1.47071(2) \times 10^{-2}$
 & $2.65711(2) \times 10^{-3}$ \\
 $3\pi/4$ 
 & $7.0011(1) \times 10^{-4}$
 & $2.396343(8) \times 10^{-4}$
 & $1.13351(1) \times 10^{-2}$
 & $7.76246(5) \times 10^{-4}$ \\
 $7\pi/8$ 
 & $3.49621(1) \times 10^{-4}$
 & $6.31089(3) \times 10^{-5}$
 & $9.28115(1) \times 10^{-3}$
 & $2.04288(2) \times 10^{-4}$ \\
 $\pi$ 
 & $0$
 & $2.92505(1) \times 10^{-5}$
 & $8.581533(5) \times 10^{-3}$
 & $0$ \\
\hline\hline
\end{tabular}
\caption{The gravitational SF at selected points along an eccentric
geodesic with $(p,e)=(7,0.4)$. The structure of the table is the same
as that of Table.~\ref{table:F_p7e02}.}
\label{table:F_p7e04}
\end{table}

\begin{table}[htb]
\begin{tabular}{c||c|c|c|c}
\hline\hline
$\chi$ & $\eta^{-2}\,F^t_{\rm cons}$ & $\eta^{-2}\,F^t_{\rm diss}$
       & $\eta^{-2}\,F^r_{\rm cons}$ & $\eta^{-2}\,F^r_{\rm diss}$ \\
\hline
 $0$ 
 & $0$
 & $-1.024249(7) \times 10^{-3}$
 & $2.303169(9) \times 10^{-2}$
 & $0$ \\
 $\pi/8$ 
 & $7.2278(2) \times 10^{-4}$
 & $-8.05046(5) \times 10^{-4}$
 & $2.21659(3) \times 10^{-2}$
 & $8.51208(8) \times 10^{-4}$ \\
 $\pi/4$ 
 & $1.16148(2) \times 10^{-3}$
 & $-3.67855(4) \times 10^{-4}$
 & $1.98540(2) \times 10^{-2}$
 & $1.17785(3) \times 10^{-3}$ \\
 $3\pi/8$ 
 & $1.247664(9) \times 10^{-3}$
 & $-6.1078(3) \times 10^{-5}$
 & $1.677294(8) \times 10^{-2}$
 & $9.63601(6) \times 10^{-4}$ \\
 $\pi/2$ 
 & $1.08725(1) \times 10^{-3}$
 & $3.3434(2) \times 10^{-5}$
 & $1.36220(1) \times 10^{-2}$
 & $5.65458(5) \times 10^{-4}$ \\
 $5\pi/8$ 
 & $8.11160(7) \times 10^{-4}$
 & $2.83103(3) \times 10^{-5}$
 & $1.088087(7) \times 10^{-2}$
 & $2.63807(5) \times 10^{-4}$ \\
 $3\pi/4$ 
 & $5.122807(9) \times 10^{-4}$
 & $1.10418(3) \times 10^{-5}$
 & $8.81008(2) \times 10^{-3}$
 & $1.06375(1) \times 10^{-4}$ \\
 $7\pi/8$ 
 & $2.408105(6) \times 10^{-4}$
 & $2.45312(6) \times 10^{-6}$
 & $7.537692(9) \times 10^{-3}$
 & $3.61963(8) \times 10^{-5}$ \\
 $\pi$ 
 & $0$
 & $2.836(2) \times 10^{-7}$
 & $7.110909(3) \times 10^{-3}$
 & $0$ \\
\hline\hline
\end{tabular}
\caption{The gravitational SF at selected points along an eccentric
geodesic with $(p,e)=(10,0.3)$. The structure of the table is the same
as that of Table.~\ref{table:F_p7e02}.}
\label{table:F_p10e03}
\end{table}

\begin{table}[htb]
\begin{tabular}{c||c|c|c|c}
\hline\hline
$\chi$ & $\eta^{-2}\,F^t_{\rm cons}$ & $\eta^{-2}\,F^t_{\rm diss}$
       & $\eta^{-2}\,F^r_{\rm cons}$ & $\eta^{-2}\,F^r_{\rm diss}$ \\
\hline
 $0$ 
 & $0$
 & $-1.040267(9) \times 10^{-4}$
 & $1.139648(3) \times 10^{-2}$
 & $0$ \\
 $\pi/8$ 
 & $3.03529(3) \times 10^{-4}$
 & $-8.23882(8) \times 10^{-5}$
 & $1.102460(1) \times 10^{-2}$
 & $1.03215(6) \times 10^{-4}$ \\
 $\pi/4$ 
 & $5.07593(2) \times 10^{-4}$
 & $-3.9157(1) \times 10^{-5}$
 & $1.000022(2) \times 10^{-2}$
 & $1.43970(1) \times 10^{-4}$ \\
 $3\pi/8$ 
 & $5.68296(2) \times 10^{-4}$
 & $-8.3320(6) \times 10^{-6}$
 & $8.56001(2) \times 10^{-3}$
 & $1.19734(3) \times 10^{-4}$ \\
 $\pi/2$ 
 & $5.06147(1) \times 10^{-4}$
 & $1.97928(7) \times 10^{-6}$
 & $6.99810(1) \times 10^{-3}$
 & $7.2183(3) \times 10^{-5}$ \\
 $5\pi/8$ 
 & $3.765543(7) \times 10^{-4}$
 & $2.2286(2) \times 10^{-6}$
 & $5.578997(8) \times 10^{-3}$
 & $3.4786(1) \times 10^{-5}$ \\
 $3\pi/4$ 
 & $2.340696(1) \times 10^{-4}$
 & $8.4719(9) \times 10^{-7}$
 & $4.486884(2) \times 10^{-3}$
 & $1.4418(1) \times 10^{-5}$ \\
 $7\pi/8$ 
 & $1.0842785(8) \times 10^{-4}$
 & $1.0087(2) \times 10^{-7}$
 & $3.814949(2) \times 10^{-3}$
 & $4.9887(8) \times 10^{-6}$ \\
 $\pi$ 
 & $0$
 & $-9.092(5) \times 10^{-8}$
 & $3.590240(2) \times 10^{-3}$
 & $0$ \\
\hline\hline
\end{tabular}
\caption{The gravitational SF at selected points along an eccentric
geodesic with $(p,e)=(15,0.3)$. The structure of the table is the same
as that of Table.~\ref{table:F_p7e02}.}
\label{table:F_p15e03}
\end{table}

\subsection{Dissipation of energy and angular momentum} \label{subsec:fluxes}

Given the local SF, we can calculate the (orbit-averaged) rate at which
orbital energy and angular momentum are dissipated. This information
is contained in the $t$ and $\varphi$ components of the local SF.
From Eq.~(\ref{eq:EOM}) one readily obtains
%~~~~~~~~~~~~~~~~~~~~~~~~~~~~~~~~~~~~~~~~~~~~~~~~~~~~~~~~~~~~
\begin{equation} \label{eq:EdotLdot}
\dot{\cal E}(\chi)
= -[\mu u^t(\chi)]^{-1}F_t(\chi), \quad\quad
\dot{{{\cal L}}}(\chi)
= [\mu u^t(\chi)]^{-1}F_\varphi(\chi),
\end{equation}
%~~~~~~~~~~~~~~~~~~~~~~~~~~~~~~~~~~~~~~~~~~~~~~~~~~~~~~~~~~~~
where in this section (unlike elsewhere in this work) an overdot denotes
$d/dt$. We shall assume here, in effect, that $\mu/M$ is sufficiently
small that the back-reaction effect on the orbit over a period of a
few $T_r$ can be neglected. At this ``adiabatic'' limit the functions
$\dot{{\cal E}}(\chi)$ and $\dot{{\cal L}}(\chi)$ are periodic in $t$ with period $T_r$,
and their time-average is hence given by
%~~~~~~~~~~~~~~~~~~~~~~~~~~~~~~~~~~~~~~~~~~~~~~~~~~~~~~~~~~~~
\begin{equation} \label{eq:average}
\langle \dot{{{\cal E}}} \rangle =
\frac{1}{T_r} \int_0^{T_r} \dot{{{\cal E}}}\, dt, \quad\quad
\langle \dot{{{\cal L}}} \rangle =
\frac{1}{T_r} \int_0^{T_r} \dot{{{\cal L}}}\, dt.
\end{equation}
%~~~~~~~~~~~~~~~~~~~~~~~~~~~~~~~~~~~~~~~~~~~~~~~~~~~~~~~~~~~~
These averages determine the ``secular'' dissipative drift in the values
of $\cal E$ and $\cal L$. Note that $\langle \dot{{{\cal E}}} \rangle $
and $\langle \dot{{{\cal L}}} \rangle$ depend only on the dissipative
components $F^t_{\rm diss}$ and $F^\varphi_{\rm diss}$ (respectively);
it is immediately evident from the symmetry relations (\ref{eq:trick})
that the contributions from $F^t_{\rm cons}$ and $F^\varphi_{\rm cons}$
vanish upon orbital-averaging. We also
note that the dissipative radial component $F^r_{\rm diss}$ (let alone $F^r_{\rm cons}$)
has no effect on the values of $\langle \dot{{{\cal E}}} \rangle $
and $\langle \dot{{{\cal L}}} \rangle$.
The secular drifts $\langle \dot{{{\cal E}}} \rangle$ and
$\langle \dot{{{\cal L}}} \rangle $ must be balanced by the flux of
energy and azimuthal angular momentum in the gravitational waves radiated
to infinity and down the event horizon. Denoting the respective energy fluxes
by $\langle \dot{E} \rangle_{\infty/{\rm EH}}$ and angular momentum fluxes by
$\langle \dot{L} \rangle_{\infty/{\rm EH}}$, we have the balance equations
%~~~~~~~~~~~~~~~~~~~~~~~~~~~~~~~~~~~~~~~~~~~~~~~~~~~~~~~~~~~~
\begin{eqnarray} \label{eq:Ebalance}
- \mu\langle \dot{{\cal E}} \rangle &=&
\langle \dot{E} \rangle_{\infty}
+ \langle \dot{E} \rangle_{\rm BH}\equiv
\langle \dot{E} \rangle_{\rm total}, \\
- \mu\langle \dot{{\cal L}} \rangle &=& \label{eq:Lbalance}
\langle \dot{L} \rangle_{\infty}
+ \langle \dot{L} \rangle_{\rm BH}
\equiv \langle \dot{L} \rangle_{\rm total}.
\end{eqnarray}
%~~~~~~~~~~~~~~~~~~~~~~~~~~~~~~~~~~~~~~~~~~~~~~~~~~~~~~~~~~~~

Two validation tests for our code now suggest themselves. First, we may attempt to extract the asymptotic
fluxes directly from our numerically-calculated Lorenz-gauge metric perturbation,
and compare with results in the literature. Second, using Eqs.\ (\ref{eq:EdotLdot}) we can derive the local quantities 
$\langle \dot{{{\cal E}}} \rangle$ and $\langle \dot{{{\cal L}}} \rangle$
from our SF results, and check whether the balance equations (\ref{eq:Ebalance}) and (\ref{eq:Lbalance}) are
indeed satisfied.

For both above tests we need a time-domain formulation of the asymptotic
fluxes in Schwarzschild spacetime. Such a formulation was presented by
Martel \cite{Martel:2003jj} and Poisson \cite{Poisson:2004cw}\footnote{
Although Martel's expressions for the horizon fluxes in \cite{Martel:2003jj}
are correct, his derivation of these fluxes was not; this was later
noted and corrected by Poisson in \cite{Poisson:2004cw}.}, and we shall
adopt it here. Martel--Poisson's construction is based on the
Regge--Wheeler and Zerilli--Moncrief perturbations functions
$\Psi_{\rm RW}^{lm}$ and $\Psi_{\rm ZM}^{lm}$, which are related to
our Lorenz-gauge variables through
%~~~~~~~~~~~~~~~~~~~~~~~~~~~~~~~~~~~~~~~~~~~~~~~~~~~~~~~~~~~~
\begin{eqnarray}\label{eq:RW}
\Psi_{\rm RW}^{lm} &=&
-\frac{(l-2)!}{2(l+2)!}\left[
\frac{\lambda}{r}\bar{h}^{(9)}
+ \frac{f}{r}\bar{h}^{(10)}
- \bar{h}^{(10)}_{,r_*}
\right], \\
\Psi_{\rm ZM}^{lm} &=& \label{eq:ZM}
\frac{2r}{l(l+1)(\lambda r + 6M)}\left[
\bar{h}^{(1)} - \bar{h}^{(5)} - f \bar{h}^{(6)}
+ \frac{l(l+1)r+2M}{2r} \bar{h}^{(3)}
- r \bar{h}^{(3)}_{,r_*}
+ \frac{\lambda r+6M}{2\lambda r}\bar{h}^{(7)}
\right],
\end{eqnarray}
%~~~~~~~~~~~~~~~~~~~~~~~~~~~~~~~~~~~~~~~~~~~~~~~~~~~~~~~~~~~~
with $\lambda\equiv(l+2)(l-1)$. In terms of $\Psi_{\rm RW}^{lm}$ and
$\Psi_{\rm ZM}^{lm}$, the fluxes at infinity are given by
\cite{Martel:2003jj,Poisson:2004cw}
%~~~~~~~~~~~~~~~~~~~~~~~~~~~~~~~~~~~~~~~~~~~~~~~~~~~~~~~~~~~~
\begin{eqnarray}
\langle \dot{E} \rangle_{\infty} &=&
\frac{1}{64\pi}\sum_{lm}\frac{(l+2)!}{(l-2)!}
\left\langle
4|\Psi_{\rm RW}^{lm}(u)|^2
+|\dot{\Psi}_{\rm ZM}^{lm}(u)|^2
\right\rangle, \label{eq:Edot-infinity} \\
\langle \dot{L} \rangle_{\infty} &=&
\frac{1}{64\pi}\sum_{lm}\frac{(l+2)!}{(l-2)!}
(im)\left\langle
4\Psi_{\rm RW}^{lm}(u) \int^u \Psi_{\rm RW}^{lm*}(u')du'
+ \dot{\Psi}_{\rm ZM}^{lm}(u) \Psi_{\rm ZM}^{lm*}(u)
\right\rangle, \label{eq:Ldot-infinity}
\end{eqnarray}
%~~~~~~~~~~~~~~~~~~~~~~~~~~~~~~~~~~~~~~~~~~~~~~~~~~~~~~~~~~~~
where an asterisk denotes complex conjugation, $\left\langle\cdots\right\rangle$
indicates a suitable time-average (in our case an average over a
period $T_r$ would suffice), and the functions $\Psi_{\rm RW}^{lm}$ and
$\Psi_{\rm ZM}^{lm}$ are evaluated at the ``wave zone'', $v\to\infty$ with
fixed $u$. The horizon fluxes $\langle \dot{E} \rangle_{\rm BH}$ and
$\langle \dot{L} \rangle_{\rm BH}$ are given by expressions similar to
(\ref{eq:Edot-infinity}) and (\ref{eq:Ldot-infinity}), merely replacing $u\to v$
and evaluating $\Psi_{\rm RW}^{lm}$ and $\Psi_{\rm ZM}^{lm}$ near the
horizon, i.e., at $u\to\infty$ with fixed $v$.

To calculate $\langle \dot{E} \rangle_{\infty}$ and $\langle \dot{L} \rangle_{\infty}$, we start by recording the
numerical values of the perturbation functions $\bar h^{(i)}(u)$ at $v=10000M$ (approximating null infinity), over a complete radial period.  
The desired fluxes at infinity are then calculated using Eqs.\ (\ref{eq:RW})--(\ref{eq:Ldot-infinity}), making sure that
the error from truncating the sum over $l$ is properly controlled (this is not difficult, as the mode sum converges exponentially; in none of the cases considered here we found it necessary to include modes beyond $l=12$). In a similar manner we also construct the horizon fluxes $\langle \dot{E} \rangle_{\rm EH}$ and
$\langle \dot{L} \rangle_{\rm EH}$, starting by extracting the numerical
values $\bar h^{(i)}(v)$ at very large $u$ (approximating the horizon;
in practice we used $u=10000M$), and then using Eq.\ (\ref{eq:RW}) and
(\ref{eq:ZM}) with the horizon-version of Eqs.\ (\ref{eq:Edot-infinity})
and (\ref{eq:Ldot-infinity}). We estimate the error in our flux values by comparing results obtained at two different grid resolutions ($h=0.1$ against $h=0.2$).

The eccentric-orbit fluxes $\langle \dot{E} \rangle_{\infty}$ and $\langle \dot{L} \rangle_{\infty}$ were computed independently in the past by several authors, including Tanaka {\it et al.}~\cite{Tanaka} and Cutler {\it et al.}~\cite{Cutler:1994pb}
[using frequency-domain (FD) analyses based on Teukolsky's formalism], and later by Martel \cite{Martel:2003jj} [using a time-domain (TD) analysis based on the Regge-Wheeler-Zerilli formalism]. Very recently, Fujita {\it et al.}~\cite{Fujita:2009us} developed a highly accurate FD algorithm for flux calculations. In Table \ref{table:compare_flux} we compare our flux data with the TD data of Martel and the FD data of Fujita {\it et al.}~\cite{FujitaPrivate}.
We look at two strong-field orbits, one with moderate eccentricity ($e\simeq 0.19$) and the other with a rather high eccentricity ($e\simeq 0.76$). All the results shown agree with ours to within $1\%$. The FD results agree with ours to within at least $0.01\%$, and in all cases the FD results fall well within our estimated error bars. Martel's TD results were presented without error bars, but they are likely less accurate than the FD ones.

\begin{table}
{\tabcolsep = 5mm
\begin{tabular}{lccc}
\hline\hline
 & This paper (TD) & Martel (TD)
 & Fujita {\it et al.}~(FD) \\
\hline
\multicolumn{4}{l}{$p=7.50478, \  e=0.188917$} \\
$\langle \dot{E} \rangle_{\infty}\times 10^{4}\eta^{-2}$
 & $3.169(1)$
 & $3.1770$ 
 & $3.16899989184$  \\
$\langle \dot{L} \rangle_{\infty}\times 10^{3}\eta^{-2}M^{-1}$
 & $5.96760(8)$
 & $5.9329$  
 & $5.96755215608$  \\
\hline
\multicolumn{4}{l}{$p=8.75455, \  e=0.764124$} \\
$\langle \dot{E} \rangle_{\infty}\times 10^{4}\eta^{-2}$
 & $2.124(3)$
 & $2.1484$ 
 & $2.12360313326$  \\
$\langle \dot{L} \rangle_{\infty}\times 10^{3}\eta^{-2}M^{-1}$
 & $2.7774(6)$
 & $2.7932$ 
 & $2.77735938996$  \\
\hline\hline
\end{tabular}}
\caption{The flux of energy and angular momentum in the gravitational
waves radiated to infinity: comparison with results in the literature.
The second column shows the values of the radiative fluxes
$\langle \dot{E} \rangle_{\infty}$ and $\langle \dot{L} \rangle_{\infty}$,
evaluated from our numerical results using Eq.\ (\ref{eq:Edot-infinity})
and (\ref{eq:Ldot-infinity}) for two sample values of $p,e$. Values in parentheses
estimate the uncertainty in the last displayed figure. The subsequent columns display, for comparison, the corresponding values obtained by Martel~\cite{Martel:2003jj} and Fujita {\it et al.}~\cite{Fujita:2009us}.
TD/FD indicate time/frequency-domain methods. Fujita {\it et al.}~claim all their displayed figures are significant.  %Values in square bracket show relative differences with respect to our results.
}
\label{table:compare_flux}
\end{table}

In Tables~\ref{table:Edot} and \ref{table:Ldot} we carry out the second
test mentioned above, i.e., we check whether our numerical SF data and
flux data satisfy the balance equations (\ref{eq:Ebalance}) and (\ref{eq:Lbalance}).
The tables compare the time-averaged rates of loss of orbital energy and angular
momentum, $\mu\langle \dot{{\cal E}} \rangle$ and $\mu\langle \dot{{\cal L}} \rangle$,
with the corresponding total (horizon$+$infinity) fluxes $\langle \dot{E} \rangle_{\rm total}$
and $\langle \dot{L} \rangle_{\rm total}$. We find that the two are consistent with
each other to within the numerical accuracy. This constitutes a significant, highly
non-trivial validation test for our code.

Tables~\ref{table:Edot} and \ref{table:Ldot} also display the
partial contributions to the fluxes $\langle \dot{E} \rangle_{\rm total}$
and $\langle \dot{L} \rangle_{\rm total}$ coming form black
hole absorption, i.e., $\langle \dot{E} \rangle_{\rm BH}$ and
$\langle \dot{L} \rangle_{\rm BH}$. Note that our numerical
accuracy is sufficient to confidently resolve these (relatively
minute) horizon fluxes in the examples considered. The consistency
of the local dissipative SF with the asymptotic fluxes, at the
accuracy level maintained here, is evident only when black hole
absorption is correctly accounted for.
The data in Tables~\ref{table:Edot} and \ref{table:Ldot}
represent a first numerical test of the Martel--Poisson TD
horizon-flux formula.
\begin{table}
\begin{tabular}{c|c||c|c|c|c}
\hline\hline
$p$ & $e$
 & $-\mu\langle \dot{{\cal E}} \rangle\times 10^{4}\eta^{-2}$
 & $\langle \dot{E} \rangle_{\rm total}\times 10^{4}\eta^{-2}$
 & $\langle \dot{E} \rangle_{\rm BH}
          / \langle \dot{E} \rangle_{\rm total}$
 & $1+\mu\langle \dot{{{\cal E}}} \rangle
          / \langle \dot{E} \rangle_{\rm total}$ \\
\hline
$7.0$ & $0.0$
 & $4.00163(5)$
 & $4.00166(4)$
 & $1.3\times 10^{-3}$
 & $7\times 10^{-6}$ \\
$7.0$ & $0.1$
 & $4.21696(9)$
 & $4.217(2)$
 & $1.6\times 10^{-3}$
 & $9\times 10^{-6}$ \\
$7.0$ & $0.2$
 & $4.8983(2)$
 & $4.898(3)$
 & $2.7\times 10^{-3}$
 & $-6\times 10^{-5}$ \\
$7.0$ & $0.3$
 & $6.1852(3)$
 & $6.185(4)$
 & $4.6\times 10^{-3}$
 & $-3\times 10^{-5}$ \\
$7.0$ & $0.4$
 & $8.5436(5)$
 & $8.544(5)$
 & $7.8\times 10^{-3}$
 & $5\times 10^{-5}$ \\
\hline\hline
\end{tabular}
\caption{Testing our code using the global energy balance
equation (\ref{eq:Ebalance}).
The third column displays the (negative of the) average rate of loss of
orbital energy over one radial period, $-\mu\langle \dot{{\cal E}}\rangle$,
for strong-field orbits with semi-latus rectum $p=7$ and a range of
eccentricities $e$. This quantity is calculated from local SF data using
Eqs.\ (\ref{eq:EdotLdot}) and (\ref{eq:average}). For comparison, we give
in the fourth column the corresponding values of the total energy fluxes
$\langle \dot{E} \rangle_{\rm total}$ in the gravitational waves radiated
to infinity and down the event horizon. These fluxes are extracted from
our numerically-derived  Lorenz-gauge metric perturbation, evaluated at
the corresponding asymptotic domains. Values in parentheses are estimates
of the numerical error in the last displayed figure. The last column shows
the relative difference between $-\mu\langle \dot{{\cal E}}\rangle$ and
$\langle \dot{E} \rangle_{\rm total}$, confirming that the balance
equations are satisfied to within our working precision.
For reference, the fifth column shows the relative contribution to
$\langle \dot{E} \rangle_{\rm total}$ from horizon absorption alone,
denoted $\langle \dot{E} \rangle_{\rm BH}$. Manifestly, for the
strong-field orbits considered here and with our working precision
of $<10^{-4}$, black hole absorption cannot be neglected.
}
\label{table:Edot}
\end{table}

\begin{table}
\begin{tabular}{c|c||c|c|c|c}
\hline\hline
$p$ & $e$
 & $-\langle \dot{{\cal L}} \rangle\times 10^{3}\eta^{-1}$
 & $\langle \dot{L} \rangle_{\rm total}\times 10^{3}(M/\mu^2)$
 & $\langle \dot{L} \rangle_{\rm BH}
          / \langle \dot{L} \rangle_{\rm total}$
 & $1+\mu\langle \dot{{{\cal L}}} \rangle
          / \langle \dot{L} \rangle_{\rm total}$ \\
\hline
$7.0$ & $0.0$
 & $7.41112(9)$
 & $7.41124(7)$
 & $1.3\times 10^{-3}$
 & $2\times 10^{-5}$ \\
$7.0$ & $0.1$
 & $7.5896(2)$
 & $7.58961(7)$
 & $1.6\times 10^{-3}$
 & $1\times 10^{-6}$ \\
$7.0$ & $0.2$
 & $8.1517(4)$
 & $8.15137(7)$
 & $2.3\times 10^{-3}$
 & $-4\times 10^{-5}$ \\
$7.0$ & $0.3$
 & $9.2090(5)$
 & $9.2092(2)$
 & $4.0\times 10^{-3}$
 & $2\times 10^{-5}$ \\
$7.0$ & $0.4$
 & $11.1627(7)$
 & $11.164(2)$
 & $7.0\times 10^{-3}$
 & $1\times 10^{-4}$ \\
\hline\hline
\end{tabular}
\caption{
Testing our code using the global angular-momentum balance equation
(\ref{eq:Lbalance}). The structure of this table is similar to that
of Table \ref{table:Edot}. The table compares between the average rate
of change of orbital angular momentum, $\langle \dot{{\cal L}} \rangle$,
as inferred from local SF data, and the flux of angular momentum carried
away by the gravitational waves, $\langle \dot{L} \rangle_{\rm total}$,
as inferred from the asymptotic waveforms. We also indicate the
relative contribution to $\langle \dot{L} \rangle_{\rm total}$
from angular momentum absorbed by the black hole, denoted
$\langle \dot{L} \rangle_{\rm BH}$.}
\label{table:Ldot}
\end{table}

\subsection{Zoom-whirl orbits}

An interesting family of eccentric geodesics, so called
``zoom-whirl'' orbits~\cite{Glampedakis:2002ya}, has
$p-6-2e\equiv\epsilon\ll 1$. These geodesics correspond to points in
the $p,e$ plane lying very close to the separatrix
(see Fig.\ \ref{fig:pe}), possessing energy-squared ${\cal E}^2$
only slightly smaller than the maximum of the effective potential
$R(r,{\cal L}^2)$. A particle on a zoom-whirl orbit spends most of
the radial period ``whirling'' around the central hole in a
nearly circular orbit near periapsis, before ``zooming out'' back
to apoapsis distance. During the whirl episode the particle may complete
many revolutions in $\varphi$---the $\varphi$-phase accumulated
over one radial period scales as $\Delta\varphi\propto \ln(64e/\epsilon)$
[see Eq.\ (2.25) in \cite{Cutler:1994pb}].

\begin{figure}[htb]
\includegraphics[width=9cm]{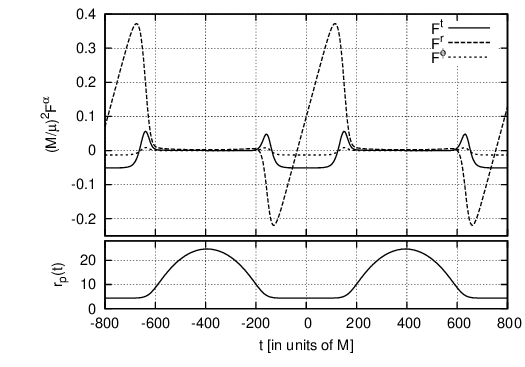}
\caption{The gravitational SF for a zoom-whirl orbit
with parameters $(p,e)=(7.4001, 0.7)$. In the upper panel we show the
(contravariant) $t$, $r$, and $\varphi$ components of the SF as functions
of time $t$ along the orbit. The lower panel shows the radial motion of
the particle, for reference. $t=0$ is periapsis, and the radial period
is $\sim 789M$. While $F^t$ and $F^{\varphi}$ quickly settle to a
constant value during the quasi-circular whirl episode (as one would
expect), the radial component exhibits a peculiar linear behavior.
This behavior is analyzed in the text (cf.\ Fig.\ \ref{fig:SF-zm-split}).
}
\label{fig:SF-zm}
\end{figure}

In Fig.~\ref{fig:SF-zm} we show the SF along a sample
zoom-whirl orbit with parameters $(p,e)=(7.4001,0.7)$.
All components of the SF are relatively very small near the apoapsis, as would be expected in virtue of the large distance from the central black hole. During the brief ``zoom-in'' and ``zoom out''
episodes the particle has a large radial velocity component, and the
SF changes rapidly. During the whirl phase the particle moves on
a nearly-circular orbit, and we expect the SF to settle to a
constant value (this expectation is indeed confirmed in studies of
the scalar-field and EM SFs \cite{Haas:2007kz,Haas:Capra}).
We find that while this is true for the $t$ and $\varphi$ components,
the $r$-component of the SF shows an unexpected linear-in-$t$
behavior during the whirl. Other zoom-whirl orbits we examined showed
a similar behavior.

\begin{figure}[htb]
\includegraphics[width=8cm]{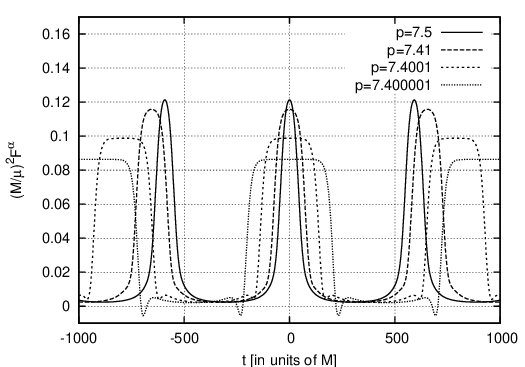}
\includegraphics[width=8cm]{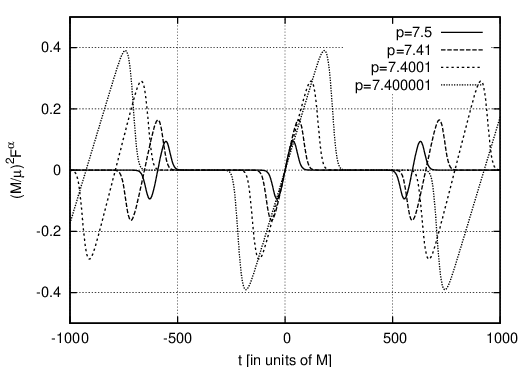}
\caption{Conservative and dissipative pieces of the radial SF for
zoom-whirl orbits--- left and right panels, respectively. Shown are
results for four orbits with the same eccentricity ($e=0.7$) but
with increasing proximity to the separatrix $p=6+2e$.
}
\label{fig:SF-zm-split}
\end{figure}

In order to understand the above peculiarity we conducted several
experiments. The following points summarize the information gained.
(i) The linearly growing piece of $F^r$ is entirely dissipative;
the conservative piece $F^r_{\rm cons}$ shows no such linear growth
(see Fig.\ \ref{fig:SF-zm-split}).
(ii) The linearly growing piece of $F^r_{\rm diss}$ is entirely attributed to
the time-dependent piece of the {\em monopole} ($l=0$) contribution
to the SF; other modes show no such behavior.
(iii) When examining a sequence of zoom-whirl orbits approaching the
separatrix ($\epsilon\to 0$) with fixed $e$---see Fig.\ \ref{fig:SF-zm-split}---we
observe that the rate of linear growth decreases with $\epsilon$, although rather 
slowly (slower than $\propto\epsilon$). It is possible that our
numerical results are in fact consistent with $F^r_{\rm diss}\to 0$
as $\epsilon\to 0$ (as expected), but this is difficult to verify
numerically.
(iv) We have applied our numerical algorithm to study the zoom-whirl
behavior of the scalar-field SF. As in \cite{Haas:2007kz}, we observed
no linear behavior during the whirl.\footnote{Interestingly, we reproduced the linear zoom-whirl behavior with a manipulated scalar-field model, where the peak of the scalar-field's effective potential is artificially ``shifted'' to radii below the whirl radius. This behavior deserves further investigation, but we have not attempted such an investigation here.} 

%also found that linear behavior occurs even , manipulating the shape of the scalar-field potential  when we %(artificially) modified the effective potential in the scalar-field equation so as
%to resemble the effective potential in the gravitational problem,
%we found that the linear behavior reappears.\footnote{More precisely,
%it appears that the linear zoom-whirl behavior occurs whenever the
%whirl radius falls below the peak of the effective potential in the
%relevant field equation. While in the scalar-field case the potential
%peaks well below the whirl orbit, the potential functions for some
%of the metric-perturbation components (in the Lorenz gauge) peak
%above the whirl orbit. We can artificially reproduce a linear zoom-whirl behavior
%in the scalar-field model by shifting the peak of the scalar-field potential to larger radii. The physical reason %behind
%this empirical observation is not clear to us and deserves further
%investigation.
%}

The last point (iv) serves to reaffirm our trust in the numerical code.
The combination of points (i) and (ii) (together with the fact that no linear
growth is observed in the scalar and EM cases) implies
that the linear behavior is purely gauge-related, and suggests that it
should have no observational (gauge invariant) consequences. In particular,
the culprit monopole piece of $F^r_{\rm diss}$ is non-radiative and
has no secular physical effect on the orbit. Finally, the observation
made in (iii) suggests there is nothing wrong with our choice of
gauge either. In fact, it may be argued that the observed linear-in-$t$ behavior,
with a weak dependence on $\epsilon$, is perfectly consistent with the
theoretical expectation based on a local analysis of $F^r_{\rm diss}$ near
the separatrix. We explain this in the following.

Consider the behavior of $F^r_{\rm diss}$ for a zoom-whirl orbit, $\epsilon\ll 1$.
During the whirl the radial phase $\chi$ changes very little, so, taking
$\chi=0$ at periapsis as usual, we can assume $\chi\ll 1$. In the following
analysis we fix the eccentricity $e(>0)$ and consider the limit
$\epsilon\to 0$ by taking $p\to 6+2e$ (from above), inspecting the
behavior of $F^r_{\rm diss}$ at leading order in both $\chi$ and $\epsilon$.
A convenient starting point for this analysis is the orthogonality condition
$u_{\alpha}F^{\alpha}=0$, whose dissipative part can be rearranged to give
\begin{equation}\label{W}
\mu^{-1}F^r_{\rm diss} =
\frac{({\cal E}^2/f_{\rm p})
      (\dot{\cal E}_{\rm diss}-\Omega\dot{\cal L}_{\rm diss})}
     {u^r}
\equiv
\frac{({\cal E}^2/f_{\rm p})\Psi}{u^r}.
\end{equation}
Here we have used the dissipative part of Eqs.\ (\ref{eq:EdotLdot}), and
denoted $\Omega\equiv d\varphi_{\rm p}/dt$. Recall $f_{\rm p}=1-2M/r_{\rm p}$
and an overdot denotes $d/dt$.
The factor $({\cal E}^2/f_{\rm p})$ is regular at $\chi=\epsilon=0$ and hence
uninteresting. For the radial velocity, Eq.\ (\ref{ur}) gives $u^r\propto
\epsilon^{1/2}\chi$, where throughout our present discussion a proportionately
symbol implies the leading-order scaling with $\chi$ and $\epsilon$.
The function $\Psi$ is symmetric in $\chi$ [recall Eq.\ (\ref{eq:trick})]
and clearly $\Psi(\chi=0)=0$; hence we write
$\Psi=\chi^2 \psi(e,\epsilon)+O(\chi^4)$,
where $\psi$ is $\chi$-independent. Taking now $t(\chi=0)=0$,
Eq.\ (\ref{eq:dt_dchi}) gives $\chi\propto \epsilon^{1/2}t$,
and we thus obtain $u^r\propto \epsilon t$
and $\Psi\propto \epsilon t^2 \psi(e,\epsilon)$. We conclude that
\begin{equation}
F^r_{\rm diss}\propto t\, \psi(e,\epsilon).
\end{equation}
It is now essential to understand the scaling of $\psi$ with
$\epsilon$ as $\epsilon\to 0$. Clearly, $\psi\to 0$ at this limit,
since for circular orbits we have
$\dot{\cal E}_{\rm diss}=\Omega \dot{\cal L}_{\rm diss}$.
It may be argued \cite{Cutler:1994pb,Glampedakis:2002ya} that
$\psi$ scales like the (small) fraction of the radial period that
the particle is spending in the zoom phase, which, in turn,
is proportional to
$2\pi/\Delta\varphi\propto [\ln(64e/\epsilon)]^{-1}$. If this argument
is to be trusted, we obtain
\begin{equation}\label{scaling}
F^r_{\rm diss}\propto t\times[\ln(64e/\epsilon)]^{-1},
\end{equation}
which may explain the very weak $\epsilon$-dependence of the slope in
Fig.\ \ref{fig:SF-zm-split}. It is difficult in practice to test our numerical
results more quantitatively against the scaling relation (\ref{scaling})
(precisely because the $\epsilon$-dependence is so weak), but we cannot
rule out the possibility that our results are in fact consistent with
this scaling.

It remains to understand why the linear mode does not exhibit itself
so pronouncedly in the scalar and EM cases, and how the amplitude of
this mode might depend on the choice of gauge in the gravitational case.
%and also how
%the amplitude of this mode might relate to the properties of the
%effective potential in the field equations [recall our point (iv) above].
%This, however, goes beyond what we have set ourselves to do here.
%We remark only that, although the scaling relation (\ref{scaling})
%is likely to be quite generic, 
This requires further analysis, which we do not attempt here.  
We remind that, from a practical point of view, the linear mode
should not cause any real concern, as it cannot affect any gauge-invariant quantity derived from the SF.

%%%%%%%%%%%%%%%%%%%%%%%%%%%%%%%%%%%%%%%%%%%%%%%%%%%%%%%%%%
\section{ISCO shift}\label{Sec:ISCO}
%%%%%%%%%%%%%%%%%%%%%%%%%%%%%%%%%%%%%%%%%%%%%%%%%%%%%%%%%%

Using our SF code we can start to explore the $O(\mu)$ ``post-geodesic''
dynamics of the orbit, and quantify the physical effects resulting from
the finiteness of $\mu$. As a first concrete application of the code,
we calculated the $O(\mu)$ shift in the location and frequency of
the ISCO due to the conservative piece of the SF. We recently reported
the results of this calculation in a Letter \cite{Barack:2009ey}. Here
(and in Appendix~\ref{app:e-exp-form}) we provide full details of this analysis.

The radiative transition across the ISCO, from a slow quasi-circular
inspiral to a rapid plunge, has been studied in the past and is now
well understood. The transition occurs not at a well-defined radius
but rather in a gradual manner, across a ``transition regime'',
whose width (in terms of the corresponding azimuthal frequency) is
proportional to a low power of the mass ratio:
$\Delta\Omega_{\rm trans} \propto(\mu/M)^{2/5}$
\cite{Buonanno:2000ef,Ori:2000zn}.
However, apart from the dominant radiative effect which drives the
inspiral (and the eventual transition to plunge), the gravitational
SF also has a conservative effect, which shifts the location of the
ISCO away from $r=6M$ [by an amount of $O(\mu)$]. Unlike the radiative
transition, this conservative shift is precisely quantifiable.
Moreover, the value of the azimuthal frequency at the shifted ISCO,
$\Omega_{\rm isco}$, is essentially gauge invariant, and hence provides
a useful handle on the strong-field conservative dynamics. Indeed,
the value of $\Omega_{\rm isco}$  (for mass ratios not necessarily
extreme) has long been utilized in testing and calibrating various
approximate treatments of the general-relativistic binary problem
(see., e.g., \cite{Damour:2000we,Blanchet:2002mb}, and the very recent
\cite{Damour:2009sm}). Our SF code allows us, for the first time, to
obtain a {\it precise} value for $\Omega_{\rm isco}$
(modulo a controlled numerical error) at $O(\mu)$.

In what follows we first derive a formula for the ISCO frequency
$\Omega_{\rm isco}$ including $O(\mu)$ conservative SF corrections,
and then describe the numerical method used to obtain the necessary
SF data, and the results. Many of the details are relegated to
Appendix~\ref{app:e-exp-form}.

\subsection{Formulation}

We first review the notion of ISCO in the unperturbed [geodesic, $O(\mu^0)$] case.
The radial geodesic equation [obtained by differentiating Eq.~(\ref{eq:geodesic2})
with respect to $\tau$] reads
%~~~~~~~~~~~~~~~~~~~~~~~~~~~~~~~~~~~~~~~~~~~~~~~~~~~~~~~~~~~~
\begin{equation}
\frac{d^2r_{\rm p}}{d\tau^2}
={\cal F}_{\rm eff}(r_{\rm p},{\cal L}^2); \quad\quad
{\cal F}_{\rm eff}(r,{\cal L}^2)
\equiv -\frac{1}{2}\frac{\partial R(r,{\cal L}^2)}{\partial r}.
\label{eq:EOM-bg}
\end{equation}
%~~~~~~~~~~~~~~~~~~~~~~~~~~~~~~~~~~~~~~~~~~~~~~~~~~~~~~~~~~~~
We consider a slightly eccentric orbit representing an $e$-perturbation of
a circular orbit with radius $r_0$. We write
%~~~~~~~~~~~~~~~~~~~~~~~~~~~~~~~~~~~~~~~~~~~~~~~~~~~~~~~~~~~~
\begin{equation}\label{r1}
r_{\rm p}(\tau) = r_0 + e r_1(\tau) + O(e^2),
\end{equation}
%~~~~~~~~~~~~~~~~~~~~~~~~~~~~~~~~~~~~~~~~~~~~~~~~~~~~~~~~~~~~
where $e\ll 1$ and $r_1(\tau)$ is $e$-independent.
Substituting this in Eq.\ (\ref{eq:EOM-bg}) and reading
the $O(e)$ term, we obtain
%~~~~~~~~~~~~~~~~~~~~~~~~~~~~~~~~~~~~~~~~~~~~~~~~~~~~~~~~~~~~
\begin{equation}
e\frac{d^2r_1}{d\tau^2} =
\left.
\frac{\partial{\cal F}_{\rm eff}(r_{\rm p},{\cal L}^2)}
{\partial r_{\rm p}}
\right|_{e=0} e r_1
+ \left.
\frac{\partial{\cal F}_{\rm eff}(r_{\rm p},{\cal L}^2)}
{\partial {\cal L}^2}
\right|_{e=0} \delta_e({\cal L}^2),
\label{eq:EOM_e1_geodesic}
\end{equation}
%~~~~~~~~~~~~~~~~~~~~~~~~~~~~~~~~~~~~~~~~~~~~~~~~~~~~~~~~~~~~
where $\delta_e$ denotes a linear variation with respect to
$e$ (holding $r_0$ fixed). Since $\delta_e({\cal L}^2)=0$ for geodesics
[see Eq.~(\ref{eq:EL_geodesic})], we obtain
%~~~~~~~~~~~~~~~~~~~~~~~~~~~~~~~~~~~~~~~~~~~~~~~~~~~~~~~~~~~~
\begin{equation}
\frac{d^2 r_1}{d\tau^2} = -\omega_r^2 r_1,
\label{eq:EOM-bg-r1}
\end{equation}
%~~~~~~~~~~~~~~~~~~~~~~~~~~~~~~~~~~~~~~~~~~~~~~~~~~~~~~~~~~~~
with
%~~~~~~~~~~~~~~~~~~~~~~~~~~~~~~~~~~~~~~~~~~~~~~~~~~~~~~~~~~~~
\begin{equation}
\omega_r^2 =
-\left.\frac{\partial {\cal F}_{\rm eff}(r_{\rm p},{\cal L}^2)}
      {\partial r_{\rm p}} \right|_{e=0}
=
\frac{M(r_0-6M)}{r_0^3(r_0-3M)}.
\end{equation}
%~~~~~~~~~~~~~~~~~~~~~~~~~~~~~~~~~~~~~~~~~~~~~~~~~~~~~~~~~~~~
%where ${\cal L}_0$ is the circular-orbit value of ${\cal L}$.
Equation~(\ref{eq:EOM-bg-r1}) tells us that the orbit is stable under
small-$e$ perturbations whenever $\omega_r^2>0$, and is unstable under
such perturbations when $\omega_r^2<0$. The innermost stable circular
orbit is identified by the condition $\omega_r^2=0$, giving
$r_{\rm isco}=6M$.
We also find that, at $O(e)$, the radial motion is harmonic
in $\tau$: Integrating Eq.~(\ref{eq:EOM-bg-r1}) with the assumption of
a periapsis passage at $\tau=0$, we obtain $r_1=-r_0\cos\omega_r\tau$
and hence
%~~~~~~~~~~~~~~~~~~~~~~~~~~~~~~~~~~~~~~~~~~~~~~~~~~~~~~~~~~~~
\begin{equation}
r_{\rm p}(\tau) =
r_0 ( 1 - e\cos\omega_r\tau ) + O(e^2).
\label{eq:rp_expansion}
\end{equation}
%~~~~~~~~~~~~~~~~~~~~~~~~~~~~~~~~~~~~~~~~~~~~~~~~~~~~~~~~~~~~

Next, we consider the $O(\mu)$ correction to the orbit caused by
the conservative piece of the SF. The equations of motion become
%~~~~~~~~~~~~~~~~~~~~~~~~~~~~~~~~~~~~~~~~~~~~~~~~~~~~~~~~~~~~
\begin{equation}
\frac{d{\tilde{\cal E}}}{d\tilde\tau} = -\mu^{-1}F_t^{\rm cons}, \quad\quad
\frac{d{\tilde{\cal L}}}{d\tilde\tau} = \mu^{-1}F_\varphi^{\rm cons},
\label{eq:EOM-perturbEL}
\end{equation}
%~~~~~~~~~~~~~~~~~~~~~~~~~~~~~~~~~~~~~~~~~~~~~~~~~~~~~~~~~~~~
%~~~~~~~~~~~~~~~~~~~~~~~~~~~~~~~~~~~~~~~~~~~~~~~~~~~~~~~~~~~~
\begin{equation}
\frac{d^2 \tilde{r}_{\rm p}}{d\tilde\tau^2}
=
{\cal F}_{\rm eff}(\tilde{r}_{\rm p},\tilde{\cal L}^2)
+ \mu^{-1}F^r_{\rm cons},
\label{eq:EOM-perturbFr}
\end{equation}
%~~~~~~~~~~~~~~~~~~~~~~~~~~~~~~~~~~~~~~~~~~~~~~~~~~~~~~~~~~~~
where hereafter overtildes indicate quantities associated with the SF-corrected
orbit (which is no longer a geodesic). Here we have defined the SF-corrected
energy and angular momentum parameters $\tilde{{\cal E}}(\tilde\tau)$ and
$\tilde{{\cal L}}(\tilde\tau)$ (in general no longer constants of the motion)
through
%~~~~~~~~~~~~~~~~~~~~~~~~~~~~~~~~~~~~~~~~~~~~~~~~~~~~~~~~~~~~
\begin{equation}
\frac{d\tilde{t}_{\rm p}}{d\tilde\tau} = \frac{\tilde{{\cal E}}}{f(\tilde{r}_{\rm p})},
\quad\quad
\frac{d\tilde{\varphi}_{\rm p}}{d\tilde\tau} = \frac{\tilde{{\cal L}}}{\tilde{r}_{\rm p}^2},
\label{eq:def-purterbed-EL}
\end{equation}
%~~~~~~~~~~~~~~~~~~~~~~~~~~~~~~~~~~~~~~~~~~~~~~~~~~~~~~~~~~~~
in analogy with Eqs.\ (\ref{eq:geodesic}). We assume that the orbit remains
bound under the effect of the conservative SF, with
$\tilde r_{\rm min}\leq\tilde r_{\rm p}(\tilde\tau)\leq\tilde r_{\rm max}$, and once
again define $p$ and $e$ as in Eq.\ (\ref{eq:def-pe}), replacing $r_{\rm min}\to
\tilde r_{\rm min}$ and $r_{\rm max}\to \tilde r_{\rm max}$ (we leave $e$ and $p$
untilded for notational brevity). Without loss of generality we take
$\tilde r_{\rm p}(\tilde\tau=0)=\tilde r_{\rm min}$.

The radial component $F^r_{\rm cons}$ is an even, periodic function of $\tau$
along the geodesic $x_{\rm p}(\tau)$ [recall Eq.\ (\ref{eq:trick})], and hence also,
at leading order in $\mu$, an even, periodic function of $\tilde\tau$
along the perturbed orbit $\tilde x_{\rm p}(\tau)$. [This is because
$\tilde x_{\rm p}(\tau)-x_{\rm p}(\tau)\propto O(\mu)$ while $F^r_{\rm cons}$ is
already $O(\mu^2)$.] Since $\tilde r_{\rm p}(\tau)$ too is even and periodic in
$\tilde\tau$ (and monotonically increasing between $\tilde r_{\rm min}$ and
$\tilde r_{\rm max}$), we may express $F^r_{\rm cons}$ as a function of
$\tilde r_{\rm p}$ only, for given $p,e$:
$F^r_{\rm cons}=F^r_{\rm cons}(\tilde{r}_{\rm p};p,e)$.
In Eq.\ (\ref{eq:EOM-perturbFr}) the quantities $\tilde r_{\rm p}(\tau)$,
$d^2 \tilde{r}_{\rm p}/d\tilde\tau^2$ and $F^r_{\rm cons}$ are all periodic
and even in $\tau$, and we conclude that $\tilde{\cal L}$, too, is periodic and
even. Hence, we may write $\tilde{\cal L}=\tilde{\cal L}(\tilde{r}_{\rm p};p,e)$.

Let us now specialize to a slightly eccentric (SF-perturbed) orbit.
Working through $O(e)$, we write $r_{\rm min}=r_0(1-e)$ and $r_{\rm max}=
r_0(1+e)$, where $r_0[=pM+O(e^2)]$ is the radius of the circular orbit
about which we perturb. For this orbit we write
$\tilde r_{\rm p}(\tau) = r_0 + e \tilde r_1(\tau) + O(e^2)$
[as in Eq.\ (\ref{r1})], and we have
$F^r_{\rm cons}=F^r_{\rm cons}(\tilde{r}_{\rm p};r_0,e)$ and
$\tilde{\cal L}=\tilde{\cal L}(\tilde{r}_{\rm p};r_0,e)$.
At $O(e^0)$ (i.e., at the circular-orbit limit) $\tilde{\cal L}$ is constant
along the orbit from symmetry, and solving Eq.\ (\ref{eq:EOM-perturbFr}) with
$d^2\tilde{r}_{\rm p}/d\tilde\tau^2=0$ immediately gives
%~~~~~~~~~~~~~~~~~~~~~~~~~~~~~~~~~~~~~~~~~~~~~~~~~~~~~~~~~~~~
\begin{equation} \label{eq:purterbed-EL0}
%\tilde{\cal E}_0^2 =
%\frac{(r_0-2M)^2}{r_0(r_0-3M)}
%\left[
%1 - \frac{r_0^2}{r_0-2M}F^r_{{\rm con},0}
%\right], \quad
\tilde{\cal L}_0^2 =
\frac{Mr_0^2}{r_0-3M}
\left[
1 - \frac{r_0^2}{\mu M} F^r_{0}
\right],
\end{equation}
%~~~~~~~~~~~~~~~~~~~~~~~~~~~~~~~~~~~~~~~~~~~~~~~~~~~~~~~~~~~~
where hereafter subscripts ``0'' denote circular-orbit values.
In particular, we denote by $F^r_{0}$ the circular-orbit value
of $F^r_{{\rm cons}}$ (omitting the label `cons' for brevity).
Then, at $O(e)$, Eq.\ (\ref{eq:EOM-perturbFr}) yields
%~~~~~~~~~~~~~~~~~~~~~~~~~~~~~~~~~~~~~~~~~~~~~~~~~~~~~~~~~~~~
\begin{equation}
e\frac{d^2\tilde{r}_1}{d\tilde\tau^2} =
\left.
\frac{\partial
{\cal F}_{\rm eff}(\tilde{r}_{\rm p},\tilde{\cal L}_0^2)}
{\partial \tilde{r}_{\rm p}}
\right|_{\tilde{r}_{\rm p}=r_0} e \tilde{r}_1
+ \left.
\frac{\partial
{\cal F}_{\rm eff}(r_0,\tilde{\cal L}^2)}
{\partial (\tilde{\cal L}^2)}
\right|_{\tilde{\cal L}=\tilde{\cal L}_0}
\delta_e\tilde{\cal L}^2
+ \mu^{-1}\delta_e F^r_{\rm cons},
\label{eq:EOM_e1}
\end{equation}
%~~~~~~~~~~~~~~~~~~~~~~~~~~~~~~~~~~~~~~~~~~~~~~~~~~~~~~~~~~~~
where we have used $\delta_e \tilde{r}_{\rm p}=e\tilde r_1$. To evaluate
$\delta_e \tilde{\cal L}^2$ and $\delta_e F_{\rm cons}^r$, we note that the two
quantities depend on $e$ both implicitly, thought $r_{\rm p}(\tau;r_0,e)$,
and explicitly. However, as we showed in Ref.\ \cite{Barack:2009ey},
the explicit linear variation of $\tilde{\cal L}^2$ and $F_{\rm cons}^r$
with respect to $e$ (with fixed $r_0$ and $r_{\rm p}$) vanishes at $e=0$.
Hence we may write
$\delta_e \tilde{\cal L}^2=e\tilde{r}_1 d\tilde{{\cal L}}^2/d{\tilde{r}_{\rm p}}$
and
$\delta_e F_{\rm cons}^r=e\tilde{r}_1 dF_{\rm cons}^r/d{\tilde{r}_{\rm p}}$
(where the derivatives are evaluated at $e=0$),
resulting in that Eq.\ (\ref{eq:EOM_e1}) takes the form
%~~~~~~~~~~~~~~~~~~~~~~~~~~~~~~~~~~~~~~~~~~~~~~~~~~~~~~~~~~~~
\begin{equation} \label{tildeomega}
\frac{d^2 \tilde{r}_1}{d\tilde\tau^2} = -\tilde{\omega}_r^2 \tilde{r}_1,
\end{equation}
%~~~~~~~~~~~~~~~~~~~~~~~~~~~~~~~~~~~~~~~~~~~~~~~~~~~~~~~~~~~~
with
%~~~~~~~~~~~~~~~~~~~~~~~~~~~~~~~~~~~~~~~~~~~~~~~~~~~~~~~~~~~~
\begin{equation}
\tilde{\omega}_r^2 =
-\frac{d}{d\tilde{r}_{\rm p}} \left[
{\cal F}_{\rm eff}(\tilde{r}_{\rm p},\tilde{\cal L}(\tilde{r}_{\rm p})^2)
+ \mu^{-1}F^r_{\rm cons}(\tilde{r}_{\rm p})
\right]_{\tilde{r}_{\rm p}=r_0}.
\label{eq:EOM-perturb-r1}
\end{equation}
%~~~~~~~~~~~~~~~~~~~~~~~~~~~~~~~~~~~~~~~~~~~~~~~~~~~~~~~~~~~~

To obtain a more explicit expression for the shifted radial frequency
$\tilde\omega_r$, we next expand $\tilde{\cal L}$ and
$F^r_{\rm cons}$ in $e$ through $O(e)$. First, Solving Eq.\ (\ref{tildeomega})
with the initial condition $\tilde r_{\rm p}=\tilde r_{\rm min}$, we
find $\tilde r_1 = - r_0\cos\tilde\omega_r\tilde\tau$.
Then we expand $F_{\rm cons}^r=
F^r_0+e\tilde{r}_1 \left(dF^r_{\rm cons}/d\tilde r_{\rm p}\right)_{\tilde r_{\rm p}=r_0}
+O(e^2)$, giving
%~~~~~~~~~~~~~~~~~~~~~~~~~~~~~~~~~~~~~~~~~~~~~~~~~~~~~~~~~~~~
\begin{equation} \label{eq160}
F_{\rm cons}^r=F^r_0+e F^r_1\cos\tilde\omega_r\tilde\tau+O(e^2),
\end{equation}
%~~~~~~~~~~~~~~~~~~~~~~~~~~~~~~~~~~~~~~~~~~~~~~~~~~~~~~~~~~~~
where we have denoted $F^r_1\equiv
-r_0 \left(dF^r_{\rm cons}/d\tilde r_{\rm p}\right)_{\tilde r_{\rm p}=r_0}$.
In a similar manner we expand
$\tilde{\cal L}=
\tilde{\cal L}_0+e\tilde{r}_1 (d\tilde{\cal L}/d\tilde r_{\rm p})_{\tilde r_{\rm p}=r_0}
+O(e^2)$, which, in conjunction with Eq.\ (\ref{eq:EOM-perturbEL}),
gives
%~~~~~~~~~~~~~~~~~~~~~~~~~~~~~~~~~~~~~~~~~~~~~~~~~~~~~~~~~~~~
\begin{equation} \label{eq150}
F^{\rm cons}_\varphi=e\tilde\omega_r
F^1_{\varphi}\sin\tilde\omega_r\tilde\tau  +O(e^2),
\end{equation}
%~~~~~~~~~~~~~~~~~~~~~~~~~~~~~~~~~~~~~~~~~~~~~~~~~~~~~~~~~~~~
where $F^1_{\varphi}\equiv \mu r_0 (d\tilde{\cal L}/d\tilde r_{\rm p})
_{\tilde r_{\rm p}=r_0}$. Finally, using Eqs.\ (\ref{eq160}) and (\ref{eq150})
and substituting for $\tilde{\cal L}_0$ from Eq.~(\ref{eq:purterbed-EL0}),
Eq.~(\ref{eq:EOM-perturb-r1}) becomes
%~~~~~~~~~~~~~~~~~~~~~~~~~~~~~~~~~~~~~~~~~~~~~~~~~~~~~~~~~~~~
\begin{eqnarray} \label{eq170}
\tilde{\omega}_r^2 &=& \omega_{r}^2
- \frac{3(r_0-4M)}{r_0(r_0-3M)}\mu^{-1}F^r_0
+ \frac{1}{r_0}\mu^{-1} F^r_1
- \frac{2}{r_0^4}\sqrt{M(r_0-3M)}\mu^{-1} F^1_{\varphi}
\nonumber \\ &=&
\frac{M}{r_0^3(r_0-3M)}\left[
r_0 - 6M
- \frac{3r_0^2(r_0-4M)}{M\mu}F_0^r
+ \frac{r_0^2(r_0-3M)}{M\mu}F_1^r
- \frac{2(r_0-3M)\sqrt{M(r_0-3M)}}{Mr_0\mu}F_\varphi^1
\right].
\end{eqnarray}
%~~~~~~~~~~~~~~~~~~~~~~~~~~~~~~~~~~~~~~~~~~~~~~~~~~~~~~~~~~~~
This formula describes the $O(\mu)$ conservative shift in the radial
frequency off its geodesic value. Note that it requires knowledge
of the SF through $O(e)$ (knowledge of the circular-orbit SF is not
sufficient).

The perturbed ISCO radius, $r=\tilde{r}_{\rm isco}$, is now obtained
from the condition $\tilde{\omega}_r^2(r_0=\tilde{r}_{\rm isco})=0$.
Namely, $\tilde{r}_{\rm isco}$ is the value of $r_0$ that nullifies the
expression in square brackets in the second line of Eq.\ (\ref{eq170})
through $O(\mu)$. Note that in this expression we are allowed to substitute
$r_0=r_{\rm isco}=6M$ in all SF terms, since such terms are already $O(\mu)$ [so
the error introduced affects $\tilde r_{\rm isco}$ only at $O(\mu^2)$].
Thus, we readily obtain
%~~~~~~~~~~~~~~~~~~~~~~~~~~~~~~~~~~~~~~~~~~~~~~~~~~~~~~~~~~~~
\begin{eqnarray} \label{eq180}
\Delta r_{\rm isco}
&\equiv&
\tilde{r}_{\rm isco} - 6M
%\left[
%\frac{3r_0^2(r_0-4M)}{M}F_0^r
%- \frac{r_0^2(r_0-3M)}{M}F_1^r
%+ \frac{2(r_0-3M)\sqrt{M(r_0-3M)}}{Mr_0}F_\varphi^1
%\right]_{r_0=6M}
\nonumber \\ &=&
(M^2/\mu)\left(216F^r_{0\rm is} -108 F^r_{1\rm is} +
\sqrt{3}\,M^{-2}F^1_{\varphi\rm is}\right)
\end{eqnarray}
%~~~~~~~~~~~~~~~~~~~~~~~~~~~~~~~~~~~~~~~~~~~~~~~~~~~~~~~~~~~~
through $O(\mu)$, where we have denoted
$F^r_{0\rm is}\equiv F^r_0(r_0=6M)$ and similarly for
$F^r_{1\rm is}$, $F^1_{\varphi\rm is}$.

Since the coordinate ISCO shift $\Delta r_{\rm isco}$ is gauge dependent
(just like the SF itself), it is not very useful as a benchmark for
comparisons. Instead, we now consider the (SF-corrected) circular-orbit
azimuthal frequency,
\begin{equation}
\tilde \Omega \equiv
\frac{d\tilde\varphi_{\rm p}}{d\tilde t} =
\frac{d\tilde\varphi_{\rm p}/d\tilde\tau}
     {d\tilde t_{\rm p}/d\tilde\tau},
\end{equation}
which, as discussed in \cite{Sago:2008id}, is invariant under
all $O(\mu)$ gauge transformations whose generators respect the
helical symmetry of the circular-orbit configuration.
Using Eqs.~(\ref{eq:def-purterbed-EL}) and (\ref{eq:purterbed-EL0})
we obtain
%~~~~~~~~~~~~~~~~~~~~~~~~~~~~~~~~~~~~~~~~~~~~~~~~~~~~~~~~~~~~
\begin{equation}  \label{eq190}
\tilde{\Omega} = \Omega\left[1-\frac{r_0(r_0-3M)}{2 M\mu f_0}
F_0^r\right],
\end{equation}
%~~~~~~~~~~~~~~~~~~~~~~~~~~~~~~~~~~~~~~~~~~~~~~~~~~~~~~~~~~~~
where $\Omega\equiv (M/r_0^3)^{1/2}$ is the geodesic (no SF)
value. Evaluated at $r_0=\tilde{r}_{\rm isco}$, the SF-induced frequency
shift $\Delta\Omega\equiv\tilde\Omega-\Omega$ reads
{[}through $O(\mu)$]
%~~~~~~~~~~~~~~~~~~~~~~~~~~~~~~~~~~~~~~~~~~~~~~~~~~~~~~~~~~~~
\begin{equation} \label{eq200}
\Delta\Omega_{\rm isco}
= -\frac{1}{6^{3/2}\,M} \left[\frac{\Delta r_{\rm isco}}{4M}
  + \frac{27M}{2\mu}F_{0\rm is}^r
\right].
\end{equation}
%~~~~~~~~~~~~~~~~~~~~~~~~~~~~~~~~~~~~~~~~~~~~~~~~~~~~~~~~~~~~
Equations analogous to (\ref{eq180}) and (\ref{eq200}) were
obtained by Diaz-Rivera {\em et al.}\ \cite{DiazRivera:2004ik}
in their study of scalar SF effects.

Despite the fact that $\Delta\Omega_{\rm isco}$ is gauge invariant
(in the above sense), care must be exercised in interpreting the
quantity expressed in Eq.\ (\ref{eq200}). As pointed out in \cite{Barack:2005nr},
the Lorenz-gauge metric perturbation has the somewhat peculiar feature
that its $tt$ component (in Schwarzschild coordinates) does not fall
to zero as $r\to\infty$, but instead $h_{tt}\to -2\alpha(=\rm const)$,
with $\alpha=\mu[r_0(r_0-3M)]^{-1/2}$. This peculiarity can be removed
simply by ``rescaling'' the time coordinate as $t\to \hat t=
(1+2\alpha)^{1/2}t\cong (1+\alpha)t$ [neglecting terms of $O(\mu^2)$].
The angular frequency $\tilde\Omega$, whose definition has an explicit
reference to $t$, will be modified under such rescaling as
$\tilde\Omega\to \hat{\tilde\Omega}=(1-\alpha)\tilde\Omega$.
As explained in Ref.\ \cite{Sago:2008id}, and recently emphasized by
Damour \cite{Damour:2009sm}, in order to compare with the frequency
derived in a gauge in which $h_{tt}$ admits the ordinary asymptotic
fall off [such as the Regge-Wheeler gauge, or the gauge associated with
the $O(\mu)$ part of EOB theory], one must not use the ``Lorenz-gauge''
frequency $\tilde\Omega$, but rather the $t$-rescaled frequency
$\hat{\tilde\Omega}$. For practical reasons, therefore, we also give
here the ``$t$-rescaled'' version of Eq.\ (\ref{eq200}):
%~~~~~~~~~~~~~~~~~~~~~~~~~~~~~~~~~~~~~~~~~~~~~~~~~~~~~~~~~~~~
\begin{equation} \label{eq200hat}
\Delta\hat\Omega_{\rm isco}
= -\frac{1}{6^{3/2}\,M}\left[\frac{\Delta r_{\rm isco}}{4M}
  + \frac{27M}{2\mu}F_{0\rm is}^r+\frac{\mu}{\sqrt{18}M}
\right],
\end{equation}
%~~~~~~~~~~~~~~~~~~~~~~~~~~~~~~~~~~~~~~~~~~~~~~~~~~~~~~~~~~~~
where we have used $\alpha(r_0=6M)=(\mu/M)/\sqrt{18}$.

\subsection{Numerical method and results}

In view of Eqs.\ (\ref{eq180}) and (\ref{eq200}), the task of calculating
$\Delta r_{\rm isco}$ and $\Delta\Omega_{\rm isco}$ amounts to
obtaining numerical values for the three coefficients $F^r_{0\rm is}$,
$F^r_{1\rm is}$ and $F^1_{\varphi\rm is}$. The first coefficient is
easily obtained: $F^r_{0\rm is}$ is just the SF along a
strictly circular geodesic with radius $r_0=6M$, and we already
computed it in Paper I using our circular-orbit code (it is one of the
values listed in Table V of \cite{Barack:2007tm}). Here we repeat this
calculation at greater numerical precision, obtaining
%~~~~~~~~~~~~~~~~~~~~~~~~~~~~~~~~~~~~~~~~~~~~~~~~~~~~~~~~~~~~
\begin{equation}\label{Fr0}
F^r_{0\rm is}=0.0244665(1)\, \mu/M^2.
\end{equation}
%~~~~~~~~~~~~~~~~~~~~~~~~~~~~~~~~~~~~~~~~~~~~~~~~~~~~~~~~~~~~
This is consistent with Berndtson's \cite{Berndtson:2009hp} result of
$0.024466497\, \mu/M^2$.

The computation of $F^1_{\varphi\rm is}$ and $F^r_{1\rm is}$ is much
more delicate, as it requires to resolve numerically the small variation
in the SF under a small-$e$ perturbation of a circular orbit [recall
Eqs.\ (\ref{eq160}) and (\ref{eq150})]. We are not helped
by the fact that this variation need be evaluated at
$(p,e)=(6,0)$, which is a singular point in the $p$--$e$ plane
(see below). The subtlety of the computation task calls for an extra
caution, so, as a safeguard measure, we devised and implemented two
completely independent strategies for evaluating $F^1_{\varphi\rm is}$
and $F^r_{1\rm is}$.
The first, more direct approach ({\it method I}) involves an evaluation of
the SF along a sequence of eccentric geodesics approaching the ISCO along
a suitable curve in the $p$--$e$ plane; the required SF coefficients are
then extracted as certain orbital integrals, extrapolated to $e\to 0$ (see below).
The second strategy ({\it method II}) involves an expansion of the field
equations themselves about a circular orbit, through $O(e)$. In what
follows we describe each of the two methods and their outcomes in turn.

\subsubsection{Method I: extrapolation in the $p$--$e$ plane}

From Eqs.\ (\ref{eq160}) and (\ref{eq150}) we obtain
%~~~~~~~~~~~~~~~~~~~~~~~~~~~~~~~~~~~~~~~~~~~~~~~~~~~~~~~~~~~~
\begin{equation}\label{Fr1}
F^r_{1\rm is}=\lim_{p\to 6}\lim_{e\to 0} \hat F^r_{1}(p,e), \quad\quad
\hat F^r_{1}(p,e)\equiv
2\omega_r(e\pi)^{-1}\int_{0}^{\pi/\omega_r}F_{\rm cons}^r \cos\omega_r\tau\, d\tau,
\end{equation}
%~~~~~~~~~~~~~~~~~~~~~~~~~~~~~~~~~~~~~~~~~~~~~~~~~~~~~~~~~~~~
and
%~~~~~~~~~~~~~~~~~~~~~~~~~~~~~~~~~~~~~~~~~~~~~~~~~~~~~~~~~~~~
\begin{equation}\label{Ff1}
F^1_{\varphi\rm is}=\lim_{p\to 6}\lim_{e\to 0} \hat F^1_{\varphi}(p,e), \quad\quad
\hat F^1_{\varphi}(p,e)\equiv
2(e\pi)^{-1}\int_{0}^{\pi/\omega_r}F^{\rm cons}_\varphi \sin\omega_r\tau\, d\tau,
\end{equation}
%~~~~~~~~~~~~~~~~~~~~~~~~~~~~~~~~~~~~~~~~~~~~~~~~~~~~~~~~~~~~
where $\omega_r$, $\tau$, $F_{\rm cons}^r$ and $F^{\rm cons}_\varphi$ are
the values corresponding to a {\em geodesic} with parameters $p,e$ (hence we
were allowed to remove the tilde symbols off $\omega_r$ and $\tau$).
It may be noticed that, formally, the quantities $\hat F_1^r(p,e)$
and $\hat F^1_{\varphi}(p,e)$ are inverse Fourier integrals describing the first
$\omega_r$-harmonic of $F_{\rm cons}^r$ and $F^{\rm cons}_\varphi$, respectively.
The latter two quantities, recall, are both periodic in $\tau$ along the geodesic,
with $F_{\rm cons}^r$ being even and $F^{\rm cons}_\varphi$ being odd in $\tau$.

The order of the limits in Eqs.\ (\ref{Fr1}) and (\ref{Ff1}) is very important:
Since $F^r_{1\rm is}$ and $F^1_{\varphi\rm is}$ are defined through an
expansion about a stable circular orbit ($e=0$), we must first take the
limit $e\to 0$ before taking $p\to 6$. In practice, however, it is more
computationally economical to approach the point $(p,e)=(6,0)$ along a
certain continuous curve in the $p$--$e$ plane, rather than having to
extrapolate to $e\to 0$ along several
different $p$=const lines and then extrapolate the resulting data
again to $p\to 6$. In doing so, however, we must choose our curve
carefully, since the limiting point (6,0) is known to be a singular
one. [A simple manifestation of this singularity is the
fact that the rate at which the radial frequency $\omega_r$ vanishes
at the limit $(p,e)\to(6,0)$ depends upon the direction in the $p$--$e$
plane from which this limit is taken---see, e.g., Eq.\ (2.36) of
\cite{Cutler:1994pb}.] In particular, the curve must always ``stay
away'' from the separatrix $p=6+2e$, where the $e$-expansions (\ref{eq160})
and (\ref{eq150}) [on which Eqs.\ (\ref{Fr1}) and (\ref{Ff1}) rely]
are meaningless. As discussed by Cutler {\it et al.}~in \cite{Cutler:1994pb}
(in a slightly different context), in order for the $e$-expansion to
hold, one must require not only $e\ll 1$ but also $e\ll p-6$.
For our purpose, the point $(6,0)$ must be approached keeping
$e\ll \min\{1,p-6\}$.

\begin{figure}[htb]
\includegraphics[width=8cm]{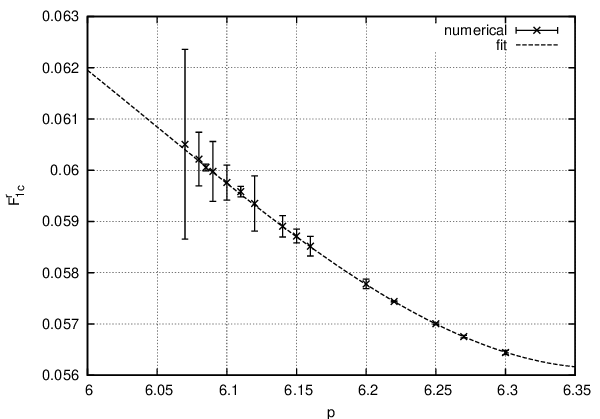}
\includegraphics[width=8cm]{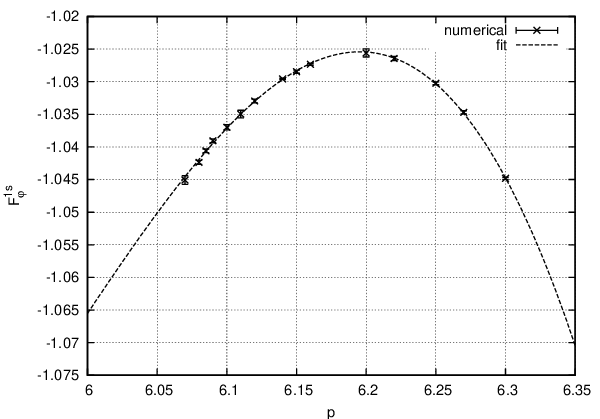}
\caption{Derivation of the SF coefficients $F^r_{1\rm is}$ and
$F^1_{\varphi \rm is}$ entering the ISCO-shift formula (\ref{eq180}),
using {\it method I}.
Plotted are the numerical values of $\hat F^r_{1}(p,e)$
and $\hat F^1_{\varphi}(p,e)$ [see Eqs.\ (\ref{Fr1}) and (\ref{Ff1})]
for a sequence of points in the $p$--$e$ plane approaching the ISCO along
the curve $p=6+\sqrt{e}$. Error bars indicate the estimated numerical error,
which is evaluated as explained in the text. The solid curves are weighted
polynomial fits used to extrapolate the values of $\hat F^r_{1}$
and $\hat F^1_{\varphi}$ to $p\to 6$. The extrapolated values
are the desired coefficients $F^r_{1\rm is}$ and $F^1_{\varphi \rm is}$.}
\label{fig:methodI}
\end{figure}

Here we pick the curve $p=6+\sqrt{e}$, for which $e\ll 1$ automatically
implies $e\ll p-6$. We select a set of $p$ values approaching $p\to 6$,
and for each value we use our code to calculate the SF along an eccentric
geodesic with parameters $(p,e_c(p))$, where $e_c(p)=(p-6)^2$.
For each of these parameter-space points we then construct the
quantities
$\hat F^r_{1}(p,e_c(p))$ and
$\hat F^1_{\varphi}(p,e_c(p))$ defined in Eqs.\ (\ref{Fr1}) and (\ref{Ff1}),
by numerically integrating the SF data along the orbit. The results are
shown in Fig.\ \ref{fig:methodI}. Finally, we obtain the desired coefficients
$F^r_{1\rm is}$ and $F^1_{\varphi\rm is}$ by extrapolating the numerical
values of $\hat F^r_{1}$ and $\hat F^1_{\varphi}$ to $p=6$. We do this, in
practice, by fitting a cubic polynomial to the numerical data, writing
(for example) $\hat F^r_{1}=a_0+a_1(p-6)+\cdots+a_3(p-6)^3$ and taking
the value of the fitting coefficient $a_0$ as our approximation for
$F^r_{1\rm is}$. The results are
%~~~~~~~~~~~~~~~~~~~~~~~~~~~~~~~~~~~~~~~~~~~~~~~~~~~~~~~~~~~~
\begin{eqnarray}\label{coeffI}
F^r_{1\rm is} &=& 0.0620(5)\,\mu/M^2 ,    \nonumber\\
F^1_{\varphi \rm is}&=&-1.066(1)\,\mu
\quad\quad \text{({\it method I})}.
%%Bare data
%%Fr0_isco (bare) : 2.4448318840e-02 [4.8e-06]
%%Fr1c_isco       : 6.1950845758e-02 [4.5e-04]
%%F_p1s_isco      : -1.0655096637e+00 [1.2e-03]
\end{eqnarray}
%~~~~~~~~~~~~~~~~~~~~~~~~~~~~~~~~~~~~~~~~~~~~~~~~~~~~~~~~~~~~
Here, as elsewhere in this work, a parenthetical figure indicates
the uncertainty in the last displayed decimal place due to numerical
error [so, e.g., $0.0620(5)$ stands for $0.0620 \pm 0.0005$].

%As elsewhere in this work, the parenthetical figures indicate
%the uncertainty in the last displayed figures
%due to numerical error.
The error in the above calculation is estimated as follows.
For each value of $e$ considered, we first estimate the numerical error
in each of $\hat F^r_{1}$ and $\hat F^1_{\varphi}$
as the difference between the value obtained with the finest numerical
grid used and the value obtained with a coarser grid (of four times
the cell area). We calculate this difference for each of the $l$-mode
contributions to $\hat F^r_{1}$ and $\hat F^1_{\varphi}$,
and conservatively take the total ($l$-summed) error as the sum over
the moduli of the individual $l$-mode errors (adding to this the large-$l$ tail extrapolation
error). The ``error bars'' thus
obtained are those displayed in Fig.\ \ref{fig:methodI}.
In the final step we fit a cubic polynomial curve to each of the
$\hat F^r_{1},\hat F^1_{\varphi}$ data sets, using the above
error bars as a fitting weight. We take the standard fitting error in the
constant term (the above coefficient $a_0$) as our estimate for the
error in $F^r_{1\rm is},F^1_{\varphi \rm is}$. It is this estimated
error that we indicate in Eqs.\ (\ref{coeffI}).

\begin{figure}[htb]
\includegraphics[width=8cm]{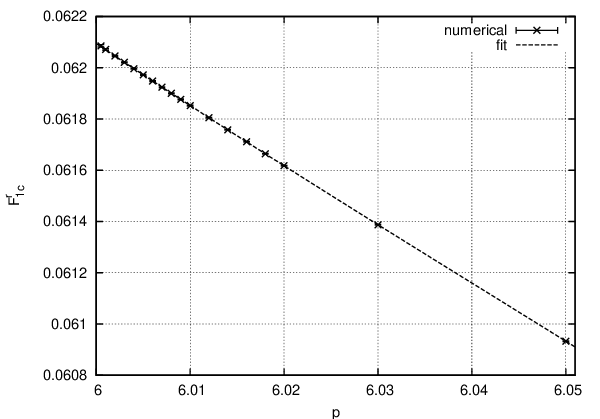}
\includegraphics[width=8cm]{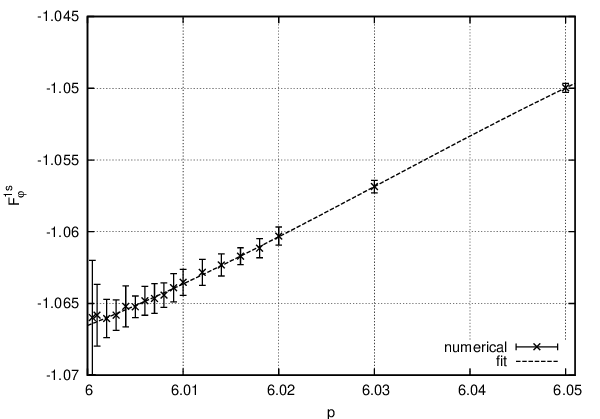}
\caption{Derivation of $F^r_{1\rm is}$ and $F^1_{\varphi \rm is}$,
using {\it method II}.
Plotted are the numerical values of $F^r_{\cos}(p)$
and $F^{\sin}_{\varphi}(p)$ [see Eqs.\ (\ref{Fcos}) and (\ref{Fsin})]
for a sequence of points approaching the ISCO along the $p$-axis
in the $p$--$e$ plane. Error bars indicate the estimated numerical error,
which is evaluated as explained in the text. The solid curves are weighted
polynomial fits used to extrapolate the values of $F^r_{\cos}$
and $F^{\sin}_{\varphi}$ to $p\to 6$. The extrapolated values
are the desired coefficients $F^r_{1\rm is}$ and $F^1_{\varphi \rm is}$.}
\label{fig:methodII}
\end{figure}

%With the coefficient values given in Eq.\ (\ref{Fr0}) and (\ref{coeffI}),
%Eqs.\ (\ref{eq180}), (\ref{eq200}) and (\ref{eq200hat}) yield
%%~~~~~~~~~~~~~~~~~~~~~~~~~~~~~~~~~~~~~~~~~~~~~~~~~~~~~~~~~~~~
%\begin{eqnarray}
%\Delta r_{\rm isco}&=&-3.25(5)\,\mu ,
%\nonumber\\
%\Delta\Omega_{\rm isco}&=& \Omega_{\rm isco}\times 0.48(1)\, \mu/M,
%\nonumber\\
%\Delta\hat\Omega_{\rm isco}&=& \Omega_{\rm isco}\times 0.25(1)\, \mu/M
%\quad\quad \text{({\it method I})},
%%%Bare data
%%%dRisco       : -3.2514442153e+00 [5.1e-02]
%%%dWisco/Wisco : 4.8256330383e-01 [1.3e-02]
%%%dhWis/hWis   : 2.4686104344e-01 [1.3e-02]
%\end{eqnarray}
%%~~~~~~~~~~~~~~~~~~~~~~~~~~~~~~~~~~~~~~~~~~~~~~~~~~~~~~~~~~~~
%where $\Omega_{\rm isco}=(6^{3/2}\,M)^{-1}$ is the unperturbed (geodesic)
%value of of $\Omega$ at $r=6M$.

\subsubsection{Method II: $e$-expansion of the field equations}

Our second procedure for evaluating $F^r_{1\rm is}$ and $F^1_{\varphi \rm is}$
is based on a systematic expansion of the field equations (and the SF)
through $O(e)$, and is similar, in principle, to the method used in Ref.\
\cite{DiazRivera:2004ik} for the scalar-field SF. The idea is to consider
a slightly eccentric geodesic, with $r_{\rm p}(\tau)=r_0 ( 1 - e\cos\omega_r\tau)$
where $e\ll 1$ and $r_0$=const [as in Eq.\ (\ref{eq:rp_expansion})],
and for this geodesic expand the source term in the perturbation equations
(\ref{eq:field-eqs}) through $O(e)$ as
\begin{equation}\label{esource}
S^{(i)}(t,r)\delta(r-r_{\rm p}) = S^{(i)}_{0}(t,r_0)\delta(r-r_{0})
+e\left[S^{(i)}_{1}(t,r_0)\delta(r-r_{0})+T^{(i)}_{1}(t,r_0)\delta'(r-r_{0})\right].
\end{equation}
Here we have omitted the indices $lm$ for brevity and used a prime to
denote $d/dr$. The various expansion coefficients are given by
%~~~~~~~~~~~~~~~~~~~~~~~~~~~~~~~~~~~~~~~~~~~~~~~~~~~~~~~~~~~~
\begin{equation} \label{S0S1}
S_0^{(i)} \equiv S^{(i)}(t,r_0), \quad\quad
S_1^{(i)} \equiv
\frac{\partial S^{(i)}(t,r_{\rm p})}{\partial e}\bigg|_{e=0},
\quad\quad
T_1^{(i)} \equiv
- S^{(i)}(t,r_0)
\frac{\partial r_{\rm p}}{\partial e}\bigg|_{e=0}=
S_0^{(i)}r_0\cos\omega_r\tau.
\end{equation}
%~~~~~~~~~~~~~~~~~~~~~~~~~~~~~~~~~~~~~~~~~~~~~~~~~~~~~~~~~~~~
The coefficients $S_1^{(i)}$ (for $i=1,\ldots,10$) are evaluated by
substituting the above expression for $r_{\rm p}(\tau)$ in Eqs.\
(\ref{eq:source-term1})--(\ref{eq:source-term10}),
along with $u^r=e r_0 \omega_r \sin\omega_r\tau + O(e^2)$,
${\cal E}=(r_0-2M)[r_0(r_0-3M)]^{-1/2} + O(e^2)$
and ${\cal L}=r_0(r_0/M-3)^{-1/2} + O(e^2)$,
and then expanding through $O(e)$. Explicit expressions for the
$S_1^{(i)}$'s are given in Appendix \ref{app:e-exp-form}.
%we note that each of the $S_1^{(i)}$'s is proportional to either
%$\cos\omega_r\tau$ or $\sin\omega_r\tau$.
We also formally expand the metric perturbation functions $\bar{h}^{(i)lm}$ in the form
%~~~~~~~~~~~~~~~~~~~~~~~~~~~~~~~~~~~~~~~~~~~~~~~~~~~~~~~~~~~~
\begin{equation}\label{hexp}
\bar{h}^{(i)}(t,r) = \bar{h}_0^{(i)}(t,r) + e\bar{h}_1^{(i)}(t,r)+O(e^2)
\end{equation}
%~~~~~~~~~~~~~~~~~~~~~~~~~~~~~~~~~~~~~~~~~~~~~~~~~~~~~~~~~~~~
(again omitting the indices $lm$ for brevity), and consequently single out the $O(e^0)$ and $O(e^1)$ pieces of the perturbation equations (\ref{eq:field-eqs}) as
%~~~~~~~~~~~~~~~~~~~~~~~~~~~~~~~~~~~~~~~~~~~~~~~~~~~~~~~~~~~~
\begin{eqnarray}
\square \bar h_0^{(i)} +
{\cal M}^{(i)}_{\;(j)}\bar h_0^{(j)}
&=& S_0^{(i)}\delta(r-r_0), \label{h_0} \\
\square \bar h_1^{(i)} +
{\cal M}^{(i)}_{\;(j)}\bar h_1^{(j)}
&=& S_1^{(i)}\delta(r-r_0)
+ T_1^{(i)}\delta'(r-r_0). \label{h_1}
\end{eqnarray}
%~~~~~~~~~~~~~~~~~~~~~~~~~~~~~~~~~~~~~~~~~~~~~~~~~~~~~~~~~~~~
Notice that the source terms in these equations are evaluated along
a {\it circular} geodesic (of radius $r_0$). The function $\bar{h}_0^{(i)}$
is just the physical Lorenz-gauge perturbation from this circular geodesic.
The function $\bar{h}_1^{(i)}$ is not a physical perturbation; in
particular, it is discontinuous across the orbit (due to the $\delta'$
term in its source).

For our purpose we need to solve Eqs.\ (\ref{h_0}) and (\ref{h_1}) for $r_0=6M$. 
While there is no problem solving for $\bar{h}_0^{(i)}$ (as we in fact already did in Paper I), solving for $\bar{h}_1^{(i)}$ is numerically subtle, since the source functions $S_1^{(i)}$ become singular as $\omega_r\to 0$ [cf.\ Eqs.\ (\ref{S1})--(\ref{S12}) in Appendix \ref{app:e-exp-form}]. Hence, as in method I, we use extrapolation: we first solve Eq.\ (\ref{h_1}) for a sequence of $r_0$ values approaching $6M$, and then extrapolate our data to $r_0=6M$.

A second difficulty arises because our TD algorithm assumes that the numerical variables are continuous on the particle's orbit. This is no longer the case for the functions $\bar h_1^{(i)}$, which suffer finite jump discontinuities across the orbit. This requires an adaptation of the junction conditions implemented in the finite-difference scheme. We describe the suitably modified junction conditions in Appendix~\ref{app:e-exp-form}.
We use our amended TD algorithm to solve for the functions $\bar h_1^{(i)}$ via time evolution.

Once the functions $\bar{h}_0^{(i)}(t,r)$ and $\bar{h}_1^{(i)}(t,r)$ are at hand we
can proceed to construct the SF along the slightly eccentric orbit through $O(e)$. Here we need to be cautious:
When we evaluate $\bar{h}^{(i)}(t,r)$ (and its derivatives) along the slightly eccentric orbit $r_{\rm p}(\tau)$, we must take proper account of the $O(e)$ contribution coming from the term $\bar{h}_0^{(i)}(t,r_{\rm p})$
via the expansion of $r_{\rm p}$ in $e$. Recalling Eq.\ (\ref{hexp}) we have, in fact,
\begin{equation} \label{eq:MP-small-e}
\bar{h}^{(i)}(t,r_{\rm p}) =
\bar{h}_0^{(i)}(t,r_0) + e \left[\bar{h}_1^{(i)}(t,r_0)-
\bar{h}_{0,r}^{(i)}(t,r_0) r_0\cos\omega_r\tau\right] + O(e^2),
\end{equation}
%where
%\begin{eqnarray}
%H_0^{(i)}(t) &\equiv& \bar{h}_0^{(i)}(t,r_0), \nonumber\\
%H_1^{(i)}(t) &\equiv&
%\bar{h}_1^{(i)}(t,r_0)- \bar{h}_{0,r}^{(i)}(t,r_0) r_0\cos\omega_r\tau.
%\end{eqnarray}
with similar expansions applying to the $r$ and $t$ derivatives of the perturbation
along $r_{\rm p}(\tau)$:
\begin{equation} \label{eq:DMP-small-e}
\bar{h}^{(i)}_{,\sigma}(t,r_{\rm p}) =
 \bar{h}_{0,\sigma}^{(i)}(t,r_0) + e\left[ \bar{h}_{1,\sigma}^{(i)}(t,r_0)-
 \bar{h}_{0,r\sigma}^{(i)}(t,r_0) r_0\cos\omega_r\tau\right]+ O(e^2),
\end{equation}
where $\sigma=r$ or $t$.
%\begin{eqnarray}
%H_{0,\sigma}^{(i)}(t) &\equiv& \bar{h}_{0,\sigma}^{(i)}(t,r_0), \nonumber\\
%H_{1,\sigma}^{(i)}(t) &\equiv&
%\bar{h}_{1,\sigma}^{(i)}(t,r_0)- \bar{h}_{0,r\sigma}^{(i)}(t,r_0) r_0\cos\omega_r\tau.
%\end{eqnarray}
Recall that the derivatives of $\bar{h}^{(i)}_0$ and $\bar{h}^{(i)}_1$, as well as the
function $\bar{h}^{(i)}_1$ itself, are discontinuous at $r_0$ and hence
defined only in the sense of a directional limit $r\to r_0^{\pm}$ (however, for simplicity we choose here not to reflect this with a suitable notation). Notice also that the first derivatives
of the metric perturbation $\bar{h}^{(i)}$ along the slightly eccentric geodesic
involve {\em second} derivatives of the function $\bar{h}_{0}^{(i)}$ along the circular
geodesic (an $rr$ derivative for $\bar{h}^{(i)}_{,r}$ and an $rt$ derivative
for $\bar{h}^{(i)}_{,t}$). The necessary data for our calculation therefore includes
the functions $\bar{h}_{0}^{(i)}$ and $\bar{h}_{1}^{(i)}$ themselves, as well
as the derivatives $\bar{h}_{0,r}^{(i)}$, $\bar{h}_{0,t}^{(i)}$
$\bar{h}_{1,r}^{(i)}$, $\bar{h}_{1,t}^{(i)}$, $\bar{h}_{0,rr}^{(i)}$
and $\bar{h}_{0,rt}^{(i)}$, all evaluated (in a one-sided fashion) at the
circular orbit with radius $r_0=6M$.

To construct the SF through $O(e)$, we substitute Eq.\ (\ref{hexp}) in
Eq.\ (\ref{eq:Flfull}), and evaluate the outcome along the slightly eccentric
orbit $r_{\rm p}(\tau)$. We remind that the functions ${\cal F}$ appearing
in Eq.\ (\ref{eq:Flfull}) are given explicitly in Appendix \ref{app:Ffull}
in terms of the perturbation $\bar{h}^{(i)}(t,r)$ and its $r$ and $t$ derivatives.
We next expand Eq.\ (\ref{eq:Flfull}) in $e$ through $O(e)$ using
the expansions (\ref{eq:MP-small-e}) and (\ref{eq:DMP-small-e}) and the
above $e$-expansions for $u^r$, $\cal E$ and $\cal L$. Keeping only the $O(e)$
terms, we proceed through the mode-sum regularization procedure as
described in Sec.\ \ref{subsec:consdiss}, and construct the $O(e)$ pieces
of the conservative SF components $F^r_{\rm cons}$ and $F_t^{\rm diss}$.
We find that these $O(e)$ pieces contain only terms proportional to either
$\cos\omega_r\tau$ or $\sin\omega_r\tau$. More precisely, we obtain
\begin{eqnarray}\label{Fcos}
O(e) \text{ piece of } F^r_{\rm cons}&=& eF^r_{\rm cos}\cos\omega_r\tau,
\\
O(e) \text{ piece of } F_t^{\rm cons}&=& e\omega_r F_t^{\rm sin}\sin\omega_r\tau,
\end{eqnarray}
where the coefficients $F^r_{\rm cos}$ and $F_t^{\rm sin}$ are constructed
from $\bar{h}_{1}^{(i)}$, $\bar{h}_{1,\sigma}^{(i)}$, $\bar{h}_{0,\sigma}^{(i)}$
and $\bar{h}_{0,r\sigma}^{(i)}$. (The explicit form of these coefficients is
rather complicated and will not be shown here. The coefficients can be readily
evaluated using computer algebra tools.) The $O(e)$ part of the
orthogonality condition $u^\alpha F_{\alpha}=0$ then also gives
\begin{equation}\label{Fsin}
O(e) \text{ piece of } F_\varphi^{\rm cons}=
e\omega_r F_\varphi^{\rm sin}\sin\omega_r\tau,
\end{equation}
where $F_\varphi^{\rm sin}$ is related to the numerically-computed
coefficients $F_t^{\rm sin}$ and $F^r_0$ through
\begin{equation}
F_\varphi^{\rm sin}=
-\frac{r_0^2}{{\cal L}_0 f_0} \left(
{\cal E}_0 F_t^{{\rm sin}} + r_0 F_0^r
\right).
\end{equation}
Comparing Eqs.\ (\ref{Fcos}) and (\ref{Fsin}) with Eqs.\ (\ref{eq160})
and (\ref{eq150}), one finally identifies $F_1^r=F^r_{\rm cos}$ and
$F^1_{\varphi}=F_\varphi^{\rm sin}$.

We have implemented the above procedure for a dozen or so radii $r_0$ near $6M$, and for each radius obtained 
the values of $F_1^r$ and $F^1_{\varphi}$. We then extrapolated these values to $r_0=6M$ by fitting a cubic polynomial
(see Fig.~\ref{fig:methodII}). We obtained
%~~~~~~~~~~~~~~~~~~~~~~~~~~~~~~~~~~~~~~~~~~~~~~~~~~~~~~~~~~~~
\begin{eqnarray}\label{coeffII}
F^r_{1\rm is}&=&0.062095(1)\,\mu/M^2 ,    \nonumber\\
F^1_{\varphi \rm is}&=&-1.0665(8)\,\mu  \quad\quad \text{({\it method II})}.
%%Bare data
%%Fr1c_isco  : 6.2094705321e-02 [1.4e-06]
%%F_p1s_isco : -1.0665315625e+00 [8.0e-04]
\end{eqnarray}
%~~~~~~~~~~~~~~~~~~~~~~~~~~~~~~~~~~~~~~~~~~~~~~~~~~~~~~~~~~~~
These values are in agreement, within the estimated numerical accuracy,
with the values obtained using method I [compare with Eqs.\ (\ref{coeffI})].
Using method II we were able to explore orbits much nearer to the ISCO,
resulting in a much improved accuracy. 
Of course, the agreement between our two independent calculations 
provides significant reassurance.

With the SF coefficient values given in Eqs.\ (\ref{Fr0}) and (\ref{coeffII}),
formulas (\ref{eq180}), (\ref{eq200}) and (\ref{eq200hat}) finally yield
%~~~~~~~~~~~~~~~~~~~~~~~~~~~~~~~~~~~~~~~~~~~~~~~~~~~~~~~~~~~~
\begin{eqnarray}\label{resultsII}
\Delta r_{\rm isco}&=&-3.269(2)\,\mu ,
\nonumber\\
\Delta\Omega_{\rm isco}&=& \Omega_{\rm isco}\times 0.4869(4)\, \mu/M,
\nonumber\\
\Delta\hat\Omega_{\rm isco}&=& \Omega_{\rm isco}\times 0.2512(4)\, \mu/M ,
%\quad\quad \text{({\it method II})}.
%%Bare data
%%dRis       : -3.2687316236e+00 [1.5e-03]
%%dWis/Wis   : 4.8688394307e-01 [3.9e-04]
%%dhWis/hWis : 2.5118168267e-01 [3.9e-04]
\end{eqnarray}
%~~~~~~~~~~~~~~~~~~~~~~~~~~~~~~~~~~~~~~~~~~~~~~~~~~~~~~~~~~~~
where $\Omega_{\rm isco}=(6^{3/2}\,M)^{-1}$ is the unperturbed (geodesic)
value of $\Omega$ at $r=6M$. The values obtained here are slightly more accurate than the ones we give in our Letter 
\cite{Barack:2009ey} [$\Delta r_{\rm isco}=-3.269(3)\,\mu$ and $\Delta\Omega_{\rm isco}= \Omega_{\rm isco}\times 0.4870(6)\, \mu/M$], an improvement made possible by the inclusion of additional numerical data points in our analysis.

%%%%%%%%%%%%%%%%%%%%%%%%%%%%%%%%%%%%%%%%%%%%%%%%%%%%%%%%%%
\section{Concluding remarks and future applications}
%%%%%%%%%%%%%%%%%%%%%%%%%%%%%%%%%%%%%%%%%%%%%%%%%%%%%%%%%%

This work marks a new frontline in the program to model realistic
two-body inspirals in the extreme mass-ratio regime. For the first
time we are able to calculate the full [$O(\mu^2)$] gravitational
SF across (essentially) the entire parameter space of strong-field
bound geodesics in Schwarzschild spacetime. This work also represents
a first complete end-to-end implementation of a range of computational
techniques which were developed gradually over the past decade: mode-sum
regularization scheme \cite{Barack:2001gx}, the 1+1D Lorenz-gauge
perturbation formalism \cite{Barack:2005nr}, and the method of
extended homogeneous solutions \cite{Barack:2008ms}. As the reader
may appreciate, the underlying computational challenge is rather daunting,
given the complexity of the field equations, the technical subtleties involved
in dealing with the delta-function source, the high computational cost,
and the need to patch together different techniques in both the time
and frequency domains. Our eventual working code is of considerable
complexity and took over two years to develop and test.

Following is a summary of the various tests which helped us establish
confidence in our code's performance. (i) The mode-sum regularization
procedure is self-validating, in the sense that it is extremely
sensitive to errors in the computation of the perturbation multipoles
(especially the high-$l$ ones, which are most computationally demanding).
If the regularized mode sum shows the expected fall-off behavior at large
$l$, this by itself is a strong indication that the high-$l$ modes were
calculated correctly.
(ii) The code reproduces the known results in the circular-orbit case;
these results are now confirmed by 3 independent analyses
\cite{Barack:2007tm,Detweiler:2008ft,Berndtson:2009hp}.
(iii) Our evolution code reproduces the correct asymptotic fluxes of
gravitational-wave energy and angular momentum, as verified by comparing
with results in the literature.
(iv) The work done by the dissipative piece of the computed SF is found
to precisely balance these fluxes.
(v) The value of the ISCO frequency shift derived from our SF seems
consistent with the value derived in EOB at 3rd post-Newtonian (PN) order:
Damour recently showed that the latter is about 72.5\% of the SF value,
with the difference likely attributed to higher-order PN terms \cite{Damour:2009sm}.

In principle, our code can return the SF along any bound geodesic
in Schwarzschild geometry, although in practice computational cost
may becomes prohibitive when the orbital period is too large
(i.e., for very large $p$ and/or $e$ close to unity).
The ``workable domain'' of our code using a current-day standard
single-processor desktop computer is  roughly
$0\lesssim e\lesssim 0.5$ and $p\lesssim 20M$ if a fractional
accuracy of $<10^{-4}$ in the SF is sought. The current algorithm
incorporates an explicit reference to the radial frequency parameter,
so it cannot be used to tackle unbound orbits. However, it may be
adapted with moderate effort to handle unbound orbits (including orbits
below the last stable orbit) as well.

%By ``{\em almost} any geodesic'' we refer to a certain
%yet-to-be-addressed difficulty with our frequency domain code: Our present
%algorithm for the dipole mode ($l=1$) cannot handle geodesics for which
%one (or more) of the frequency harmonics $\omega=\Omega+n\omega_r$
%(where $n$ is integer) is very small. Specifically, for points in the
%$p$--$e$ space for which there is a frequency $|\omega|\lesssim 10^{-4}M^{-1}$,
%we can no longer meet our accuracy standard of $<10^{-4}$ (fractionally) in
%the final SF. This difficulty has to do with the fact that the dimensionality
%of the set of Lorenz-gauge dipole equations degenerates at $\omega\to 0$.
%The issue will be discussed in a forthcoming paper \cite{BGS:lowL}, where we
%shall also proposed ways in which the problem could be mitigated.
%In practice, the problem singles out certain narrow ``bands'' in the
%$p$--$e$ plane, where our code becomes less accurate. However, even with
%the current version of the code, this should not pose a serious difficulty
%in any foreseeable application, as the narrow ``gaps'' in parameter space could
%easily be interpolated across.

Even within the above ``workable domain'', the current code is
discouragingly slow. It takes a few days to compute the SF along
a single strong-field geodesic, which makes it impractical to cover
the entire parameter space of inspirals at sufficient resolution
(having in mind the development of theoretical gravitational
waveform templates for astrophysical inspirals).
There are, however, various ways in which one may improve the
computational efficiency and speed. Most obvious, one can use
distributed computing---our algorithm should be easily amenable
for distribution on a cluster, since different $l$ modes can be
calculated in parallel. Other natural approaches include the use
of mesh refinement (see Thornburg's recent report \cite{Thornburg:2009mw})
and/or higher-order finite-difference schemes. One may also seek to reduce
the computational cost attached to the initial stage of the numerical
evolution (when spurious initial waves dominate) by iteratively improving
the initial conditions for the evolution---this idea is
already being implemented successfully in a 2+1D framework \cite{BDprep}. Other ideas represent
a more significant deviation from our approach: (i) work entirely in the
frequency domain, making full use of the method of extended homogeneous
solutions \cite{Barack:2008ms} (this is likely to prove most efficient
with smaller eccentricities); or (ii) abandon finite differencing altogether and
instead use finite elements or other pseudospectral techniques
\cite{Canizares:2009ay,Field:2009kk}, benefiting from their natural
flexibility in accommodating multiple lengthscales.

Nonetheless, some interesting applications are already possible
with the current version of our code, and we have presented one of them
in Sec.\ \ref{Sec:ISCO}. Our computation of the ISCO frequency shift
represents the first physically-meaningful new result coming out of the
SF program, and it has already informed both Numerical-Relativistic
calculations \cite{Lousto:2009mf,Lousto:2009ka} and EOB/PN theory
\cite{Damour:2009sm}. Further ideas for SF/EOB synergy were recently discussed
by Damour in \cite{Damour:2009sm}. In particular, Damour showed that a computation of the (gauge invariant)
conservative SF correction to the precession rate of the periapsis, for slightly eccentric orbits, will give access to the presently unknown 4PN (and possibly higher-order) parameters of EOB theory.  We are currently working to extract the necessary SF data to facilitate this calculation [these are essentially the coefficients $F_0^r$, $F^r_{1}$ and $F^1_{\varphi}$ of Eqs.\ (\ref{eq160}) and (\ref{eq150}), as functions of $r_0$]. Our preliminary results show excellent agreement with the analytic EOB predictions at 2PN and 3PN, and we are hoping to publish these findings elsewhere \cite{EOBGSF}.

More generally, our code allows to tackle the calculation of
post-geodesic [$O(\mu)$] precession effects at {\em any} eccentricity,
not necessarily small. This would not only provide a handle on the `Q'
function of EOB theory, but can also directly inform the computation
of conservative effects in inspiral trajectories. Information on the
periastron advance as a function of $p,e$ could, for example, be
incorporated into an orbital evolution scheme \`{a} la Pound \& Poisson
\cite{Pound:2007th}. We are currently investigating this direction.

Finally, we comment briefly on the possibility of extending this work
to the Kerr case (a more elaborate discussion of possible strategies
for attacking the Kerr problem can be found in \cite{Barack:2009ux}).
The framework of the current code, i.e., numerical evolution in 1+1D
is not directly applicable in Kerr spacetime, because the perturbation
equations in Kerr cannot be separated into harmonics in the time domain.
It is possible to tackle the field equations in 2+1D (i.e, separating
the perturbation into azimuthal $m$-modes only) or in full 3+1D,
and there has been considerable progress in that direction in the
past two years---although work so far has been restricted to the
toy problem of a scalar field in Schwarzschild geometry
\cite{Barack:2007jh,Lousto:2008mb,Vega:2009qb,BDprep}.
Schemes for regularizing the SF directly in 2+1D or 3+1D have been
proposed and recently implemented \cite{Barack:2007we,Vega:2009qb,BDprep}.
Alternatively, one may attempt to tackle the field equations in
1+1D by properly accounting for the coupling between different
$l$-harmonics, and a similar strategy may be applicable in the
frequency domain too. A first calculation of the SF in the Kerr case
(using the frequency-domain approach)---for a scalar charge in a
circular equatorial orbit---will be presented in a forthcoming
paper \cite{WBprep}.

%%%%%%%%%%%%%%%%%%%%%%%%%%%%%%%%%%%%%%%%%%%%%%%%%%%%%%%%%%
\section*{ACKNOWLEDGEMENTS}
%%%%%%%%%%%%%%%%%%%%%%%%%%%%%%%%%%%%%%%%%%%%%%%%%%%%%%%%%%
We would like to thank Ryuichi Fujita for providing us (on request)
with the accurate flux data shown in Table~\ref{table:compare_flux}
and for permitting us to include these data here. We are grateful to
Thibault Damour for useful comments on the manuscript.
LB acknowledges support from STFC through Grant No.~PP/E001025/1,
and wishes to thank Misao Sasaki and the Yukawa Institute for
Theoretical Physics in Kyoto (where part of this work took place)
for their hospitality. NS acknowledges support from Monbukagaku-sho
Grant-in-Aid for the global COE program ``The Next Generation of
Physics, Spun from Universality and Emergence''. 

\appendix

%%%%%%%%%%%%%%%%%%%%%%%%%%%%%%%%%%%%%%%%%%%%%%%%%%%%%%%%%%
\section{Algebraic reconstruction of the metric components in the
Lorenz gauge}
\label{app:reconst}
%%%%%%%%%%%%%%%%%%%%%%%%%%%%%%%%%%%%%%%%%%%%%%%%%%%%%%%%%%

We prescribe here the construction of the various
components of the (trace-reversed) Lorenz-gauge metric perturbation
$\bar h_{\alpha\beta}$ in terms of the 10 time-radial scalar-like
functions $\bar h^{(i)lm}$ introduced in Eq.\ (\ref{eq:decomp}). In the following
$Y^{lm}=Y^{lm}(\theta,\varphi)$ are the standard spherical harmonics,
$f=1-2M/r$, and for brevity we suppress the multipolar indices $l,m$ in $\bar h^{(i)lm}$.
Recall that the functions $\bar h^{(i)lm}$---our basic numerical evolution
variables---are obtained by solving the coupled set (\ref{eq:field-eqs}).

The metric perturbation is reconstructed through
%~~~~~~~~~~~~~~~~~~~~~~~~~~~~~~~~~~~~~~~~~~~~~~~~~~~~~~~~~~~~~~~~~~~~~~~
\begin{equation}\label{h construction1}
h_{\alpha\beta}=\frac{\mu}{2r}\sum_{l=0}^{\infty}
\sum_{m=-l}^{l} h^{lm}_{\alpha\beta},
\end{equation}
%~~~~~~~~~~~~~~~~~~~~~~~~~~~~~~~~~~~~~~~~~~~~~~~~~~~~~~~~~~~~~~~~~~~~~~~
where the various Schwarzschild components are given by
%~~~~~~~~~~~~~~~~~~~~~~~~~~~~~~~~~~~~~~~~~~~~~~~~~~~~~~~~~~~~~~~~~~~~~~~
\begin{eqnarray}\label{h construction2}
h^{lm}_{tt}&=& \left(\bar h^{(1)}+f\bar h^{(6)}\right)Y^{lm}, \nonumber\\
h^{lm}_{tr}&=& f^{-1}\bar h^{(2)}Y^{lm}, \nonumber\\
h^{lm}_{rr}&=& f^{-2}\left(\bar h^{(1)}-f\bar h^{(6)}\right)Y^{lm}, \nonumber\\
h^{lm}_{t\theta}&=& r\left(\bar h^{(4)}Y^{lm}_{\rm V1}
                   +\bar h^{(8)}Y^{lm}_{\rm V2}\right),\nonumber\\
h^{lm}_{t\varphi}&=& r\sin\theta\left(\bar h^{(4)}Y^{lm}_{\rm V2}
                   -\bar h^{(8)}Y^{lm}_{\rm V1}\right),\nonumber\\
h^{lm}_{r\theta}&=& rf^{-1}\left(\bar h^{(5)}Y^{lm}_{\rm V1}
                   +\bar h^{(9)}Y^{lm}_{\rm V2}\right), \nonumber\\
h^{lm}_{r\varphi}&=& rf^{-1}\sin\theta\left(\bar h^{(5)}Y^{lm}_{\rm V2}
                   -\bar h^{(9)}Y^{lm}_{\rm V1}\right), \nonumber\\
h^{lm}_{\theta\theta}&=& r^2\left(\bar h^{(3)}Y^{lm}
 +\bar h^{(7)}Y^{lm}_{\rm T1}+\bar h^{(10)}Y^{lm}_{\rm T2}\right), \nonumber\\
h^{lm}_{\theta\varphi}&=& r^2\sin\theta\left(\bar h^{(7)}Y^{lm}_{\rm T2}
                   -\bar h^{(10)}Y^{lm}_{\rm T1}\right), \nonumber\\
h^{lm}_{\varphi\varphi}&=&
             r^2\sin^2\theta\,\left(\bar h^{(3)}Y^{lm}
             -\bar h^{(7)}Y^{lm}_{\rm T1}-\bar h^{(10)}Y^{lm}_{\rm T2}\right).
\end{eqnarray}
%~~~~~~~~~~~~~~~~~~~~~~~~~~~~~~~~~~~~~~~~~~~~~~~~~~~~~~~~~~~~~~~~~~~~~~~
The angular functions appearing in these relations are defined as
\begin{eqnarray}
Y^{lm}_{\rm V1}(\theta,\varphi) &\equiv & \frac{1}{l(l+1)}\, Y^{lm}_{,\theta}
    \quad \text{(for $l>0$)}, \nonumber\\
Y^{lm}_{\rm V2}(\theta,\varphi)&\equiv & \frac{1}{l(l+1)}\,\sin^{-1}\theta\,Y^{lm}_{,\varphi}
     \quad \text{(for $l>0$)},\nonumber\\
Y^{lm}_{\rm T1}(\theta,\varphi) &\equiv & \frac{(l-2)!}{(l+2)!} \left[
\sin\theta \left(\sin^{-1}\theta\, Y^{lm}_{,\theta}\right)_{,\theta}
-\sin^{-2}\theta\, Y^{lm}_{,\varphi\varphi}\right]
     \quad \text{(for $l>1$)}, \nonumber\\
Y^{lm}_{\rm T2}(\theta,\varphi) &\equiv & \frac{2(l-2)!}{(l+2)!}\,
\left(\sin^{-1}\theta\, Y^{lm}_{,\varphi}\right)_{,\theta}
 \quad \text{(for $l>1$)}.
\end{eqnarray}
We note that for $l=0,1$ we have $h^{(7,10)}=0$ identically,
and that for $l=0$ we have additionally $h^{(4,5,8,9)}=0$.

%%%%%%%%%%%%%%%%%%%%%%%%%%%%%%%%%%%%%%%%%%%%%%%%%%%%%%%%%%
\section{Field equations and gauge conditions for the
perturbation functions $\bar{h}^{(i)lm}(r,t)$}
\label{app:field-eqs}
%%%%%%%%%%%%%%%%%%%%%%%%%%%%%%%%%%%%%%%%%%%%%%%%%%%%%%%%%%

We give here explicit expressions for the various terms appearing
in our 1+1D field equation (\ref{eq:field-eqs}). In what follows
$f=1-2M/r$, $f'=2M/r^2$, $r_*$ is the standard tortoise radial coordinate
defined through $dr/dr_*=f(r)$, and $v=t+r_*$. We also denote
$\lambda= (l+2)(l-1)$.

The terms ${\cal M}^{(i)l}_{\;(j)}\bar h^{(j)lm}$ in
Eq.~(\ref{eq:field-eqs}) read
%~~~~~~~~~~~~~~~~~~~~~~~~~~~~~~~~~~~~~~~~~~~~~~~~~~~~~~~~~~~~~~~~~~~~~~~
\begin{equation} \label{eq:M-term1}
{\cal M}^{(1)}_{\;(j)}\bar h^{(j)}
=
\frac{\partial}{\partial r_*}
\left( \frac{1}{2}ff'\bar{h}^{(3)} \right)
+ \frac{(r-4M)f}{2r^3} (\bar{h}^{(1)} - \bar{h}^{(5)})
- \frac{(r^2-10Mr+20M^2)f}{2r^4} \bar{h}^{(3)}
- \frac{(r-6M)f^2}{2r^3} \bar{h}^{(6)},
\end{equation}
\begin{eqnarray} \label{eq:M-term2}
{\cal M}^{(2)}_{\;(j)}\bar h^{(j)}
&=&
\frac{\partial}{\partial r_*}
\left( \frac{1}{2}ff'\bar{h}^{(3)} \right)
+ \frac{\partial}{\partial v}
  \left[ f'(\bar{h}^{(2)} - \bar{h}^{(1)})\right]
- \frac{3Mf}{r^3} \bar{h}^{(1)}
+ \frac{(r+2M)f}{2r^3} \bar{h}^{(2)}
+ \frac{(3r-8M)Mf}{r^4} \bar{h}^{(3)}
\nonumber \\ &&
- \frac{f^2}{2r^2} \bar{h}^{(4)}
+ \frac{ff'}{2r} \bar{h}^{(5)}
+ \frac{f^2f'}{r} \bar{h}^{(6)},
\end{eqnarray}
\begin{equation} \label{eq:M-term3}
{\cal M}^{(3)}_{\;(j)}\bar h^{(j)}
=
-\frac{f}{2r^2}\left[\bar h^{(1)}-\bar h^{(5)}
-\left(1-\frac{4M}{r}\right)
 \left(\bar h^{(3)}+\bar h^{(6)}\right)\right],
\end{equation}
\begin{equation} \label{eq:M-term4}
{\cal M}^{(4)}_{\;(j)}\bar h^{(j)}
=
\frac{\partial}{\partial v}
\left[ \frac{f'}{2}(\bar{h}^{(4)} - \bar{h}^{(5)})\right]
- \frac{l(l+1)f}{2r^2} \bar{h}^{(2)}
- \frac{Mf}{2r^3} \bar{h}^{(4)}
- \frac{2Mf}{r^3} \bar{h}^{(5)}
- \frac{l(l+1)ff'}{4r} \bar{h}^{(6)}
+ \frac{ff'}{4r} \bar{h}^{(7)},
\end{equation}
\begin{equation} \label{eq:M-term5}
{\cal M}^{(5)}_{\;(j)}\bar h^{(j)}
=
\frac{f}{r^2}\left[
\left( 1-\frac{9M}{2r} \right) \bar h^{(5)}
- \frac{1}{2}l(l+1)\left(\bar h^{(1)}-f\bar h^{(3)}\right)
+\frac{1}{2}\left(1-\frac{3M}{r}\right)
 \left(l(l+1)\bar h^{(6)}-\bar h^{(7)}\right)
\right],
\end{equation}
\begin{equation} \label{eq:M-term6}
{\cal M}^{(6)}_{\;(j)}\bar h^{(j)}
=
-\frac{f}{2r^2}\left[
 \bar h^{(1)}-\bar h^{(5)}
 -\left(1-\frac{4M}{r}\right)
  \left(\bar h^{(3)}+\bar h^{(6)}\right)
\right],
\end{equation}
\begin{equation} \label{eq:M-term7}
{\cal M}^{(7)}_{\;(j)}\bar h^{(j)}
=
-\frac{f}{2r^2}\left(
 \bar h^{(7)} + \lambda\,\bar h^{(5)}
\right),
\end{equation}
\begin{equation} \label{eq:M-term8}
{\cal M}^{(8)}_{\;(j)}\bar h^{(j)}
=
\frac{\partial}{\partial v}
\left[ \frac{f'}{2}(\bar{h}^{(8)} - \bar{h}^{(9)}) \right]
- \frac{Mf}{2r^3} \bar h^{(8)}
- \frac{2Mf}{r^3} \bar h^{(9)}
+ \frac{Mf}{2r^3} \bar h^{(10)},
\end{equation}
\begin{equation} \label{eq:M-term9}
{\cal M}^{(9)}_{\;(j)}\bar h^{(j)}
=
\frac{f}{r^2}\left(1-\frac{9M}{2r}\right)\bar h^{(9)}
-\frac{f}{2r^2}\left(1-\frac{3M}{r}\right)\,\bar h^{(10)},
\end{equation}
\begin{equation} \label{eq:M-term10}
{\cal M}^{(10)}_{\;(j)}\bar h^{(j)}
=
-\frac{f}{2r^2}\left(\bar h^{(10)}+\lambda\,\bar h^{(9)}\right).
\end{equation}
These expression are the same as those given in Appendix A of paper I,
although we write them here in a slightly different form, more amenable
to discretization in $v,u$ coordinates.

The various source terms $S^{(i)lm}$ in Eq.~(\ref{eq:field-eqs}) are
given (referring to Sec.~\ref{subsec:orbit} for notation) by
%~~~~~~~~~~~~~~~~~~~~~~~~~~~~~~~~~~~~~~~~~~~~~~~~~~~~~~~~~~~~~~~~~~~~~~~
\begin{eqnarray}
S_{lm}^{(1)} &=&
\frac{4\pi f_{\rm p}^2}{{\cal E}r_{\rm p}^3}
\left( 2{\cal E}^2 r_{\rm p}^2 - f_{\rm p} r_{\rm p}^2
- {\cal L}^2 f_{\rm p} \right)
Y_{lm}^*(\pi/2, \varphi_{\rm p}),
\label{eq:source-term1} \\
S_{lm}^{(2)} &=&
-\frac{8\pi f_{\rm p}^2}{r_{\rm p}}u^r Y_{lm}^*(\pi/2, \varphi_{\rm p}),
\label{eq:source-term2} \\
S_{lm}^{(3)} &=&
\frac{4\pi}{{\cal E}r_{\rm p}^3}f_{\rm p}^2(r_{\rm p}^2+{\cal L}^2)
Y_{lm}^*(\pi/2, \varphi_{\rm p}),
\label{eq:source-term3} \\
S_{lm}^{(4)} &=&
\frac{8\pi imf_{\rm p}^2{\cal L}}{r_{\rm p}^2}
Y_{lm}^*(\pi/2, \varphi_{\rm p}),
\label{eq:source-term4} \\
S_{lm}^{(5)} &=&
-\frac{8\pi imf_{\rm p}^2 u^r {\cal L}}{r_{\rm p}^2 {\cal E}}
Y_{lm}^*(\pi/2, \varphi_{\rm p}),
\label{eq:source-term5} \\
S_{lm}^{(6)} &=&
\frac{4\pi f_{\rm p}^2 {\cal L}^2}{r_{\rm p}^3{\cal E}}
Y_{lm}^*(\pi/2, \varphi_{\rm p}),
\label{eq:source-term6} \\
S_{lm}^{(7)} &=&
\left[ l(l+1) - 2m^2 \right] S_{lm}^{(6)},
\label{eq:source-term7} \\
S_{lm}^{(8)} &=&
\frac{8\pi f_{\rm p}^2 {\cal L}}{r_{\rm p}^2}
Y_{lm,\theta}^*(\pi/2, \varphi_{\rm p}),
\label{eq:source-term8} \\
S_{lm}^{(9)} &=&
-\frac{8\pi f_{\rm p}^2 u^r {\cal L}}{r_{\rm p}^2 {\cal E}}
Y_{lm,\theta}^*(\pi/2, \varphi_{\rm p}),
\label{eq:source-term9} \\
S_{lm}^{(10)} &=&
\frac{8\pi imf_{\rm p}^2 {\cal L}^2}{r_{\rm p}^3{\cal E}}
Y_{lm,\theta}^*(\pi/2, \varphi_{\rm p}).
\label{eq:source-term10}
\end{eqnarray}
%%~~~~~~~~~~~~~~~~~~~~~~~~~~~~~~~~~~~~~~~~~~~~~~~~~~~~~~~~~~~~~~~~~~~~~~~

The 1+1D field equations (\ref{eq:field-eqs}) are supplemented by 4 elliptic
``constraints'' stemming from the Lorenz-gauge conditions (\ref{eq:Lgauge-condition}).
These constraint equations read
%%~~~~~~~~~~~~~~~~~~~~~~~~~~~~~~~~~~~~~~~~~~~~~~~~~~~~~~~~~~~~~~~~~~~~~~~
\begin{mathletters} \label{gauge}
\begin{equation}\label{gauge1}
%H_{1}^{lm}(r,t)\equiv
-\bar h^{(1)}_{,t}
+ f \left( -\bar h^{(3)}_{,t} + \bar h^{(2)}_{,r}
+ \frac{\bar h^{(2)} - \bar h^{(4)}}{r}\right)
= 0,
\end{equation}
\begin{eqnarray}\label{gauge2}
%H_{2}^{lm}(r,t)\equiv
\bar h^{(2)}_{,t}
-f \bar h^{(1)}_{,r}+f^2\bar h^{(3)}_{,r}
-\frac{f}{r}\left(
 \bar h^{(1)}-\bar h^{(5)}-f\bar h^{(3)}-2f \bar h^{(6)}\right)
= 0,
\end{eqnarray}
\begin{eqnarray}\label{gauge3}
%H_3^{lm}(r,t)\equiv
\bar h^{(4)}_{,t}
-\frac{f}{r} \left(
r \bar h^{(5)}_{,r} + 2\bar h^{(5)}
+ l(l+1)\,\bar h^{(6)} - \bar h^{(7)} \right)
= 0,
\end{eqnarray}
\begin{eqnarray}\label{gauge4}
%H_4^{lm}(r,t)\equiv
\bar h^{(8)}_{,t}
-\frac{f}{r}\left(
r \bar h^{(9)}_{,r}
+ 2\bar h^{(9)} - \bar h^{(10)} \right)
= 0.
\end{eqnarray}
\end{mathletters}
%%~~~~~~~~~~~~~~~~~~~~~~~~~~~~~~~~~~~~~~~~~~~~~~~~~~~~~~~~~~~~~~~~~~~~~~~

\nopagebreak
%%%%%%%%%%%%%%%%%%%%%%%%%%%%%%%%%%%%%%%%%%%%%%%%%%%%%%%%%%
\section{Construction of the full-force spherical-harmonic modes}
\label{app:Ffull}
%%%%%%%%%%%%%%%%%%%%%%%%%%%%%%%%%%%%%%%%%%%%%%%%%%%%%%%%%%

Following are the explicit values of the functions ${\cal F}^{\alpha lm}_{(n)}$
appearing in Eq.\ (\ref{eq:Flfull}):
%~~~~~~~~~~~~~~~~~~~~~~~~~~~~~~~~~~~~~~~~~~~~~~~~~~~~~~~~~~~~~~~~~~~~~~~
\begin{eqnarray} \label{calF}
{\cal F}^{\alpha lm}_{(-3)} &=&
\zeta_{(+3)}^{lm} f_{6\pm}^{\alpha lm}
+ \xi_{(+3)}^{lm}f_{7\pm}^{\alpha lm},
\nonumber \\
{\cal F}^{\alpha lm}_{(-2)} &=&
\alpha_{(+2)}^{lm} f_{1\pm}^{\alpha lm}
+ \beta_{(+2)}^{lm} f_{2\pm}^{\alpha lm}
+ \gamma_{(+2)}^{lm} f_{3\pm}^{\alpha lm},
\nonumber \\
{\cal F}^{\alpha lm}_{(-1)} &=&
\epsilon_{(+1)}^{lm} f_{4\pm}^{\alpha lm}
+ \delta_{(+1)}^{lm} f_{5\pm}^{\alpha lm}
+ \zeta_{(+1)}^{lm} f_{6\pm}^{\alpha lm}
+ \xi_{(+1)}^{lm} f_{7\pm}^{\alpha lm},
\nonumber\\
{\cal F}^{\alpha lm}_{(0)} &=&
f_{0\pm}^{\alpha lm}
+ \alpha_{(0)}^{\alpha lm} f_{1\pm}^{\alpha lm}
+ \beta_{(0)}^{lm} f_{2\pm}^{\alpha lm}
+ \gamma_{(0)}^{lm} f_{3\pm}^{\alpha lm},
\nonumber \\
{\cal F}^{\alpha lm}_{(+1)} &=&
\epsilon_{(-1)}^{lm} f_{4\pm}^{\alpha lm}
+ \delta_{(-1)}^{lm} f_{5\pm}^{\alpha lm}
+ \zeta_{(-1)}^{lm} f_{6\pm}^{\alpha lm}
+ \xi_{(-1)}^{lm} f_{7\pm}^{\alpha lm},
\nonumber \\
{\cal F}^{\alpha lm}_{(+2)} &=&
\alpha_{(-2)}^{lm} f_{1\pm}^{\alpha lm}
+ \beta_{(-2)}^{lm} f_{2\pm}^{\alpha lm}
+ \gamma_{(-2)}^{lm} f_{3\pm}^{\alpha lm},
\nonumber \\
{\cal F}^{\alpha lm}_{(+3)} &=&
\zeta_{(-3)}^{lm} f_{6\pm}^{\alpha lm}
+ \xi_{(-3)}^{lm}f_{7\pm}^{\alpha lm}.
\end{eqnarray}
%~~~~~~~~~~~~~~~~~~~~~~~~~~~~~~~~~~~~~~~~~~~~~~~~~~~~~~~~~~~~~~~~~~~~~~~
The various functions $f_{n\pm}^{\alpha lm}$ are those appearing in Eq.\
(\ref{eq:Ffull2}), and below we give these functions explicitly
for $\alpha=t,r$ (the values for $\varphi,\theta$ are not needed in this work).
The values of the various $l,m$-dependent coefficients $\alpha,\beta,\gamma,\epsilon,\zeta,\xi$
in the above expressions are given in Eqs.\ (\ref{alphacoeff})--(\ref{xicoeff}) of Appendix \ref{app:iden}.
In what follows we use the notation $\hat{\cal L}\equiv{\cal L}/r_{\rm p}$, and for brevity
we suppress the indices $l,m$ as well as the subscript $\pm$; it is to be
understood that the $t$ and $r$ derivatives in the following expressions are taken from either
``outside'' or ``inside'' the orbit, yielding, in general, two different
one-sided values which correspond to the suppressed $+$ or $-$ subscripts.

\nopagebreak
%\subsection{$t$-component of $f_{n\pm}^{\alpha lm}$}

For the $t$ component we have---

\begin{eqnarray}
f_{0}^t &=&
\frac{{\cal E}}{4f_{\rm p}^4} \left[
 (u^r)^3
 +\left( {\cal E}^2 \frac{r_{\rm p}+4M}{r_{\rm p}}
        - 2f_{\rm p} \right) u^r
 -imf_{\rm p} \hat{\cal L}
 \left( (u^r)^2 + {\cal E}^2 - 2f_{\rm p} \right)
\right] \bar{h}^{(1)}
\nonumber \\ &&
-\frac{r_{\rm p}}{4f_{\rm p}^4} \left[
 ({\cal E}^2+f_{\rm p})(u^r)^2 + {\cal E}^2({\cal E}^2-f_{\rm p})
\right] \bar{h}_{,t}^{(1)}
-\frac{r_{\rm p}}{4f_{\rm p}^4} u^r{\cal E} \left[
 (u^r)^2 + {\cal E}^2 -2f_{\rm p}
\right] \bar{h}_{,r_*}^{(1)}
\nonumber \\ &&
+\frac{1}{2f_{\rm p}^4} \left[
 (u^r)^2\left( {\cal E}^2 \frac{r_{\rm p}+M}{r_{\rm p}}
           - f_{\rm p} \frac{r_{\rm p}-M}{r_{\rm p}}
    \right)
  +\frac{M}{r_{\rm p}}{\cal E}^2({\cal E}^2-f_{\rm p})
  -im f_{\rm p} u^r \hat{\cal L} ({\cal E}^2-f_{\rm p})
 \right] \bar{h}^{(2)}
\nonumber \\ &&
-\frac{r_{\rm p} u^r}{2f_{\rm p}^4}
 \left[ {\cal E}^3 \bar{h}_{,t}^{(2)}
  +u^r ({\cal E}^2-f_{\rm p}) \bar{h}_{,r_*}^{(2)}
 \right]
%\nonumber \\ &&
+\frac{{\cal E} \hat{\cal L}^2}{4f_{\rm p}}
 \left( u^r - im\hat{\cal L} \right) \bar{h}^{(3)}
-\frac{r_{\rm p} \hat{\cal L}^2}{4f_{\rm p}^2}
 \left[
 ({\cal E}^2 + f_{\rm p}) \bar{h}_{,t}^{(3)}
 + u^r {\cal E} \bar{h}_{,r_*}^{(3)}
 \right]
\nonumber \\ &&
+\frac{m}{2l(l+1)f_{\rm p}^3}\hat{\cal L} \left[
 2iu^r \left( {\cal E}^2\frac{r_{\rm p}-M}{r_{\rm p}} - f_{\rm p}^2 \right)
 + mf_{\rm p}\hat{\cal L}({\cal E}^2-f_{\rm p})
\right] \bar{h}^{(4)}
\nonumber \\ &&
-\frac{imr_{\rm p}\hat{\cal L}}{2l(l+1)f_{\rm p}^3}
 \left[
 {\cal E}^3 \bar{h}_{,t}^{(4)}
 + u^r ({\cal E}^2 - f_{\rm p}) \bar{h}_{,r_*}^{(4)}
 \right]
%\nonumber \\ &&
+\frac{m{\cal E}\hat{\cal L}}{2l(l+1)f_{\rm p}^3} \left[
 i(u^r)^2\left(2-\frac{3M}{r_{\rm p}}\right)
 + \frac{iM}{r_{\rm p}}{\cal E}^2
 + mf_{\rm p} u^r\hat{\cal L}
\right] \bar{h}^{(5)}
\nonumber \\ &&
-\frac{imr_{\rm p} u^r \hat{\cal L}}{2l(l+1)f_{\rm p}^3}
 \left[
 ({\cal E}^2 + f_{\rm p}) \bar{h}_{,t}^{(5)}
 + u^r {\cal E} \bar{h}_{,r_*}^{(5)}
\right]
%\nonumber \\ &&
+\frac{{\cal E}}{4f_{\rm p}}
(-u^r+im\hat{\cal L}) \bar{h}^{(6)}
+\frac{r_{\rm p}}{4f_{\rm p}^2}\left[
  ({\cal E}^2-f_{\rm p}) \bar{h}_{,t}^{(6)}
  + u^r{\cal E} \bar{h}_{,r_*}^{(6)}
 \right]
\nonumber \\ &&
-\frac{m{\cal E}\hat{\cal L}^2}{4l(l+1)\lambda f_{\rm p}}
 \left[
  3mu^r - i\hat{\cal L}(4+m^2)
 \right] \bar{h}^{(7)}
+\frac{m^2r_{\rm p}\hat{\cal L}^2}{4l(l+1)\lambda f_{\rm p}^2}
 \left[
 ({\cal E}^2+f_{\rm p}) \bar{h}_{,t}^{(7)}
 + u^r{\cal E} \bar{h}_{,r_*}^{(7)}
 \right],
\end{eqnarray}

\begin{eqnarray}
f_{1}^t &=&
-\frac{u^r{\cal E}\hat{\cal L}^2}{2f_{\rm p}^2} \bar{h}^{(1)}
-\frac{\hat{\cal L}^2}{2f_{\rm p}^2}
 ({\cal E}^2-f_{\rm p}) \bar{h}^{(2)}
+\frac{u^r{\cal E}\hat{\cal L}^2}{2f_{\rm p}} \bar{h}^{(3)}
-\frac{im{\cal E}\hat{\cal L}^3}{2l(l+1)f_{\rm p}}
 \bar{h}^{(5)}
\nonumber \\ &&
+\frac{{\cal E}\hat{\cal L}^2}{4f_{\rm p}}
 (3u^r-im\hat{\cal L}) \bar{h}^{(6)}
-\frac{r_{\rm p}}{4f_{\rm p}^2}\hat{\cal L}^2
 ({\cal E}^2+f_{\rm p}) \bar{h}_{,t}^{(6)}
-\frac{r_{\rm p}}{4f_{\rm p}^2}u^r{\cal E}\hat{\cal L}^2
 \bar{h}_{,r_*}^{(6)}
-\frac{im{\cal E}\hat{\cal L}^3}{l(l+1)\lambda f_{\rm p}}
 \bar{h}^{(7)},
\end{eqnarray}

\begin{eqnarray}
f_{2}^t &=&
-\frac{\hat{\cal L}^2({\cal E}^2-f_{\rm p})}
      {2l(l+1)f_{\rm p}^2} \bar{h}^{(4)}
-\frac{u^r{\cal E}\hat{\cal L}^2}
      {2l(l+1)f_{\rm p}^2} \bar{h}^{(5)}
\nonumber \\ &&
+\frac{\hat{\cal L}^2}
      {4l(l+1)\lambda f_{\rm p}^2} \left[
  f_{\rm p}{\cal E}(3u^r-5im\hat{\cal L}) \bar{h}^{(7)}
  -r_{\rm p}({\cal E}^2+f_{\rm p}) \bar{h}_{,t}^{(7)}
  -r_{\rm p} u^r{\cal E} \bar{h}_{,r_*}^{(7)}
 \right],
\end{eqnarray}

\begin{eqnarray}
f_{3}^t &=&
\frac{\hat{\cal L}^2}
     {4l(l+1)\lambda f_{\rm p}^2} \left[
  -f_{\rm p}{\cal E}(3u^r-im\hat{\cal L}) \bar{h}^{(7)}
  +r_{\rm p}({\cal E}^2+f_{\rm p}) \bar{h}_{,t}^{(7)}
  +r_{\rm p} u^r{\cal E} \bar{h}_{,r_*}^{(7)}
 \right],
\end{eqnarray}

\begin{eqnarray}
f_{4}^t &=&
-\frac{im\hat{\cal L}^2}{2l(l+1)f_{\rm p}^2}
 \left[
  ({\cal E}^2-f_{\rm p}) \bar{h}^{(8)}
  +u^r{\cal E} \bar{h}^{(9)}
 \right]
\nonumber \\ &&
+\frac{im\hat{\cal L}^2}{2l(l+1)\lambda f_{\rm p}^2}
 \left[
  f_{\rm p}{\cal E}(3u^r-2im\hat{\cal L}) \bar{h}^{(10)}
  -r_{\rm p}({\cal E}^2+f_{\rm p}) \bar{h}_{,t}^{(10)}
  -r_{\rm p} u^r{\cal E} \bar{h}_{,r_*}^{(10)}
 \right],
\end{eqnarray}

\begin{eqnarray}
f_{5}^t &=&
-\frac{u^r\hat{\cal L}}{l(l+1)f_{\rm p}^3}
 \left({\cal E}^2\frac{r_{\rm p}-M}{r_{\rm p}} - f_{\rm p}^2\right)
 \bar{h}^{(8)}
+\frac{r_{\rm p}{\cal E}^3\hat{\cal L}}
      {2l(l+1)f_{\rm p}^3} \bar{h}_{,t}^{(8)}
+\frac{r_{\rm p} u^r\hat{\cal L}}{2l(l+1)f_{\rm p}^3}
 ({\cal E}^2 - f_{\rm p}) \bar{h}_{,r_*}^{(8)}
\nonumber \\ &&
-\frac{{\cal E}\hat{\cal L}}{2l(l+1)f_{\rm p}^3}
 \left[
  (u^r)^2\frac{2r_{\rm p}-3M}{r_{\rm p}} + \frac{M}{r_{\rm p}}{\cal E}^2
 \right] \bar{h}^{(9)}
+\frac{r_{\rm p} u^r\hat{\cal L}}{2l(l+1)f_{\rm p}^3}
 ({\cal E}^2+f_{\rm p}) \bar{h}_{,t}^{(9)}
+\frac{r_{\rm p}(u^r)^2{\cal E}\hat{\cal L}}
      {2l(l+1)f_{\rm p}^3} \bar{h}_{,r_*}^{(9)}
\nonumber \\ &&
+\frac{(m^2-1){\cal E}\hat{\cal L}^3}
      {2l(l+1)\lambda f_{\rm p}} \bar{h}^{(10)},
\end{eqnarray}

\begin{equation}
f_{6}^t =
\frac{{\cal E}\hat{\cal L}^3}{2l(l+1)\lambda f_{\rm p}}
(\lambda\bar{h}^{(9)}+\bar{h}^{(10)}),
\end{equation}

\begin{equation}
f_{7}^t =
\frac{{\cal E}\hat{\cal L}^3}{2l(l+1)\lambda f_{\rm p}}
\bar{h}^{(10)}.
\end{equation}

For the $r$ component we have---
\begin{eqnarray}
f_{0}^r &=&
\frac{1}{4f_{\rm p}^3}\left[
 (u^r)^4 - imf_{\rm p}(u^r)^3\hat{\cal L}
 +(u^r)^2\left(\frac{r_{\rm p}+4M}{r_{\rm p}}{\cal E}^2 + f_{\rm p}\right)
 -im u^r f_{\rm p}\hat{\cal L}({\cal E}^2+2f_{\rm p})
 -f_{\rm p}{\cal E}^2\left(1-\frac{4M}{r_{\rm p}}\right)
\right] \bar{h}^{(1)}
\nonumber \\ &&
-\frac{r_{\rm p} u^r{\cal E}}{4f_{\rm p}^3}
 [(u^r)^2+{\cal E}^2+2f_{\rm p}] \bar{h}_{,t}^{(1)}
-\frac{r_{\rm p}}{4f_{\rm p}^3}
 [(u^r)^4 + (u^r)^2({\cal E}^2+f_{\rm p})-f_{\rm p}{\cal E}^2]
 \bar{h}_{,r_*}^{(1)}
\nonumber \\ &&
+\frac{{\cal E}}{2f_{\rm p}^3}
 \left[
  (u^r)^3\left(1+\frac{M}{r_{\rm p}}\right)
  -imf_{\rm p}(u^r)^2\hat{\cal L}
  +\frac{M}{r_{\rm p}}u^r({\cal E}^2+2f_{\rm p})
  -imf_{\rm p}^2\hat{\cal L}
 \right] \bar{h}^{(2)}
\nonumber \\ &&
-\frac{r_{\rm p}{\cal E}^2}{2f_{\rm p}^3}[(u^r)^2+f_{\rm p}]
 \bar{h}_{,t}^{(2)}
-\frac{r_{\rm p}(u^r)^3{\cal E}}{2f_{\rm p}^3} \bar{h}_{,r_*}^{(2)}
\nonumber \\ &&
+\frac{1}{4f_{\rm p}}\left[
  -(u^r)^4 + im\hat{\cal L}(u^r)^3
  +{\cal E}^2(u^r)^2
  - (imu^r\hat{\cal L} + f_{\rm p})({\cal E}^2 - f_{\rm p})
 \right] \bar{h}^{(3)}
\nonumber \\ &&
+\frac{r_{\rm p} u^r{\cal E}}{4f_{\rm p}^2}
 [(u^r)^2-{\cal E}^2+f_{\rm p}] \bar{h}_{,t}^{(3)}
+\frac{r_{\rm p}}{4f_{\rm p}^2}
 [(u^r)^4-{\cal E}^2(u^r)^2+f_{\rm p}({\cal E}^2-f_{\rm p})]
 \bar{h}_{,r_*}^{(3)}
\nonumber \\ &&
+\frac{imu^r{\cal E}\hat{\cal L}}{2l(l+1)f_{\rm p}^2}
 \left[
  2u^r\left(1-\frac{M}{r_{\rm p}}\right)
  -imf_{\rm p}\hat{\cal L}
 \right] \bar{h}^{(4)}
%\nonumber \\ &&
-\frac{im r_{\rm p} u^r{\cal E}^2\hat{\cal L}}
      {2l(l+1)f_{\rm p}^2} \bar{h}_{,t}^{(4)}
-\frac{im r_{\rm p}{\cal E}\hat{\cal L}}{2l(l+1)f_{\rm p}^2}
 [(u^r)^2-f_{\rm p}] \bar{h}_{,r_*}^{(4)}
\nonumber \\ &&
+\frac{im\hat{\cal L}}{2l(l+1)f_{\rm p}^2}
 \left[
  (u^r)^3\left(2-\frac{3M}{r_{\rm p}}\right)
  -im f_{\rm p}(u^r)^2\hat{\cal L}
  +u^r\left(\frac{M}{r_{\rm p}}{\cal E}^2 + 2f_{\rm p}^2\right)
  -im f_{\rm p}^2 \hat{\cal L}
 \right] \bar{h}^{(5)}
\nonumber \\ &&
-\frac{imr_{\rm p}{\cal E}\hat{\cal L}}{2l(l+1)f_{\rm p}^2}
 [(u^r)^2+f_{\rm p}] \bar{h}_{,t}^{(5)}
-\frac{imr_{\rm p}(u^r)^3\hat{\cal L}}{2l(l+1)f_{\rm p}^2}
 \bar{h}_{,r_*}^{(5)}
%\nonumber \\ &&
-\frac{1}{4}[(u^r)^2-imu^r\hat{\cal L}+f_{\rm p}]
 \bar{h}^{(6)}
+\frac{r_{\rm p} u^r{\cal E}}{4f_{\rm p}} \bar{h}_{,t}^{(6)}
\nonumber \\ &&
+\frac{r_{\rm p}}{4f_{\rm p}}[(u^r)^2+f_{\rm p}] \bar{h}_{,r_*}^{(6)}
+\frac{m\hat{\cal L}^2}{4l(l+1)\lambda}
 [-3m(u^r)^2+iu^r\hat{\cal L}(4+m^2)-mf_{\rm p}]
 \bar{h}^{(7)}
%\nonumber \\ &&
+\frac{m^2r_{\rm p} u^r{\cal E}\hat{\cal L}^2}
      {4l(l+1)\lambda f_{\rm p}} \bar{h}_{,t}^{(7)}
\nonumber \\ &&
+\frac{m^2r_{\rm p}\hat{\cal L}^2}{4l(l+1)\lambda f_{\rm p}}
 [(u^r)^2-f_{\rm p}]\bar{h}_{,r_*}^{(7)},
\end{eqnarray}

\begin{eqnarray}
f_1^r &=&
-\frac{\hat{\cal L}^2}{2f_{\rm p}}
 [(u^r)^2+f_{\rm p}] \bar{h}^{(1)}
-\frac{u^r{\cal E}\hat{\cal L}^2}{2f_{\rm p}}
 \bar{h}^{(2)}
+\frac{\hat{\cal L}^2}{2}
 [(u^r)^2+f_{\rm p}] \bar{h}^{(3)}
-\frac{imu^r\hat{\cal L}^3}{2l(l+1)} \bar{h}^{(5)}
\nonumber \\ &&
+\frac{\hat{\cal L}^2}{4}
 [3(u^r)^2-imu^r\hat{\cal L}+f_{\rm p}] \bar{h}^{(6)}
-\frac{r_{\rm p} u^r{\cal E}\hat{\cal L}^2}{4f_{\rm p}}
 \bar{h}_{,t}^{(6)}
-\frac{r_{\rm p}\hat{\cal L}^2}{4f_{\rm p}}
 [(u^r)^2-f_{\rm p}] \bar{h}_{,r_*}^{(6)}
-\frac{imu^r\hat{\cal L}^3}{l(l+1)\lambda}
 \bar{h}^{(7)},
\end{eqnarray}

\begin{eqnarray}
f_2^r &=&
-\frac{\hat{\cal L}^2}{2l(l+1)f_{\rm p}}
 \left[
  u^r{\cal E} \bar{h}^{(4)}
  +((u^r)^2+f_{\rm p}) \bar{h}^{(5)}
 \right]
\nonumber \\ &&
+\frac{\hat{\cal L}^2}{4l(l+1)\lambda f_{\rm p}}
 \left[
  f_{\rm p}(3(u^r)^2-5imu^r\hat{\cal L}+f_{\rm p}) \bar{h}^{(7)}
  -r_{\rm p} u^r{\cal E} \bar{h}_{,t}^{(7)}
  -r_{\rm p} ( (u^r)^2-f_{\rm p} ) \bar{h}_{,r_*}^{(7)}
 \right],
\end{eqnarray}

\begin{eqnarray}
f_3^r &=&
\frac{\hat{\cal L}^2}{4l(l+1)\lambda f_{\rm p}}
\left[
 -f_{\rm p}(3(u^r)^2-imu^r\hat{\cal L}+f_{\rm p}) \bar{h}^{(7)}
 +r_{\rm p} u^r{\cal E} \bar{h}_{,t}^{(7)}
 +r_{\rm p} ( (u^r)^2-f_{\rm p} ) \bar{h}_{,r_*}^{(7)}
\right],
\end{eqnarray}

\begin{eqnarray}
f_4^r &=&
-\frac{im\hat{\cal L}^2}{2l(l+1)f_{\rm p}}
 \left[
  u^r{\cal E} \bar{h}^{(8)}
  + ((u^r)^2+f_{\rm p}) \bar{h}^{(9)}
 \right]
\nonumber \\ &&
+\frac{im\hat{\cal L}^2}{2l(l+1)\lambda f_{\rm p}}
 \left[
  f_{\rm p}(3(u^r)^2-2imu^r\hat{\cal L}+f_{\rm p}) \bar{h}^{(10)}
  - r_{\rm p} u^r{\cal E} \bar{h}_{,t}^{(10)}
  - r_{\rm p} ((u^r)^2-f_{\rm p}) \bar{h}_{,r_*}^{(10)}
 \right],
\end{eqnarray}

\begin{eqnarray}
f_5^r &=&
-\frac{(u^r)^2{\cal E}\hat{\cal L}}
      {l(l+1)f_{\rm p}^2}
 \left(1-\frac{M}{r_{\rm p}}\right) \bar{h}^{(8)}
+\frac{r_{\rm p} u^r{\cal E}^2\hat{\cal L}}
      {2l(l+1)f_{\rm p}^2} \bar{h}_{,t}^{(8)}
+\frac{r_{\rm p}{\cal E}\hat{\cal L}}
      {2l(l+1)f_{\rm p}^2}
 [(u^r)^2-f_{\rm p}] \bar{h}_{,r_*}^{(8)}
\nonumber \\ &&
-\frac{u^r\hat{\cal L}}{2l(l+1)f_{\rm p}^2}
 \left[
  \frac{2r_{\rm p}-3M}{r_{\rm p}}(u^r)^2
  +\frac{M}{r_{\rm p}}{\cal E}^2 + 2f_{\rm p}^2
 \right] \bar{h}^{(9)}
+\frac{r_{\rm p}{\cal E}\hat{\cal L}}
      {2l(l+1)f_{\rm p}^2}
 [(u^r)^2+f_{\rm p}] \bar{h}_{,t}^{(9)}
\nonumber \\ &&
+\frac{r_{\rm p}(u^r)^3\hat{\cal L}}{2l(l+1)f_{\rm p}^2}
 \bar{h}_{,r_*}^{(9)}
+\frac{(m^2-1)u^r\hat{\cal L}^3}{2l(l+1)\lambda}
 \bar{h}^{(10)},
\end{eqnarray}

\begin{equation}
f_6^r =
\frac{u^r\hat{\cal L}^3}{2l(l+1)\lambda}
 [\lambda \bar{h}^{(9)}+\bar{h}^{(10)}],
\end{equation}

\begin{equation}
f_7^r =
\frac{u^r\hat{\cal L}^3}{2l(l+1)\lambda}
 \bar{h}^{(10)}.
\end{equation}

%%%%%%%%%%%%%%%%%%%%%%%%%%%%%%%%%%%%%%%%%%%%%%%%%%%%%%%%%%
\section{Useful identities}
\label{app:iden}
%%%%%%%%%%%%%%%%%%%%%%%%%%%%%%%%%%%%%%%%%%%%%%%%%%%%%%%%%%

The following identities are used in deriving Eq.\ (\ref{eq:Flfull})
for the full-force modes.
In these relations $Y^{lm}$ are the standard spherical harmonics,
and the identities are valid for any values of $l,m$.

\begin{eqnarray} \label{eq:Ylm-relation}
\sin^2\theta Y^{lm} &=&
\alpha^{lm}_{(+2)}Y^{l+2,m}+\alpha^{lm}_{(0)} Y^{lm}
+\alpha^{lm}_{(-2)}Y^{l-2,m},
\nonumber \\
\cos\theta\sin\theta Y^{lm}_{,\theta} &=&
\beta^{lm}_{(+2)}Y^{l+2,m}+\beta^{lm}_{(0)} Y^{lm}
+\beta^{lm}_{(-2)}Y^{l-2,m},
\nonumber \\
\sin^2\theta Y^{lm}_{,\theta\theta} &=&
\gamma^{lm}_{(+2)}Y^{l+2,m}+\gamma^{lm}_{(0)} Y^{lm}
+\gamma^{lm}_{(-2)}Y^{l-2,m},
\nonumber \\
\sin\theta Y^{lm}_{,\theta} &=&
\delta^{lm}_{(+1)}Y^{l+1,m}+\delta^{lm}_{(-1)}Y^{l-1,m},
\nonumber \\
\cos\theta Y^{lm}-\sin\theta Y^{lm}_{,\theta} &=&
\epsilon^{lm}_{(+1)}Y^{l+1,m}+\epsilon^{lm}_{(-1)}Y^{l-1,m},
\nonumber \\
\sin^3\theta Y^{lm}_{,\theta} &=&
\zeta^{lm}_{(+3)}Y^{l+3,m}+\zeta^{lm}_{(+1)}Y^{l+1,m}
+\zeta^{lm}_{(-1)}Y^{l-1,m}+\zeta^{lm}_{(-3)}Y^{l-3,m},
\nonumber \\
\cos\theta\sin^2\theta Y^{lm}_{,\theta\theta} &=&
\xi^{lm}_{(+3)}Y^{l+3,m}+\xi^{lm}_{+}Y^{l+1,m}
+\xi^{lm}_{(-1)}Y^{l-1,m}+\xi^{lm}_{(-3)}Y^{l-3,m}.
\end{eqnarray}
Here the various coefficients are all constructed from
\begin{equation}
C_{lm}=\left[\frac{l^2-m^2}{(2l+1)(2l-1)}\right]^{1/2}
\end{equation}
using
\begin{mathletters}
\begin{equation}\label{alphacoeff}
\alpha^{lm}_{(+2)}=-C_{l+1,m}C_{l+2,m}, \quad\quad
\alpha^{lm}_{(0)}=1-C_{lm}^2-C_{l+1,m}^2, \quad\quad
\alpha^{lm}_{(-2)}=-C_{lm} C_{l-1,m},
\end{equation}
\begin{equation}\label{eq:beta}
\beta^{lm}_{(+2)}=lC_{l+1,m}C_{l+2,m}, \quad\quad
\beta^{lm}_{(0)}=lC_{l+1,m}^2-(l+1)C_{lm}^2, \quad\quad
\beta^{lm}_{(-2)}=-(l+1)C_{lm} C_{l-1,m},
\end{equation}
\begin{equation}
\gamma^{lm}_{(+2)}=l^2C_{l+1,m}C_{l+2,m}, \quad\quad
\gamma^{lm}_{(0)}=m^2-l(l+1)+l^2C_{l+1,m}^2+(l+1)^2C_{lm}^2, \quad\quad
\gamma^{lm}_{(-2)}=(l+1)^2C_{lm} C_{l-1,m},
\end{equation}
\begin{equation}
\delta^{lm}_{(+1)}=lC_{l+1,m}, \quad\quad
\delta^{lm}_{(-1)}=-(l+1)C_{lm},
\end{equation}
\begin{equation}
\epsilon^{lm}_{(+1)}=(1-l)C_{l+1,m}, \quad\quad
\epsilon^{lm}_{(-1)}=(l+2)C_{lm},
\end{equation}
\begin{eqnarray}
\zeta^{lm}_{(+3)}&=&-lC_{l+1,m}C_{l+2,m}C_{l+3,m},
\nonumber \\
\zeta^{lm}_{(+1)}&=&
C_{l+1,m}\left[l\left(1-C^2_{l+1,m}-C^2_{l+2,m}\right)+(l+1)C^2_{l,m}\right],
\nonumber \\
\zeta^{lm}_{(-1)}&=&
-C_{l,m}\left[(l+1)\left(1-C^2_{l-1,m}-C^2_{l,m}\right)+lC^2_{l+1,m}\right],
\nonumber \\
\zeta^{lm}_{(-3)}&=&(l+1)C_{l,m}C_{l-1,m}C_{l-2,m},
\end{eqnarray}
\begin{eqnarray}
\xi^{lm}_{(+3)}&=&l^2C_{l+1,m}C_{l+2,m}C_{l+3,m},
\nonumber \\
\xi^{lm}_{(+1)}&=&
C_{l+1,m}
\left[m^2-l(l+1)+l^2 C^2_{l+1,m}
+(l+1)^2 C^2_{l,m}+l^2 C^2_{l+2,m}\right],
\nonumber \\
\xi^{lm}_{(-1)}&=&
C_{l,m}
\left[m^2-l(l+1)+l^2 C^2_{l+1,m}+(l+1)^2 C^2_{l,m}
+(l+1)^2 C^2_{l-1,m}\right],
\nonumber \\
\xi^{lm}_{(-3)}&=&(l+1)^2C_{l,m}C_{l-1,m}C_{l-2,m}. \label{xicoeff}
\end{eqnarray}
\end{mathletters}

%%%%%%%%%%%%%%%%%%%%%%%%%%%%%%%%%%%%%%%%%%%%%%%%%%%%%%%%%%
\section{Jump conditions for the perturbation modes and their derivatives}
\label{app:jump-condition}
%%%%%%%%%%%%%%%%%%%%%%%%%%%%%%%%%%%%%%%%%%%%%%%%%%%%%%%%%%

As explained in Sec.\ \ref{subsec:FDS}, our finite-difference scheme
makes use of formal jump conditions for the perturbation modes $\bar h^{(i)lm}$ and
their (first through fourth) derivatives across the particle's orbit.
In this appendix we derive the necessary conditions. Our derivation
refers to a specific (yet generic) point $x_0$ along the orbit, with
known coordinates $(r_0,t_0)$ or $(u_0,v_0)$, and velocity components
$\dot{u}_0\equiv \left.du_{\rm p}/d\tau\right|_{x_0}$ and $\dot{v}_0\equiv
\left.dv_{\rm p}/d\tau\right|_{x_0}$,
where $\tau$ is proper time along the orbit. We will use the notation
$\left[A\right]_0\equiv A(x_0^+)-A(x_0^-)$, where $A(x_0^\pm)$ are the values
of a 1+1D field $A(r,t)$ calculated by taking the limits $t\to t_0$
and $r\to r_0^{\pm}$. In this appendix we shall omit
multipolar indices $lm$ for brevity.

\subsection{Continuity condition for $\bar{h}^{(i)}$}

The Lorenz-gauge perturbation modes $\bar{h}^{(i)}$ are all continuous
at any point along the orbit. This can be verified, for example, by
noticing that the distributional form $\bar{h}^{(i)}=
\bar{h}_+^{(i)}(r,t)\theta[r-r_{\rm p}(t)]+\bar{h}_-^{(i)}(r,t)\theta[r_{\rm p}(t)-r]$
is indeed a solution of the perturbation equations (\ref{eq:field-eqs})
only if the homogeneous solutions $\bar{h}_+^{(i)}$ and $\bar{h}_-^{(i)}$
satisfy $\bar{h}_+^{(i)}=\bar{h}_-^{(i)}$ along the worldline. Hence, for
any $i$ we have
\begin{equation} \label{eq:jc-0th}
\left[ \bar{h}^{(i)} \right]_0 = 0.
\end{equation}

%%~~~~~~~~~~~~~~~~~~~~~~~~~~~~~~~~~~~~~~~~~~~~~~~~~~~~~~~~~~~~~~~~~~~~~~~~~~
%\begin{figure}[htb]
%\includegraphics[width=6cm]{jc_point.eps}
%\caption{The point in which the jump conditions are derived.}
%\label{fig:jc-point}
%\end{figure}
%%~~~~~~~~~~~~~~~~~~~~~~~~~~~~~~~~~~~~~~~~~~~~~~~~~~~~~~~~~~~~~~~~~~~~~~~~~~

\subsection{Jump conditions for the 1st derivatives}

Let us we re-express the field equations (\ref{eq:field-eqs}) in the form
%~~~~~~~~~~~~~~~~~~~~~~~~~~~~~~~~~~~~~~~~~~~~~~~~~~~~~~~~~~~~~~~~~~~~~~~~~~
\begin{equation} \label{eq:field-eqs2}
\bar{h}_{,uv}^{(i)} + P^{(i)}
=
\int d\tau \tilde{S}^{(i)}(x_{\rm p}(\tau))
\delta(u-u_{\rm p}(\tau)) \delta(v-v_{\rm p}(\tau)),
\end{equation}
%~~~~~~~~~~~~~~~~~~~~~~~~~~~~~~~~~~~~~~~~~~~~~~~~~~~~~~~~~~~~~~~~~~~~~~~~~~
where
%~~~~~~~~~~~~~~~~~~~~~~~~~~~~~~~~~~~~~~~~~~~~~~~~~~~~~~~~~~~~~~~~~~~~~~~~~~
\begin{equation}
P^{(i)} \equiv
V(r)\bar{h}^{(i)}
+ {\cal M}^{(i)}_{\;(j)}\bar h^{(j)}, \quad\quad
\tilde{S}^{(i)} \equiv 2{\cal E} f_{\rm p}^{-2}S^{(i)},
\label{eq:def_P_jc}
\end{equation}
%~~~~~~~~~~~~~~~~~~~~~~~~~~~~~~~~~~~~~~~~~~~~~~~~~~~~~~~~~~~~~~~~~~~~~~~~~~
with ${\cal M}^{(i)}_{\;(j)}$ and $S^{(i)}$ being the quantities given
in Appendix \ref{app:field-eqs}, and $V(r)$ being defined in Eq.\ (\ref{eq:box}).
Now consider formally integrating Eq.\ (\ref{eq:field-eqs2}) along the ray
$v=v_0$ over an infinitesimal interval $(u_0-\epsilon,u_0+\epsilon)$
across $x_0$. The integral of the left-hand side yields
%~~~~~~~~~~~~~~~~~~~~~~~~~~~~~~~~~~~~~~~~~~~~~~~~~~~~~~~~~~~~~~~~~~~~~~~~~~
\begin{equation}
\left.  \bar{h}_{,v}^{(i)}
\right|_{u_0-\epsilon}^{u_0+\epsilon}
+ \int_{u_0-\epsilon}^{u_0+\epsilon} P^{(i)} du
\quad \rightarrow \quad
-\left[\bar{h}_{,v}^{(i)} \right]_0
\end{equation}
%~~~~~~~~~~~~~~~~~~~~~~~~~~~~~~~~~~~~~~~~~~~~~~~~~~~~~~~~~~~~~~~~~~~~~~~~~~
as $\epsilon\to 0$, since $P^{(i)}$ is bounded. Similarly integrating the
right-hand side of Eq.\ (\ref{eq:field-eqs2}) one finds
%~~~~~~~~~~~~~~~~~~~~~~~~~~~~~~~~~~~~~~~~~~~~~~~~~~~~~~~~~~~~~~~~~~~~~~~~~~
\begin{equation}
\int_{u_0-\epsilon}^{u_0+\epsilon}du \int d\tau
\tilde{S}^{(i)}(x_{\rm p}(\tau))
\delta[u-u_{\rm p}(\tau)] \delta[v_0-v_{\rm p}(\tau)]
=\dot v_0^{-1} \tilde{S}^{(i)}(x_0).
\end{equation}
%~~~~~~~~~~~~~~~~~~~~~~~~~~~~~~~~~~~~~~~~~~~~~~~~~~~~~~~~~~~~~~~~~~~~~~~~~~
The desired jump condition is therefore
%~~~~~~~~~~~~~~~~~~~~~~~~~~~~~~~~~~~~~~~~~~~~~~~~~~~~~~~~~~~~~~~~~~~~~~~~~~
\begin{equation}
\left[\bar{h}_{,v}^{(i)} \right]_0
=
-\dot{v}_0^{-1}\tilde{S}^{(i)}(x_0).
\end{equation}
%~~~~~~~~~~~~~~~~~~~~~~~~~~~~~~~~~~~~~~~~~~~~~~~~~~~~~~~~~~~~~~~~~~~~~~~~~~
Similarly integrating the field equations
along $u=u_0$ over an infinitesimal interval
$(v_0-\epsilon,v_0+\epsilon)$ one also obtains
%~~~~~~~~~~~~~~~~~~~~~~~~~~~~~~~~~~~~~~~~~~~~~~~~~~~~~~~~~~~~~~~~~~~~~~~~~~
\begin{equation}
\left[\bar{h}_{,u}^{(i)} \right]_0
=\dot{u}_0^{-1}\tilde{S}^{(i)}(x_0).
\end{equation}
%~~~~~~~~~~~~~~~~~~~~~~~~~~~~~~~~~~~~~~~~~~~~~~~~~~~~~~~~~~~~~~~~~~~~~~~~~~

\subsection{Jump conditions for the 2nd derivatives}

It is straightforward to derive the jump condition in the mixed $uv$-derivative.
From Eq.\ (\ref{eq:field-eqs2}) we immediately have
%~~~~~~~~~~~~~~~~~~~~~~~~~~~~~~~~~~~~~~~~~~~~~~~~~~~~~~~~~~~~~~~~~~~~~~~~~~
\begin{equation}
\left[ \partial_u\partial_v \bar{h}^{(i)} \right]_0
=-\left[ P^{(i)} \right]_0,
\end{equation}
%~~~~~~~~~~~~~~~~~~~~~~~~~~~~~~~~~~~~~~~~~~~~~~~~~~~~~~~~~~~~~~~~~~~~~~~~~~
as the source term is supported only on the worldline. Recall
$P^{(i)}$ involves the perturbation $\bar h^{(i)}$ and its first
derivatives only, and so the jump in $P^{(i)}$ can be readily calculated
from the jump conditions obtained above.

To derive a jump condition for the $vv$-derivative, we first take the
$v$ derivatives of the field equations (\ref{eq:field-eqs2}), and then
integrate with respect to $u$ along $v=v_0$ over $(u_0-\epsilon,u_0+\epsilon)$.
We thereby obtain
%~~~~~~~~~~~~~~~~~~~~~~~~~~~~~~~~~~~~~~~~~~~~~~~~~~~~~~~~~~~~~~~~~~~~~~~~~~
\begin{equation} \label{eq:dv-field-eqs}
-\left[\bar{h}_{,vv}^{(i)}\right]_0=\lim_{\epsilon\to 0}
\int_{u_0-\epsilon}^{u_0+\epsilon} du \left[ -P_{,v}^{(i)}(u,v_0)
+\int d\tau \tilde{S}^{(i)}(x_{\rm p})\delta'[v_0-v_{\rm p}(\tau)]\delta[u-u_{\rm p}(\tau)]
\right].
\end{equation}
%~~~~~~~~~~~~~~~~~~~~~~~~~~~~~~~~~~~~~~~~~~~~~~~~~~~~~~~~~~~~~~~~~~~~~~~~~~
Now, for $P^{(i)}$ we may write
%~~~~~~~~~~~~~~~~~~~~~~~~~~~~~~~~~~~~~~~~~~~~~~~~~~~~~~~~~~~~~~~~~~~~~~~~~~
\begin{equation} \label{eq:Pterm}
P^{(i)} =P_+^{(i)}(u,v)\theta[v-v_{\rm p}(u)] + P_-^{(i)}(u,v)\theta[v_{\rm p}(u)-v],
\end{equation}
%~~~~~~~~~~~~~~~~~~~~~~~~~~~~~~~~~~~~~~~~~~~~~~~~~~~~~~~~~~~~~~~~~~~~~~~~~~
in which $P_{\pm}^{(i)}$ are smooth and where $v_{\rm p}(u)$ is the value of $v$
at which the outgoing ray of retarded time $u$ intersects the worldline.
Using this form we obtain
%~~~~~~~~~~~~~~~~~~~~~~~~~~~~~~~~~~~~~~~~~~~~~~~~~~~~~~~~~~~~~~~~~~~~~~~~~~
\begin{eqnarray} \label{eq:vv-jump2}
\int_{u_0-\epsilon}^{u_0+\epsilon} P^{(i)}_{,v}(u,v_0) du
&=&
\int_{u_0-\epsilon}^{u_0+\epsilon} du
\left\{\left(P^{(i)}_+(u,v_0)-P^{(i)}_-(u,v_0)\right) \delta[v_0-v_{\rm p}(u)]
\right. \nonumber\\
&& \left.
+P_{+,v}^{(i)}(u,v_0)\theta[v_0-v_{\rm p}(u)]+P_{-,v}^{(i)}(u,v_0)\theta[v_{\rm p}(u)-v_0]
\right\} \quad\to\quad
(\dot u_0/\dot v_0)\left[ P^{(i)} \right]_0
\end{eqnarray}
%~~~~~~~~~~~~~~~~~~~~~~~~~~~~~~~~~~~~~~~~~~~~~~~~~~~~~~~~~~~~~~~~~~~~~~~~~~
as $\epsilon\to 0$, since $P_{\pm,v}^{(i)}$ are bounded.
Lastly, for the integral involving $\tilde{S}^{(i)}$ in Eq.~(\ref{eq:dv-field-eqs})
we have
%~~~~~~~~~~~~~~~~~~~~~~~~~~~~~~~~~~~~~~~~~~~~~~~~~~~~~~~~~~~~~~~~~~~~~~~~~~
\begin{eqnarray} \label{eq:vv-jump3}
&& \hspace*{-2cm}
\int_{u_0-\epsilon}^{u_0+\epsilon} du
\int d\tau
\tilde{S}^{(i)}(x_{\rm p}(\tau))\delta'[v_0-v_{\rm p}(\tau)]\delta[u-u_{\rm p}(\tau)]
\nonumber \\ &=&
\int d\tau \tilde{S}^{(i)}(x_{\rm p}(\tau))\delta'[v_0-v_{\rm p}(\tau)]
\theta[u_{\rm p}(\tau)-(u_0-\epsilon)] \theta[(u_0+\epsilon)-u_{\rm p}(\tau)]
\nonumber \\ &=&
\int _{\tau_0-\delta_-}^{\tau_0+\delta_+} d\tau
\tilde{S}^{(i)}(x_{\rm p}(\tau))\delta'[v_0-v_{\rm p}(\tau)]
\nonumber \\ &=&
-\int _{\tau_0-\delta_-}^{\tau_0+\delta_+} d\tau
\tilde{S}^{(i)}(x_{\rm p}) \dot v_{\rm p}^{-1}
\frac{d}{d\tau}\delta[v_{\rm p}(\tau)-v_0]
\nonumber \\ &=&
\int _{\tau_0-\delta_-}^{\tau_0+\delta_+} d\tau
\frac{d}{d\tau}\left[
\dot v_{\rm p}^{-1}\tilde{S}^{(i)}(x_{\rm p}(\tau))
\right]\delta[v_{\rm p}(\tau)-v_0]
\nonumber \\ &=&
\dot v_0^{-1} \frac{d}{d\tau}\left[
\dot v_{\rm p}^{-1}\tilde{S}^{(i)}(x_{\rm p})
\right]\!\!\bigg|_{x_0},
\end{eqnarray}
%~~~~~~~~~~~~~~~~~~~~~~~~~~~~~~~~~~~~~~~~~~~~~~~~~~~~~~~~~~~~~~~~~~~~~~~~~~
where $\tau_0$ is the value of $\tau$ at $x_0$, and $\delta_{\pm}$ are the
values of $\tau$ at which the two outgoing rays of constant retarded times
$u_0\pm \epsilon$ cross the worldline. Combining the above results, we
arrive at
\begin{equation}
\left[\bar{h}_{,vv}^{(i)} \right]_0
=
(\dot u_0/\dot v_0)\left[ P^{(i)} \right]_0
-
\dot v_0^{-1} \frac{d}{d\tau}\left[
\dot v_{\rm p}^{-1}\tilde{S}^{(i)}(x_{\rm p})
\right]\!\!\bigg|_{x_0}.
\end{equation}

The jump condition for the $uu$-derivatives is derived in a similar
fashion. The result is
\begin{equation}
\left[\bar{h}^{(i)}_{,uu} \right]_0
=
(\dot v_0/\dot u_0)\left[ P^{(i)} \right]_0
+
\dot u_0^{-1} \frac{d}{d\tau}\left[
\dot u_{\rm p}^{-1}\tilde{S}^{(i)}(x_{\rm p})
\right]\!\!\bigg|_{x_0}.
\end{equation}

\subsection{Jump conditions for the 3rd derivatives}

The jumps in the two mixed third derivatives are readily obtained
by considering the derivatives of the field equations (\ref{eq:field-eqs2})
with respect to $v$ and with respect to $u$:
%~~~~~~~~~~~~~~~~~~~~~~~~~~~~~~~~~~~~~~~~~~~~~~~~~~~~~~~~~~~~~~~~~~~~~~~~~~
\begin{eqnarray}
\left[\bar{h}_{,vvu}^{(i)} \right]_0
&=&
-\left[P^{(i)}_{,v} \right]_0, \\
\left[\bar{h}_{,uuv}^{(i)} \right]_0
&=&
-\left[P^{(i)}_{,u} \right]_0.
\end{eqnarray}
%~~~~~~~~~~~~~~~~~~~~~~~~~~~~~~~~~~~~~~~~~~~~~~~~~~~~~~~~~~~~~~~~~~~~~~~~~~
Here the jumps in $P^{(i)}_{,u}$ and $P^{(i)}_{,v}$ can be obtained from the
jumps in the first and second derivatives of $\bar h^{(i)}$ derived above.

To obtain the jump condition for the $vvv$ derivative, we differentiate
the field equations (\ref{eq:field-eqs2}) twice with respect to $v$, and then
integrate with respect to $u$ along $v=v_0$ over $(u_0-\epsilon,u_0+\epsilon)$.
This yields
%~~~~~~~~~~~~~~~~~~~~~~~~~~~~~~~~~~~~~~~~~~~~~~~~~~~~~~~~~~~~~~~~~~~~~~~~~~
\begin{equation} \label{eq:dvv-field-eqs}
-\left[\bar{h}_{,vvv}^{(i)}\right]_0=\lim_{\epsilon\to 0}
\int_{u_0-\epsilon}^{u_0+\epsilon} du \left[ -P_{,vv}^{(i)}(u,v_0)
+\int d\tau \tilde{S}^{(i)}(x_{\rm p})\delta''[v_0-v_{\rm p}(\tau)]\delta[u-u_{\rm p}(\tau)]
\right].
\end{equation}
%~~~~~~~~~~~~~~~~~~~~~~~~~~~~~~~~~~~~~~~~~~~~~~~~~~~~~~~~~~~~~~~~~~~~~~~~~~
Using Eq.~(\ref{eq:Pterm}) again, we obtain
%~~~~~~~~~~~~~~~~~~~~~~~~~~~~~~~~~~~~~~~~~~~~~~~~~~~~~~~~~~~~~~~~~~~~~~~~~~
\begin{eqnarray} \label{eq:vvv-jump2}
\int_{u_0-\epsilon}^{u_0+\epsilon} P^{(i)}_{,vv}(u,v_0) du
&=&
\int_{u_0-\epsilon}^{u_0+\epsilon} du
\left\{2\left(P^{(i)}_{+,v}(u,v_0)-P^{(i)}_{-,v}(u,v_0)\right) \delta[v_0-v_{\rm p}(u)]
\right. \nonumber\\
&&
+\left(P^{(i)}_{+}(u,v_0)-P^{(i)}_{-}(u,v_0)\right) \delta'[v_0-v_{\rm p}(u)]
\nonumber\\
&& \left.
+P_{+,vv}^{(i)}(u,v_0)\theta[v_0-v_{\rm p}(u)]+P_{-,vv}^{(i)}(u,v_0)\theta[v_{\rm p}(u)-v_0]
\right\}
\nonumber\\
&&\to\quad
2\left(v_0'\right)^{-1}
\left[P^{(i)}_{,v}\right]_0
+\left(v_0'\right)^{-2}
 \left[P^{(i)}_{,u}\right]_0
+u_0''\left[P^{(i)}\right]_0
\end{eqnarray}
%~~~~~~~~~~~~~~~~~~~~~~~~~~~~~~~~~~~~~~~~~~~~~~~~~~~~~~~~~~~~~~~~~~~~~~~~~~
as $\epsilon\to 0$.
Here we have introduced $v'_{\rm p}\equiv dv_{\rm p}(u)/du$ and
$v'_0\equiv v'_{\rm p}(u_0)=\dot v_0/\dot u_0$; and similarly, defining
$u_{\rm p}(v)$ as the value of $u$ at which the incoming ray with advanced
time $v$ intersects the worldline, we introduced
$u'_{\rm p}\equiv du_{\rm p}(v)/dv$ and $u''_{\rm p}\equiv d^2u_{\rm p}(v)/dv^2$, with
$u'_0\equiv u'_{\rm p}(v_0)=\dot u_0/\dot v_0$ and
$u''_0\equiv u''_{\rm p}(v_0)=(\ddot u_0 \dot v_0 - \dot u_0 \ddot v_0)/\dot v_0^3$.
To evaluate the limit in Eq.\ (\ref{eq:vvv-jump2}) we have used
%~~~~~~~~~~~~~~~~~~~~~~~~~~~~~~~~~~~~~~~~~~~~~~~~~~~~~~~~~~~~~~~~~~~~~~~~~~
\begin{equation}
\int_{u_0-\epsilon}^{u_0+\epsilon}du
\left(P^{(i)}_{+,v}(u,v_0)-P^{(i)}_{-,v}(u,v_0)\right)
\delta[v_0-v_{\rm p}(u)]\to
\left(v_0'\right)^{-1}\left[P^{(i)}_{,v} \right]_0
\end{equation}
%~~~~~~~~~~~~~~~~~~~~~~~~~~~~~~~~~~~~~~~~~~~~~~~~~~~~~~~~~~~~~~~~~~~~~~~~~~
(as $\epsilon\to 0$), along with
%~~~~~~~~~~~~~~~~~~~~~~~~~~~~~~~~~~~~~~~~~~~~~~~~~~~~~~~~~~~~~~~~~~~~~~~~~~
\begin{eqnarray}
&& \hspace*{-2cm}
\int_{u_0-\epsilon}^{u_0+\epsilon}du
\left(P^{(i)}_{+}(u,v_0)-P^{(i)}_{-}(u,v_0)\right) \delta'[v_0-v_{\rm p}(u)]
\nonumber \\ &=&
-\int_{u_0-\epsilon}^{u_0+\epsilon}du
\left\{
\left(P^{(i)}_{+}(u,v_0)-P^{(i)}_{-}(u,v_0)\right)
(v'_{\rm p})^{-1}
\frac{d}{du} \left[
(v'_{\rm p})^{-1}\delta(u-u_{\rm p}(v))
\right] \right\}
\nonumber \\ &=&
\int_{u_0-\epsilon}^{u_0+\epsilon}du
\left\{
\frac{d}{du}\left[
\left(P^{(i)}_{+}(u,v_0)-P^{(i)}_{-}(u,v_0)\right)
(v'_{\rm p})^{-1}
\right] (v'_{\rm p})^{-1}
\delta(u-u_{\rm p}(v)) \right\}
\nonumber \\ &\to&
(v'_0)^{-2} \left[P^{(i)}_{,u}\right]_0
-v_0''(v_0')^{-3}\left[P^{(i)}\right]_0
\nonumber \\ &=&
(v'_0)^{-2} \left[P^{(i)}_{,u}\right]_0
+u_0''\left[P^{(i)}\right]_0.
\end{eqnarray}
%~~~~~~~~~~~~~~~~~~~~~~~~~~~~~~~~~~~~~~~~~~~~~~~~~~~~~~~~~~~~~~~~~~~~~~~~~~
The term involving $\tilde{S}^{(i)}$ in Eq.\ (\ref{eq:dvv-field-eqs}) gives,
upon twice integrating by parts,
%~~~~~~~~~~~~~~~~~~~~~~~~~~~~~~~~~~~~~~~~~~~~~~~~~~~~~~~~~~~~~~~~~~~~~~~~~~
\begin{eqnarray} \label{SS}
\int_{u_0-\epsilon}^{u_0+\epsilon} du
\int d\tau
\tilde{S}^{(i)}(x_{\rm p})\delta''(v_0-v_{\rm p}(\tau))\delta(u-u_{\rm p}(\tau))
&\rightarrow&
\dot v_0^{-1}
\frac{d}{d\tau}\left[
\dot v_{\rm p}^{-1}
\frac{d}{d\tau}\left(
\dot v_{\rm p}^{-1}
\tilde{S}^{(i)}
\right)\right]\!\!\bigg|_{x_0}
\end{eqnarray}
%~~~~~~~~~~~~~~~~~~~~~~~~~~~~~~~~~~~~~~~~~~~~~~~~~~~~~~~~~~~~~~~~~~~~~~~~~~
as $\epsilon\to 0$.
Finally, substituting from Eqs.\ (\ref{eq:vvv-jump2}) and (\ref{SS}) in
Eq.\ (\ref{eq:dvv-field-eqs}) we arrive at
%~~~~~~~~~~~~~~~~~~~~~~~~~~~~~~~~~~~~~~~~~~~~~~~~~~~~~~~~~~~~~~~~~~~~~~~~~~
\begin{eqnarray}
\left[\bar{h}_{,vvv}^{(i)} \right]_0=
2\left(v_0'\right)^{-1}
\left[P^{(i)}_{,v}\right]_0
+\left(v_0'\right)^{-2}
 \left[P^{(i)}_{,u}\right]_0
+u_0''\left[P^{(i)}\right]_0
-\dot v_0^{-1}
\frac{d}{d\tau}\left[
\dot v_{\rm p}^{-1}
\frac{d}{d\tau}\left(
\dot v_{\rm p}^{-1}
\tilde{S}^{(i)}
\right)\right]\!\!\bigg|_{x_0}.
\end{eqnarray}
%~~~~~~~~~~~~~~~~~~~~~~~~~~~~~~~~~~~~~~~~~~~~~~~~~~~~~~~~~~~~~~~~~~~~~~~~~~
The jumps $\left[P^{(i)}_{,u}\right]_0$ and $\left[P^{(i)}_{,v}\right]_0$
can be obtained in a straightforward way from the jumps in the first and
second derivatives of the perturbation, obtained in previous steps.

In a precisely analogous manner, differentiating
the field equations (\ref{eq:field-eqs2}) twice with respect to $u$
and then integrating with respect to $v$ along $u=u_0$ over
$(v_0-\epsilon,v_0+\epsilon)$ gives
%~~~~~~~~~~~~~~~~~~~~~~~~~~~~~~~~~~~~~~~~~~~~~~~~~~~~~~~~~~~~~~~~~~~~~~~~~~
\begin{eqnarray}
\left[\bar{h}_{,uuu}^{(i)} \right]_0
&=&
2\left(u_0'\right)^{-1}
\left[P^{(i)}_{,u}\right]_0
+\left(u_0'\right)^{-2}
 \left[P^{(i)}_{,v}\right]_0
+v_0''\left[P^{(i)}\right]_0
+\dot u_0^{-1}
\frac{d}{d\tau}\left[
\dot u_{\rm p}^{-1}
\frac{d}{d\tau}\left(
\dot u_{\rm p}^{-1}
\tilde{S}^{(i)}
\right)\right]\!\!\bigg|_{x_0},
\end{eqnarray}
%~~~~~~~~~~~~~~~~~~~~~~~~~~~~~~~~~~~~~~~~~~~~~~~~~~~~~~~~~~~~~~~~~~~~~~~~~~
where $v''_0\equiv v''_{\rm p}(u_0)=(\ddot v_0 \dot u_0 - \dot v_0 \ddot u_0)/\dot u_0^3$.

\subsection{Jump conditions for the 4th derivatives}

The jumps in the five 4th-order partial derivatives of $\bar{h}^{(i)}$
are obtained in a similar manner. The results are
%~~~~~~~~~~~~~~~~~~~~~~~~~~~~~~~~~~~~~~~~~~~~~~~~~~~~~~~~~~~~~~~~~~~~~~~~~~
\begin{equation}
\left[\bar{h}_{,uvvv}^{(i)} \right]_0= -\left[P_{,vv}^{(i)}\right]_0,
\quad\quad
\left[\bar{h}_{,uuvv}^{(i)} \right]_0=-\left[P_{,uv}^{(i)} \right]_0,
\quad\quad
\left[\bar{h}_{,uuuv}^{(i)} \right]_0=-\left[P_{,uu}^{(i)} \right]_0,
\end{equation}
%~~~~~~~~~~~~~~~~~~~~~~~~~~~~~~~~~~~~~~~~~~~~~~~~~~~~~~~~~~~~~~~~~~~~~~~~~~
%~~~~~~~~~~~~~~~~~~~~~~~~~~~~~~~~~~~~~~~~~~~~~~~~~~~~~~~~~~~~~~~~~~~~~~~~~~
\begin{eqnarray}
\left[\bar{h}_{,vvvv}^{(i)} \right]_0
&=&
u_0'''\left[P^{(i)}\right]_0
+3u_0''\left[P_{,v}^{(i)}\right]_0
+3u_0''(v_0')^{-1}\left[P_{,u}^{(i)}\right]_0
+3(v_0')^{-1}\left[P_{,vv}^{(i)}\right]_0
+3(v_0')^{-2}\left[P_{,uv}^{(i)}\right]_0
\nonumber \\ &&
+(v_0')^{-3}\left[P_{,uu}^{(i)}\right]_0
-
\dot{v}_0^{-1}
\frac{d}{d\tau}\left\{
\dot{v}_{\rm p}^{-1}
\frac{d}{d\tau}\left[
\dot{v}_{\rm p}^{-1}
\frac{d}{d\tau}\left(
\dot{v}_{\rm p}^{-1}
\tilde{S}^{(i)}
\right)\right]\right\}\!\!\bigg|_{x_0},
\end{eqnarray}
%~~~~~~~~~~~~~~~~~~~~~~~~~~~~~~~~~~~~~~~~~~~~~~~~~~~~~~~~~~~~~~~~~~~~~~~~~~
%~~~~~~~~~~~~~~~~~~~~~~~~~~~~~~~~~~~~~~~~~~~~~~~~~~~~~~~~~~~~~~~~~~~~~~~~~~
\begin{eqnarray}
\left[\bar{h}_{,uuuu}^{(i)} \right]_0
&=&
v_0'''\left[P^{(i)}\right]_0
+3v_0''\left[P_{,u}^{(i)}\right]_0
+3v_0''(u_0')^{-1}\left[P_{,v}^{(i)}\right]_0
+3(u_0')^{-1}\left[P_{,uu}^{(i)}\right]_0
+3(u_0')^{-2}\left[P_{,uv}^{(i)}\right]_0
\nonumber \\ &&
+(u_0')^{-3}\left[P_{,vv}^{(i)}\right]_0
+
\dot{u}_0^{-1}
\frac{d}{d\tau}\left\{
\dot{u}_{\rm p}^{-1}
\frac{d}{d\tau}\left[
\dot{u}_{\rm p}^{-1}
\frac{d}{d\tau}\left(
\dot{u}_{\rm p}^{-1}
\tilde{S}^{(i)}
\right)\right]\right\}\!\!\bigg|_{x_0}.
\end{eqnarray}
%~~~~~~~~~~~~~~~~~~~~~~~~~~~~~~~~~~~~~~~~~~~~~~~~~~~~~~~~~~~~~~~~~~~~~~~~~~
Here $u_0'''=u_{\rm p}'''(r_0)=
(\dddot{u}_0 \dot v_0^2-3\ddot u_0 \ddot v_0 \dot v_0
+3\dot u_0 \ddot v_0^2-\dot u_0 \dddot v_0 \dot v_0)/\dot v_0^5$
and $v_0'''=v_{\rm p}'''(r_0)
=(\dddot{v}_0 \dot u_0^2-3\ddot v_0 \ddot u_0 \dot u_0
+3\dot v_0 \ddot u_0^2-\dot v_0 \dddot u_0 \dot u_0)/\dot u_0^5$.
The jumps in the various $P^{(i)}$ terms are related to the jumps in the
1st, 2nd, and 3rd derivatives of the perturbation, which were obtained
in previous steps.

%%%%%%%%%%%%%%%%%%%%%%%%%%%%%%%%%%%%%%%%%%%%%%%%%%%%%%%%%%
\section{Source terms and jump conditions for slightly
eccentric orbits}
\label{app:e-exp-form}
%%%%%%%%%%%%%%%%%%%%%%%%%%%%%%%%%%%%%%%%%%%%%%%%%%%%%%%%%%

\subsection{Source terms}

We give here explicit expressions for the $O(e)$ source coefficients
$S^{(i)}_1$ defined in Eq.\ (\ref{S0S1}). These are needed in our
discussion of the $e$-expansion method in Sec.\ \ref{Sec:ISCO}.
We use the notation
\begin{equation}
{\cal E}_0\equiv {\cal E}(p=r_0,e=0) = \frac{r_0-2M}{\sqrt{r_0(r_0-3M)}},
\quad\quad
{\cal L}_0\equiv {\cal L}(p=r_0,e=0) = \frac{\sqrt{M}r_0}{\sqrt{r_0-3M}},
\label{eq:EL-small-e}
\end{equation}
along with $f_0\equiv 1-2M/r_0$, and omit multipolar indices $l,m$ for
brevity. 

The coefficients $S^{(i)}_1$ are obtained by formally expanding the
source terms $S^{(i)}$ given in
Eqs.\ (\ref{eq:source-term1})--(\ref{eq:source-term10}) in powers of
$e$ through $O(e)$, assuming the $e$-expansion forms of the trajectory,
\[
 r_{\rm p}(\tau) = r_0(1-e\cos\omega_r\tau) + O(e^2), \quad\quad
 \varphi_{\rm p}(\tau) =
 \omega_\varphi \tau
 + \frac{2\omega_\varphi}{\omega_r} e\sin\omega_r\tau + O(e^2),
\]
where we wrote $\omega_\varphi={\cal L}_0/r_0^2$.
$S_1^{(i)}$ in then obtained as the linear variation of $S^{(i)}$ with respect to $e$
[recall Eq.\ (\ref{S0S1})]. The result can be expressed in the form
\begin{equation} \label{S1}
S_{1}^{(i)} =
\sum_{n=\pm 1}{\cal S}_{1,n}^{(i)}\,
e^{-i\omega_{nm}\tau}, %\quad
%T_{1}^{(i)} =
%\sum_{n=\pm 1}{\cal T}_{1,n}^{(i)}
%\exp\left[-i\omega_{nm}\tau\right],
\end{equation}
where $\omega_{nm}=n\omega_r + m\omega_\varphi$ and the various coefficients are 
\begin{eqnarray}
{\cal S}_{1,n}^{(1)} &=&
\frac{2\pi(r_0-2M)^3}{{\cal E}_0r_0^4(r_0-3M)}
\left[
 r_0 - 6M + 2mn(r_0-2M)\frac{\omega_\varphi}{\omega_r}
\right]Y_{lm}^*(0), \\
{\cal S}_{1,n}^{(2)} &=&
-4\pi in\omega_r f_0^2 Y_{lm}^*(0), \\
{\cal S}_{1,n}^{(3)} &=&
\frac{2\pi {\cal E}_0}{r_0^2}
\left[
r_0 - 4M + 2mn(r_0-2M)\frac{\omega_\varphi}{\omega_r}
\right]Y_{lm}^*(0), \\
{\cal S}_{1,n}^{(4)} &=&
\frac{8\pi im {\cal L}_0(r_0-2M)}{r_0^4}
\left[
r_0 - 4M + mn(r_0-2M)\frac{\omega_\varphi}{\omega_r}
\right]Y_{lm}^*(0), \\
{\cal S}_{1,n}^{(5)} &=&
\frac{4\pi mn\omega_r {\cal L}_0(r_0-2M)^2}{{\cal E}_0 r_0^3}
Y_{lm}^*(0), \\
{\cal S}_{1,n}^{(6)} &=&
\frac{2\pi {\cal L}_0^2 (r_0-2M)}{{\cal E}_0 r_0^5}
\left[
3r_0-10M + 2mn(r_0-2M)\frac{\omega_\varphi}{\omega_r}
\right]Y_{lm}^*(0), \\
{\cal S}_{1,n}^{(7)} &=&
\left[ l(l+1) -2m^2 \right] {\cal S}_{1,n}^{(6)}, \\
{\cal S}_{1,n}^{(8)} &=&
\frac{8\pi {\cal L}_0(r_0-2M)}{r_0^4}
\left[
r_0-4M + mn(r_0-2M)\frac{\omega_\varphi}{\omega_r}
\right] Y_{lm,\theta}^*(0), \\
{\cal S}_{1,n}^{(9)} &=&
-\frac{4\pi in\omega_r {\cal L}_0 (r_0-2M)^2}{{\cal E}_0 r_0^3}
Y_{lm,\theta}^*(0), \\
{\cal S}_{1,n}^{(10)} &=&
\frac{4\pi im {\cal L}_0^2 (r_0-2M)^2}{{\cal E}_0 r_0^5}
\left[
3r_0-10M + 2mn(r_0-2M)\frac{\omega_\varphi}{\omega_r}
\right] Y_{lm,\theta}^*(0), \label{S12}
\end{eqnarray}
Here $\omega_{nm}=n\omega_r + m\omega_\varphi$,
with $Y_{lm}^*(0)$ and $Y_{lm,\theta}^*(0)$ denoting the spherical
harmonics and their $\theta$-derivatives evaluated at
$\theta=\pi/2$ and $\varphi=0$.

Notice that if we also write
\begin{equation}
T_{1}^{(i)} = \sum_{n=\pm 1}{\cal T}_{1,n}^{(i)}\,e^{-i\omega_{nm}\tau}
\quad \text{with}\quad
{\cal T}_{1,n}^{(i)} = \frac{1}{2}r_0 S_0^{(i)}(\varphi_{\rm p}=0),
\end{equation}
then we may split the $O(e)$ part of the metric perturbation into two harmonic components, 
$\bar{h}_1^{(i)}=\bar{h}_{1,+1}^{(i)}+\bar{h}_{1,-1}^{(i)}$,
each of which satisfying the field equation
\begin{equation}
\square \bar h_{1,n}^{(i)} +
{\cal M}^{(i)l}_{\;(j)}\bar h_{1,n}^{(j)}
=
\left[{\cal S}_{1,n}^{(i)}\delta(r-r_0)
+ {\cal T}_{1,n}^{(i)}\delta'(r-r_0)
\right] e^{-i\omega_{nm}\tau}
\quad (n=\pm 1).
\label{eq:eq_h1_expand}
\end{equation}
Evidently, the source for $\bar h_{1,n}^{(i)}$ is simple harmonic in $\tau$, with frequency $\omega_{nm}$. It is clear that the solutions $\bar h_{1,n}^{(i)}(r,t)$ (and their derivatives) will inherit this simple harmonic dependence on $\tau$ along the orbit. Hence we may write, for example
\begin{equation}\label{harmonic}
\left.\partial_t\bar h_{1,n}^{(i)}\right|_{r_{\rm p}(\tau)}=\left.-i\omega_{nm}f_0{\cal E}_0^{-1}\bar h_{1,n}^{(i)}\right|_{r_{\rm p}(\tau)},
\end{equation}
where the factor $f_0{\cal E}_0^{-1}$ is simply $(dt_{\rm p}/d\tau)^{-1}$.

\subsection{Jump conditions}

Here we explain the derivation of the necessary jump conditions for the variables $\bar h_1^{(i)}$
associated with the $O(e)$ piece of the metric perturbation [recall Eq.\ (\ref{hexp})]. Unlike the full physical metric perturbation functions $\bar h^{(i)}$ [or the $O(e^0)$ variables $\bar h_0^{(i)}$], the functions $\bar h_1^{(i)}$ are in general discontinuous across the particle's worldline, because the field equation which defines them, Eq.\ (\ref{h_1}), is sourced by a derivative of a delta function. For our TD algorithm we need to derive analytic jump conditions for $\bar h_1^{(i)}$ itself as well as for its first to fourth derivatives along the orbit. Our task here will be somewhat simpler than in the case of the full perturbation $\bar h^{(i)}$ (analyzed in Appendix \ref{app:jump-condition}) thanks to the simple harmonic dependence of $\bar h_1^{(i)}$ on $t$ along the orbit, expressed in Eq.\ (\ref{harmonic}).

We start by reexpressing the field equations (\ref{eq:eq_h1_expand}) in the form
\begin{equation}
\left( \partial_{r_*}^2 - \partial_t^2 \right)
\bar h_{1,n}^{(i)} - 4P_{1,n}^{(i)}
=
-4 \left[{\cal S}_{1,n}^{(i)}\delta(r-r_0)
+ {\cal T}_{1,n}^{(i)}\delta'(r-r_0)
\right] e^{-i\omega_{nm}\tau}
\quad (n=\pm 1),
\label{eq:eq_h1_expand2}
\end{equation}
where the definition of $P_{1,n}^{(i)}$ is similar to that of $P^{(i)}$ in
Eq.\ (\ref{eq:def_P_jc}), simply replacing $\bar{h}^{(i)}\to \bar{h}_{1,n}^{(i)}$.
Substituting 
\begin{equation}
\bar{h}_{1,n}^{(i)} =
\bar{h}_{1,n}^{(i)-} \theta(r_0-r)
+ \bar{h}_{1,n}^{(i)+} \theta(r-r_0)
\end{equation}
and comparing the $\delta'(r-r_0)$ and $\delta(r-r_0)$-terms on both sides of the resulting equation, we readily obtain the jump formulas
\begin{eqnarray}
\left[ \bar{h}_{1,n}^{(i)} \right]_0
&=&
-\frac{4}{f_0^2}{\cal T}_{1,n}e^{-i\omega_{nm}\tau},
\label{eq:jc_e1_0th} \\
\left[ \partial_{r_*} \bar{h}_{1,n}^{(i)} \right]_0
&=&
\frac{4}{f_0} P_{\delta,n}^{(i)}
- \frac{4}{f_0^2}\left[
  f_0 {\cal S}_{1,n}^{(i)}
+ f'_0 {\cal T}_{1,n}^{(i)}
\right]e^{-i\omega_{nm}\tau},
\label{eq:jc_e1_1st}
\end{eqnarray}
where $P_{\delta,n}^{(i)}$ are the terms $\propto\delta(r-r_0)$ in $P_{1,n}^{(i)}$, which are given by
\begin{eqnarray} \label{eq:def_Pdelta}
P_{\delta,n}^{(1)} &=&
\frac{1}{2}f_0^2 f'_0
\left[ \bar{h}_{1,n}^{(3)} \right]_0, \\
P_{\delta,n}^{(2)} &=&
\frac{1}{2}f_0 f'_0
\left[ f_0 \bar{h}_{1,n}^{(3)}
+ \bar{h}_{1,n}^{(2)} - \bar{h}_{1,n}^{(1)}
\right]_0, \\
P_{\delta,n}^{(4)} &=&
\frac{1}{4}f_0 f'_0
\left[ \bar{h}_{1,n}^{(4)} - \bar{h}_{1,n}^{(5)}
\right]_0, \\
P_{\delta,n}^{(8)} &=&
\frac{1}{4}f_0 f'_0
\left[ \bar{h}_{1,n}^{(8)} - \bar{h}_{1,n}^{(9)}
\right]_0, \\
P_{\delta,n}^{(i)} &=& 0 \quad \mbox{(others)}.
\end{eqnarray}

We may now easily obtain jump conditions for the second and higher
$r_*$ derivatives in a recursive manner: Formally differentiating
Eq.~(\ref{eq:eq_h1_expand2}) $k$ times with respect to $r_*$ we get
the jump relations
\begin{eqnarray}
\left[ \partial_{r_*}^{k+2} \bar{h}_{1,n}^{(i)} \right]_0
&=&
\left[ \partial_t^2 \partial_{r_*}^{k} \bar{h}_{1,n}^{(i)}
\right]_0
+ 4\left[ \partial_{r_*}^{k} P_{1,n}^{(i)} \right]_0 \nonumber\\
&=&-\left(\omega_{nm}f_0/{\cal E}_0\right)^2
\left[ \partial_{r_*}^{k} \bar{h}_{1,n}^{(i)}
\right]_0
+ 4\left[ \partial_{r_*}^{k} P_{1,n}^{(i)} \right]_0
\quad (k\ge 0),
\end{eqnarray}
where the second equality  is due to the harmonic dependence of $\bar h_{1,n}^{(i)}$ on $t$ along the orbit, expressed in Eq.\ (\ref{harmonic}). The jumps in the $t$ derivatives are also obtained in a straightforward manner,
by writing
\begin{equation}
\left[ \partial_t^k \partial_{r_*}^{k'}
\bar{h}_{1,n}^{(i)} \right]_0
=
\left( -i\omega_{nm}f_0/{\cal E}_0 \right)^k
\left[ \partial_{r_*}^{k'} \bar{h}_{1,n}^{(i)} \right]_0.
\end{equation}

The desired jump conditions in $\bar{h}_1^{(i)}$ and their derivatives are finally obtained by adding up the jumps in the two $n=1$ harmonics:
\begin{equation}
\left[ \partial_t^k \partial_{r_*}^{k'} \bar{h}_1^{(i)} \right]_0
=
\sum_{n=\pm 1}
\left[ \partial_t^k \partial_{r_*}^{k'} \bar{h}_{1,n}^{(i)}
\right]_0
\quad (k,k' \ge 0).
\end{equation}

%%%%%%%%%%%%%%%%%%%%%%%%%%%%%%%%%%%%%%%%%%%%%%%%%%%%%%%%%%%%%%%%%%%
%                            REFERENCES
%%%%%%%%%%%%%%%%%%%%%%%%%%%%%%%%%%%%%%%%%%%%%%%%%%%%%%%%%%%%%%%%%%%

\end{document}